\documentclass[fleqn,usenatbib]{mnras}

\usepackage{newtxtext,newtxmath}
\usepackage{mathptmx}
\usepackage{txfonts}
\usepackage{amsmath}

\usepackage[T1]{fontenc}

\DeclareRobustCommand{\VAN}[3]{#2}
\let\VANthebibliography\thebibliography
\def\thebibliography{\DeclareRobustCommand{\VAN}[3]{##3}\VANthebibliography}

\usepackage{graphicx}	
\usepackage{amsmath}	
\usepackage{amssymb}	

\def\la{\raise.5ex\hbox{$<$}\kern-.8em\lower 1mm\hbox{$\sim$}}
\def\ma{\raise.5ex\hbox{$>$}\kern-.8em\lower 1mm\hbox{$\sim$}}

\def\kms{${\rm km}\,{\rm s}^{-1}$}
\def\cm3{cm$^{-3}$}

\def\n0{$n_{\rm 0}$}
\def\B0{$B_{\rm 0}$}

\def\mum{$\mu$m}

\def\L12{$L_{12\mu m}$}
\def\F12{$F_{12\mu m}$}

\def\o3{[O\,{\sc iii}]\,$\lambda$5007}
\def\fe7l{[Fe\,{\sc vii}]\,$\lambda$6087}
\def\fe7{[Fe\,{\sc vii}]}
\def\fe10a{[Fe\,{\sc x}]\,$\lambda$6374}
\def\fe10b{[Fe\,{\sc x}]}

\title[The size of the CLR]{The size of the Coronal Line Region in Active Galactic Nuclei: a MUSE perspective}

\author[A. Rodr\'{\i}guez-Ardila et al.]{
A. Rodr\'{\i}guez-Ardila,$^{1,2,3}$\thanks{E-mail: aardila@lna.br (AR-A)}
M. A. Fonseca-Faria,$^{1,2}$
L. G. Dahmer-Hahn,$^{4}$
A. Prieto,$^{5,6,7}$
R. Riffel, $^{8}$ 
\newauthor R. A. Riffel$^{9}$
\\
$^{1}$Laborat\'orio Nacional de Astrof\'{\i}sica, Rua dos Estados Unidos, 154, Itajub\'a, MG, Brazil\\
$^{2}$Instituto Nacional de Pesquisas Espaciais, Av. dos Astronautas, 1758 - Jardim da Granja S\~ao Jos\'e dos Campos/SP - CEP 12227-010, Brazil\\
$^{3}$ Observatório Nacional, Rua General José Cristino 77, CEP 20921-400, São
Cristóvão, Rio de Janeiro, RJ, Brazil. \\
$^{4}$Shanghai Astronomical Observatory, Chinese Academy of Sciences, 80 Nandan Road, Shanghai 200030, China \\
$^{5}$ Instituto de Astrofísica de Canarias (IAC), La Laguna 38205, Tenerife, Spain\\
$^{6}$ Departamento de Astrofísica, Universidad de La Laguna, La Laguna 38205, Tenerife, Spain\\
$^{7}$ Universit\"ats-Sternwarte M\"unchen, D-81679 M\"unchen, Germany \\
$^{8}$ Departamento de Astronomia, Universidade Federal do Rio Grande do Sul. Av. Bento Gonçalves 9500, 91501-970 Porto Alegre, RS, Brazil\\
$^{9}$ Departamento de Física, Universidade Federal de Santa Maria, Centro de Ci\^encias Naturais e Exatas, 97105-900 Santa Maria, RS, Brazil\\
}

\date{Accepted XXX. Received YYY; in original form ZZZ}

\pubyear{2015}

\begin{document}
\label{firstpage}
\pagerange{\pageref{firstpage}--\pageref{lastpage}}
\maketitle

\begin{abstract}

We investigated by means of MUSE/VLT observations the true size of the coronal line region (CLR) in a local sample of nine active galactic nuclei known for displaying prominent coronal emission. Our analysis show that the CLR is extended from several hundred parsecs to a few kiloparsecs in the lines of [\ion{Fe}{vii}] (IP=99 eV) and [\ion{Fe}{x}] (IP=235 eV). In all cases, the coronal emission is closely aligned along the radio-jet axis and constrained to the limits of the [O\,{\sc iii}] ionisation cone. Besides the nuclear emission, secondary emission peaks in [\ion{Fe}{vii}] and [\ion{Fe}{x}] are found along the extended emission, with a shallow decrease of the line intensity with increasing distance from the AGN. Both facts suggest the action of an additional excitation mechanism besides nuclear photoionisation for the origin of the coronal gas. This is further supported by the fact that in some sources the extended coronal emission accounts for more than 50\% of the total emission and by the high degree of gas excitation in the off-nuclear region. A positive trend between the coronal line luminosity and the jet power points to shocks induced by the jet passage as the key mechanism to produce and excite this gas.  We provide the first estimate of the [\ion{Fe}{x}] coronal  gas  size, being in  the kpc range. Our results stress the importance of the CLR as a key ingredient that should be fully considered in models trying to explain the physics of the narrow line region in AGN.     
 
\end{abstract}




\section{Introduction}

The  coronal line region (CLR) of active galactic nuclei (AGN) is traditionally regarded as the place where the high-ionisation forbidden lines (i.e., those lines with ionisation potential $\geq 100$~eV) are formed \citep {appenzeller_1991}. It was formerly linked to an intermediate region between the broad line region (BLR) and the narrow line region (NLR).  This conclusion steamed from the fact that the high ionisation lines have critical densities for the collisional de-excitation of the order of or larger than $\sim 10^7$~cm$^{-3}$ \citep{de_robertis_1984}.  \citet{erkens_1997} conducted the first study aimed at determining the location of the CLR by means of integrated emission line spectra of 15~AGN. They found that the CLR should be compact ($< 1\arcsec$, 0.0010$ < z <$ 0.0562) and located in the innermost portion of the NLR. They also suggested that the highest ionisation gas was involved in outflows because the peak centroid of the coronal lines was very often strongly blueshifted, with the blue wing of the line profiles displaying conspicuous asymmetries. These characteristics were absent in medium and low ionisation lines. 

Evidence of an extended CLR was previously reported in radio-loud sources. \citet{tadhunter_1987, tadhunter_1988} detected [\ion{Fe}{vii}]~$\lambda$6087 in the radio-galaxy PKS\,2152-69 at $\sim$8~kpc from the core of the galaxy and co-spatial to the radio jet, suggesting that the extended coronal emission is produced by the propagation of the radio jet and its interaction with the ISM. Later,  \citet{golev_1994,  golev_1995} studied the Seyfert~1 galaxy NGC\,3516. They employed imaging with a filter centred in the  line of [\ion{Fe}{vii}]~$\lambda$\,6087, detecting coronal emission at a distance of $\sim$1.5~kpc from the AGN. Recently, \citet{riffel21} reports coronal line emission seen  up to 1.75~kpc from the nucleus in the powerful radio-source Cygnus~A.

The first systematic study on the extent and morphology of coronal emission in radio-weak AGN appeared only in the early-2000s. \citet{maiolino+00} reported the first detection of the counter-cone in Circinus  by means of the [\ion{Si}{vi}]~1.963~$\mu$m coronal emission line. Later,  \citet{prieto_2005} found, by means of adaptive optics (AO) observations, extended   [\ion{Si}{vii}]~2.48~$\mu$m coronal  emission in 4 AGN (including Circinus) at scales varying from 30\,pc to 200\,pc, aligned with the ionisation cone. 
Thereafter, \citet {ardila_2006} using seeing-limited spectroscopy studied the spectra of 6 radio-weak AGN. They found, for the first time in the literature, extended coronal emission with double peak structure in two objects of their sample (NGC\,1068 and NGC\,1386). This result was interpreted in terms of gas associated with outflows. Additionally, the authors employed models that combine the effects of shocks and photoionisation by the AGN \citep{Contini_2001} to reproduce the observed emission line ratios. They confirmed the need of both mechanisms to explain the coronal emission at tens to hunderds of parsec from the centre.

Studies with integral field unit (IFU) spectrographs  brought more information about the morphology of the CLR. \citet{muller_2011} used  SINFONI/VLT assisted with AO to spatially resolve the CLR in six nearby AGN. They found high-ionisation gas extending between 80\,pc to 150\,pc from the AGN with signatures of outflows, whose kinematic and morphology were modelled, from which it was concluded that the gas is in outflow with velocities of up to 1500~\kms. The velocity field suggested a CLR of  biconic morphology. \citet{mazzalay_2013} employed AO NIFS/Gemini to study the Seyfert~2 galaxy NGC\,1068. The results indicate coronal gas emission up to 170\,pc to the northeast and southeast of the galaxy's nucleus, aligned with the radio-jet, in line with \citet{prieto_2005} results.  Other notable examples of extended coronal emission at spatial scales of a few hundred parsecs and associated to shocks were reported by \citet{riffel+21} in six local AGN, in NGC\,1386~\citep{ardila_2017_ngc_1386} and ESO\,428-G14~\citep{may_2018}.  

Previous works on radio-weak AGNs, such as those mentioned above, probed extended coronal gas emission on scales of a few hundred parsecs in the near-infrared region (NIR). We remark that they were carried out using data collected with IFUs that have field-of-view (FoV) of $\sim 3\arcsec \times 3\arcsec$. Therefore, it was not possible, at that time, to confirm if the coronal gas extends further out. The first spectroscopic evidence of resolved, extended CL gas at several hundred of parsecs was provided by \citet{ardila_2020} in the Circinus galaxy. They reported an extended outflow of high ionised gas in that AGN by means of the CL [\ion{Fe}{vii}]~$\lambda$6087 using the Multi Unit Spectroscopic Explorer (MUSE) at VLT. This result was possible thanks to the much larger FoV, of 1$\arcmin \times 1\arcmin$, of that IFU and increased the coronal line region size by an order of magnitude as compared with \citet{prieto_2005} results. The physical conditions of the gas in Circinus show that the extended coronal emission is likely the remnant of shells inflated by the passage of a radio jet. This scenario was further supported by \citet{fonseca-faria_2021}, who nicely reproduced the line ratios observed in the extended CLR of that galaxy by means of models that combine the effect of ionisation by the central source and shocks. 

\citet{negus_2021,negus+23}  point to the detection of CLs up to several kpc by applying a data binning method to a large sample of MANGA data. Yet, further analysis of the detection and location of the emission on the specific cases would be needed. Their results point out CLR reaching 1.3-23~kpc from the galactic centre, with an average distance of 6.6~kpc. They found that ionising photons emitted by the central continuum source (i.e., AGN photoionization) primarily generate the CLs, and that energetic shocks are an additional ionization mechanism that likely produce the most extended CLRs they measured.  

Despite the above results, to the best of our knowledge, no systematic studies of the coronal line extension in nearby sources using  IFUs with large FoV, of tens of arcseconds squares, have yet been made. This prompted us to investigate the size of the CLR in $bona-fide$ AGN. The detection of extended CL in AGN is important because that emission traces the footprint of X-ray
gas in AGN and therefore, can potentially
be used to measure the kinematics of the extended X-ray emission gas \citep{trindade_2022}. Moreover, the detection of extended CL emission associated with the radio emission -jets could be used as  a quantitative measurement of the jet power by means of the CL gas kinematics \citep[e.g.,][]{may_2018}.

Thus, taking advantage of the MUSE capabilities, we started a program to examine the actual extension of the CLR in different AGN environments. Here, we analyse a sample of 9 radio-weak, mid- to low-luminosity Seyfert~2 galaxies, with previous report of extended coronal emission at scales of a few hundreds parsecs. In contrast, \citet{muller_2011} sample was composed mostly of Seyfert~1 galaxies. So a comparative analysis between both types would be possible.

In Sect.~\ref{sec:obs} we will describe our sample, observations and data treatment. Sect.~\ref{sec:extension} characterises the coronal gas extension, geometry and morphology of the CLR as well as the relationship with emission at other wavelength intervals. Sect~\ref{CapSizeCoronal} discusses the size of the coronal line region of the sample and its relationship with the radio emission. Sect~\ref{sec:degree_gas} studies the ionisation degree of the emission gas. Final remarks are in Sect~\ref{sec:final}. Throughout this paper, $H_{\rm 0}$ = 70~km~s$^{-1}$~Mpc$^{-1}$, $\Omega_{\rm m}$ = 0.30, and $\Omega_{\rm vac}$ = 0.70, have been adopted.

\section{Sample and Observations}
\label{sec:obs}

The criteria employed to select the sample were as follows. (i) The AGN should be classified as Seyfert~2 in the optical region to mask the emission from the broad line region. Thus, broad components in the permitted lines should have FWHM $<$1000~\kms\ and cannot be attributed to emission from that region. (ii) The galaxies should be close enough ($z <$ 0.013) to reach angular resolutions smaller than 250~pc per arcsecond. (iii)  Previous confirmed evidence of an extended CLR in the object. (iv) Targets should have MUSE observations available. As a starting point, the AGN that are part of the MAGNUM (Measuring Active Galactic Nuclei Under MUSE Microscope) survey \citep{mingozzi_2019} were employed. Other galaxies observed under additional programs were also included. See below. 

The nine Seyfert~2 galaxies fulfilling the above criteria are listed in Table~\ref{tab:info_galaxias} along with the source coordinates (columns 2 and 3), redshift (column 4), plate scale (column 5), ESO file ID (column 6), date of observation (column 7), average seeing during the observations (column 8), exposure time (colum 9), and pipeline version employed for the data reduction (column 10). The targets are listed in increasing order of plate scale. Circinus, IC\,5063, NGC\,1068, NGC\,1386, and NGC\,5643, were from the MAGNUM Survey. It aims at probing the physical conditions and ionisation mechanism of 9 nearby ($D < 52$ Mpc) AGNs, where $D$ is the distance to the galaxy corrected by the Galactic Standard of Rest (GSR) as published in NED.

\begin{figure*}
    \includegraphics[width=17cm]{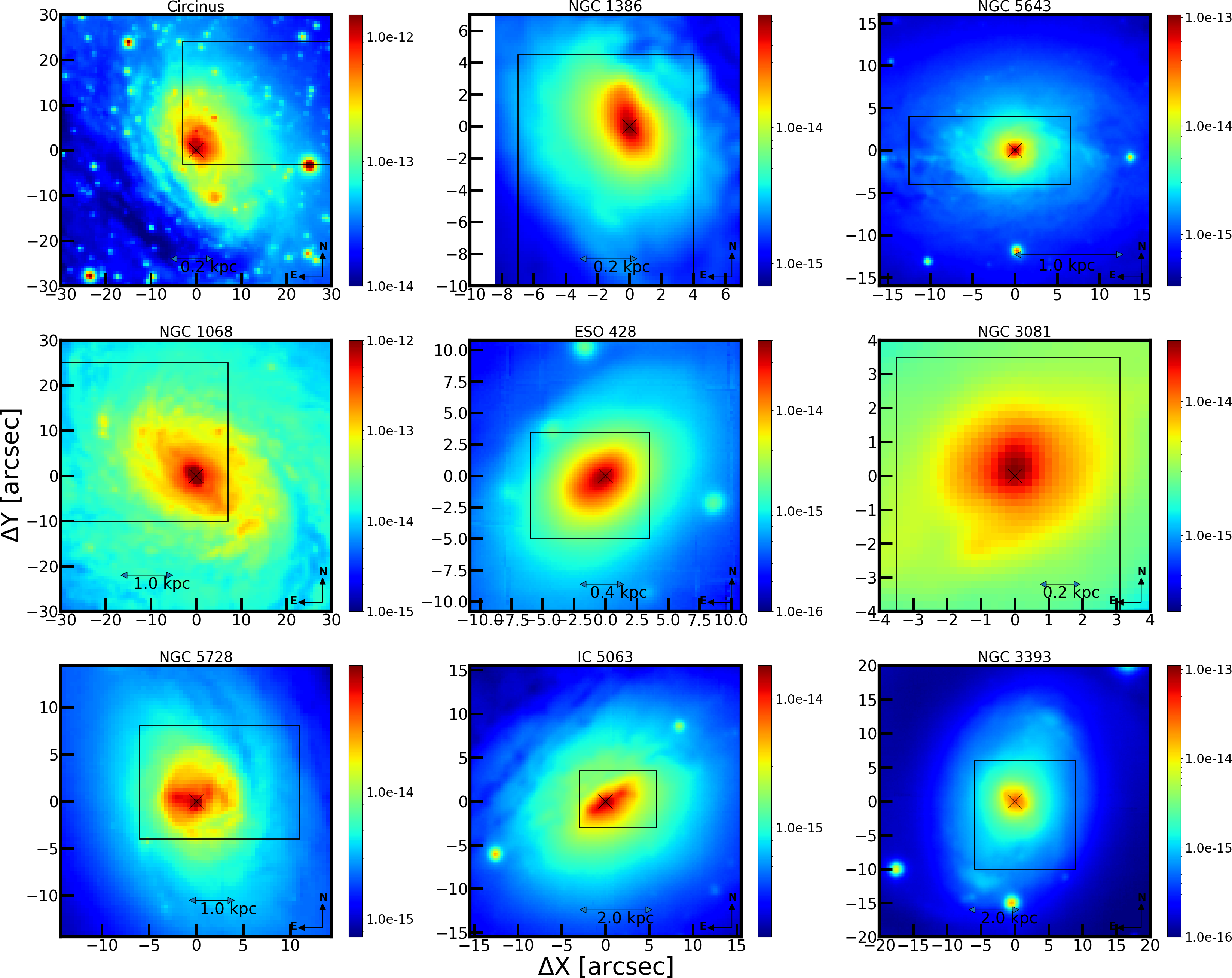}
    \caption{Continuum emission in the rest wavelength interval 5100-6000~\AA\ for the galaxy sample. The black square is the MUSE field-of-view used for the analysis of the CLR in Figures~\ref{fig:ext_circinus} to~\ref{fig:ext_NGC3393}. The colour bar is in units of erg\,cm$^{-2}$\,s$^{-1}$\,spaxel$^{-1}$}
    \label{fig:continuum}
\end{figure*}

Moreover, NGC\,3393, NGC\,5728, NGC\,3081, and ESO\,428-G14, were included because of previous reports of prominent and extended coronal emission and data availability in the MUSE database. 
All data files were downloaded from the MUSE Science Archive as detailed in Table~\ref{tab:info_galaxias}.   The observations are seeing-limited, with the MUSE Field-of-view covering 1$\arcmin \times 1\arcmin$ with a sampling of 0.2\arcsec/spaxel. Each data cube is restricted to the wavelength range 4750–9350~\AA\ in the galaxy system.  In all cases, we adopted the reduced cube provided by the ESO Quality Control Group \citep{Hanuschik_2020}.  Figure~\ref{fig:continuum} shows the continuum image of each galaxy in the sample, integrated in the wavelength region 5100-6000~\AA. 

\begin{table*} \centering  
\caption{Sample of objects, basic properties and observing details.} \label{tab:info_galaxias}
\begin{tabular}{lcccccccccc} \hline \hline

Object     &  RA        & DEC  & $z^*$ & Scale$^*$ &  ESO File (ID) & Observing  & Seeing & Exposure & Pipeline \\ 
    &  (J2000)   & (J2000)    &   &  (pc/\arcsec) & & date (dd/mm/yr) & (\arcsec) & Time (s) & version \\
\hline
Circinus   &  14:13:09   &   -65:20:21  &  0.001448  & 18   & (1) & 11/03/2015   & 0.78  & 1844  &   1.4  \\
NGC\,1386 & 03:36:6.2 & -35:59:58 & 0.003052 & 55 & (2) & 13/11/2014  &  0.61 & 2000 & 1.4 \\

  &   &   &   &  & (3) &  13/11/2014 & 0.66 & 2000 & 1.4  \\

  &   &   &   & & (4) &  13/11/2014 & 0.64 & 4000 & 1.6.1 \\
NGC\,5643   &  14:32:40    &     -44:10:27     &  0.003999    &      74     & (5) & 14/05/2015  & 0.56   & 3288  & 1.4         \\
NGC\,1068   &    02:42:40   &      -00:00:47   &  0.003793 & 79    & (6) & 01/12/2014 & 0.90    & 735  & 1.4   \\   
ESO\,428-14  &   07:16:31    &     -29:19:28     &  0.005664  &     103 & (7) & 19/04/2016   & 1.12  & 2744  & 1.6.1    \\
NGC\,3081 & 09:59:29.5 & 22:49:35 &  0.008149 & 155 & (8) & 23/04/2017  & 0.71  & 3600  & 1.6.4  \\
NGC\,5728   &   14:42:23    &     -17:15:11     &  0.009186 & 187     & (9) & 03/04/2016   & 0.69   & 4747  &  1.6.4    \\
IC\,5063    &   20:52:02     &     -57:04:07   &  0.011348   &    231    & (10) & 23/06/2014   & 0.78 & 2240  &  1.6.1   \\
NGC\,3393   &   10:48:23     &     -25:09:43  & 0.012786  &  252   & (11) & 31/01/2017 & 0.79    & 3294  &  1.6.4   \\
\hline

\multicolumn{10}{l}{$^*$ Source: NASA/IPAC Extragalactic Database (NED)} \\
\multicolumn{10}{l}{(1) ADP.2016-06-14T18:02:17.657; 
(2) ADP.2016-06-21T07:59:05.164; (3) ADP.2016-06-21T07:59:05.212; (4) ADP.2017-03-27T12:08:50.529;}\\
\multicolumn{10}{l}{(5) ADP.2016-06-17T15:46:46.093; (6) ADP.2016-06-17T08:44:56.817; (7) ADP.2016-09-22T21:00:32.895; (8) ADP.2017-05-24T11:10:28.394;}\\
\multicolumn{10}{l}{(9) ADP.2017-06-14T09:12:09.461; (10) ADP.2016-07-22T15:57:54.788; (11) ADP.2017-03-21T11:15:10.512}\\
\end{tabular} 
\end{table*} 

The data analysis were carried out using custom {\sc python} scripts, following an approach similar to that described in \citet{ardila_2020} and \citet{fonseca-faria+23}. Below, we briefly describe the procedures employed.

In order to enhance the signal-to-noise ratio (SNR), in particular in regions of weak emission, a binning was performed in some of the cubes. The binning applied was 3$\times$3 spaxels (Circinus and NGC\,1068) and 2$\times$2 spaxels (NGC\,5728). The other galaxies had their original binning preserved.
We then fit and subtracted the stellar continuum in the range 4770 -- 7200~\AA. This procedure is necessary to fully recover the fluxes 
of the H\,{\sc i} lines affected by the underlying stellar population. Moreover, some weak lines may also be strongly 
diluted by the stellar continuum. To this purpose, the stellar population synthesis code {\sc starlight} \citep{cid_2005} 
was employed, together with the set of stellar populations of E-MILES \citep{vaz_2016}.   In this work we will not analyze the stellar population of the sample. We will use {\sc starlight} just to subtract the stellar continuum.

In order to measure the flux, centroid position and the full width at half maximum (FWHM) of the emission lines at each 
spaxel we fit Gaussian functions to individual lines or to sets of blended lines. To this aim, we employed a set of custom 
scripts written in {\sc python} by our team.  
Usually, one or two Gaussian components at the most were necessary to reproduce the observed profiles. In this process, some 
constraints were applied. 
For instance, the theoretical [\ion{O}{iii}] and [\ion{N}{ii}] line  flux ratios $\lambda5007$/$\lambda4959$ and  $\lambda6583$/$\lambda6548$ were used 
for each component. Moreover, lines emitted by the same ion were constrained to have the same width and intrinsic wavelength 
separation. For example, the [\ion{S}{ii}]~$\lambda\lambda$6717,6731 doublet was constrained to have the same width, same centroid velocity and a relative theoretical wavelength separation of 14.4~\AA. Using that approach, the strongest emission lines detected in the cubes, that is, H$\beta$, [\ion{O}{iii}]$\lambda\lambda$4959,5007, [\ion{Fe}{vii}]~$\lambda$6087, [\ion{Fe}{x}]~$\lambda$6374,  H$\alpha$, [\ion{N}{ii}]$\lambda\lambda$6548,6583, and [\ion{S}{ii}]$\lambda\lambda$6716,6731, were fitted from the continuum-subtracted cube. The criterion for the best solution was the minimum value of the reduced-$\chi$2.

\begin{figure}
    \includegraphics[width=\columnwidth]{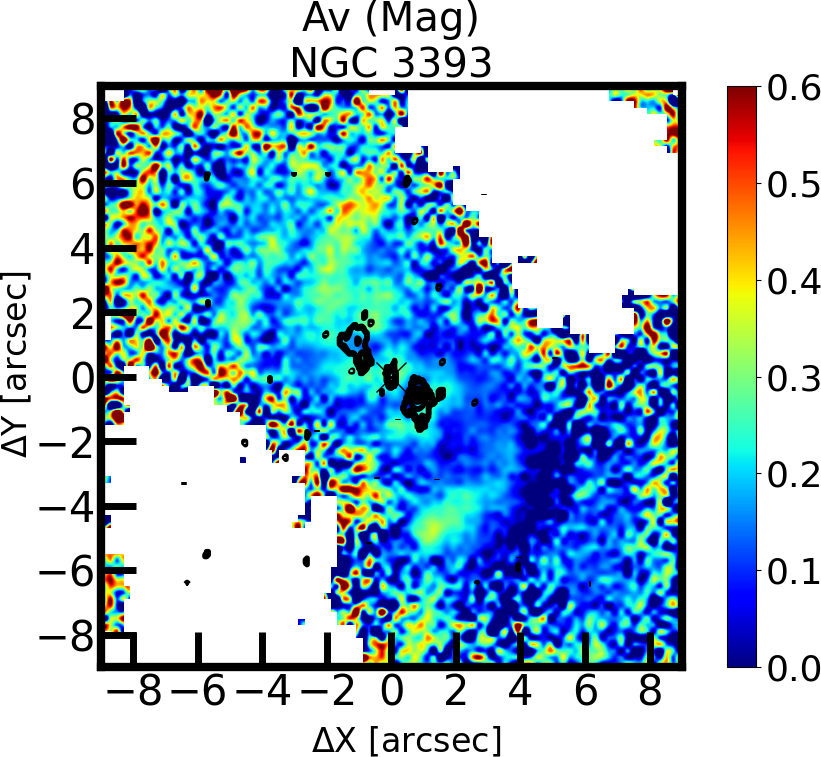}
    \caption{Extinction map for NGC\,3393 determined using the H$\alpha$/H$\beta$ ratio, the CCM extinction law \citep{cardelli_1989} and an intrinsic emission line flux ratio of 3.1. The black contours correspond to the observed radio emission at 3.6~cm, taken with the Very Large Array (VLA) and obtained from archival data. The black cross indicates the AGN position. }
    \label{fig:Av_NGC3393}
\end{figure}

After measuring the emission line fluxes of H$\alpha$ and H$\beta$, we determined the total extinction due to the Galaxy and that intrinsic to the AGN, quantified by the colour excess E(B-V). To this purpose we used Equation~\ref{eq1}, which takes into account the CCM extinction law \citep{cardelli_1989},

\begin{equation}
    E\left(B-V\right)_{\frac{H\alpha}{H\beta}} $=$ -2.31 \times log \left(\frac{3.1}{\frac{H\alpha}{H\beta}}\right)
	\label{eq1}
\end{equation} 

where $H\alpha$/$H\beta$ is the observed emission line flux ratio between these two lines. 
We measured the fluxes of these two H\,{\sc i} lines at the spaxels where they were simultaneously detected at 3$\sigma$-level. The intrinsic value adopted was 3.1 assuming case B recombination, $T$ = $10^{4}$~K, and $N_{\rm e}$ = 10$^{4}$ cm$^{-3}$. These values are more suitable to the typical gas densities found in AGN \citep{osterbrock_2006} and accounts for the effect of collisional excitation.  
At the spaxels where a narrow and a broad component were fit, for consistency we used  only the fluxes of the narrow components.  We found our extinction maps very consistent with the ones presented by \citet{mingozzi_2019} for the galaxies common to both works. For that reason they will not be shown here. For NGC\,3393 the resulting extinction map is in Figure~\ref{fig:Av_NGC3393} while those for NGC\,5728, NGC\,3081, and ESO\,428-G14, are presented in Figure~\ref{figAped:Av_NGC5728_ESO428_NGC3081}.

The above procedure allowed us to produce emission line flux maps, free of extinction, as well as maps for the parameters we are interested in (velocity, FWHM, and emission line flux ratios). They will be presented in the different sections throughout this paper.

\section{The size of the emission region of the coronal gas}
\label{sec:extension}

The study of the coronal gas extension in the galaxy sample opens the possibility to register the highest excitation component of the ionised gas.  For instance, it allows us to investigate the mechanisms that are dominating the gas ionisation of these objects (photoionisation by the AGN,  shocks or both) and the extent to which the high-ionisation gas is also involved in the feedback process. Moreover, the detection of coronal gas at scales of several hundred parsecs from the AGN and beyond reveals the presence of an underlying X-ray continuum necessary to ionise the gas. We present here the results of the coronal gas emission as traced by the lines of [Fe\,{\sc vii}]\,$\lambda6087$ (IP=99~eV, where IP is the ionisation potential required to form the ion emitting the line) and [Fe\,{\sc x}]\,$\lambda6375$ (IP=235~eV). Hereafter we will refer to these two lines as [Fe\,{\sc vii}] and [Fe\,{\sc x}], respectively. They are the strongest coronal lines detected in the wavelength interval covered by MUSE in our sample. We aimed at  characterising the maximum extension found for the high ionisation lines and the gas morphology of the coronal emission in each object.
We are aware of the presence of other coronal lines such as [\ion{Fe}{xi}]~$\lambda$7892 (IP = 262~eV) and [\ion{S}{xii}]~$\lambda$7611 (IP = 564~eV) in some of the spectra of the galaxy sample. However, in all cases, they were detected only at the nucleus or just barely resolved. For that reason the latter lines are not the focus of this work and we left their analysis for a future publication although we will report their detection if that is the case. 

Flux distribution maps for [Fe\,{\sc vii}] and [Fe\,{\sc x}] were constructed for the galaxy sample and are shown in Figures~\ref{fig:ext_circinus} to~\ref{fig:ext_NGC3393}. 

In order to detect fainter coronal emission, farther away from the one visible at the resolved spaxels, we scanned the resulting map in the search for fainter emission, not detectable in the 1$\times$1 or 3$\times$3 binning or by changing the flux scale. This was done using the {\sc qfitsview} software\footnote{https://www.mpe.mpg.de/~ott/QFitsView/index.html}, which allows one to change the integration aperture (from one to several spaxels). In all cases, CL emission beyond the one clearly visible in the maps was found further away from the nucleus. Thus, in the CL maps of Figures~\ref{fig:ext_circinus} to~\ref{fig:ext_NGC3393}, the grey circle shows the aperture employed to detect that fainter emission and the line plotted inside the circle corresponds to the CL integrated over all the region covered by the circle. In order to consider a true detection, the line must be visible at a signal-to-noise $>$ 3 in the corresponding aperture.

Below, we present the most relevant characteristics of the coronal emission in the galaxy sample from the point of view of gas morphology and extension.  

\subsection{Circinus}

\begin{figure}
    \centering
    \includegraphics[width=8.6cm]{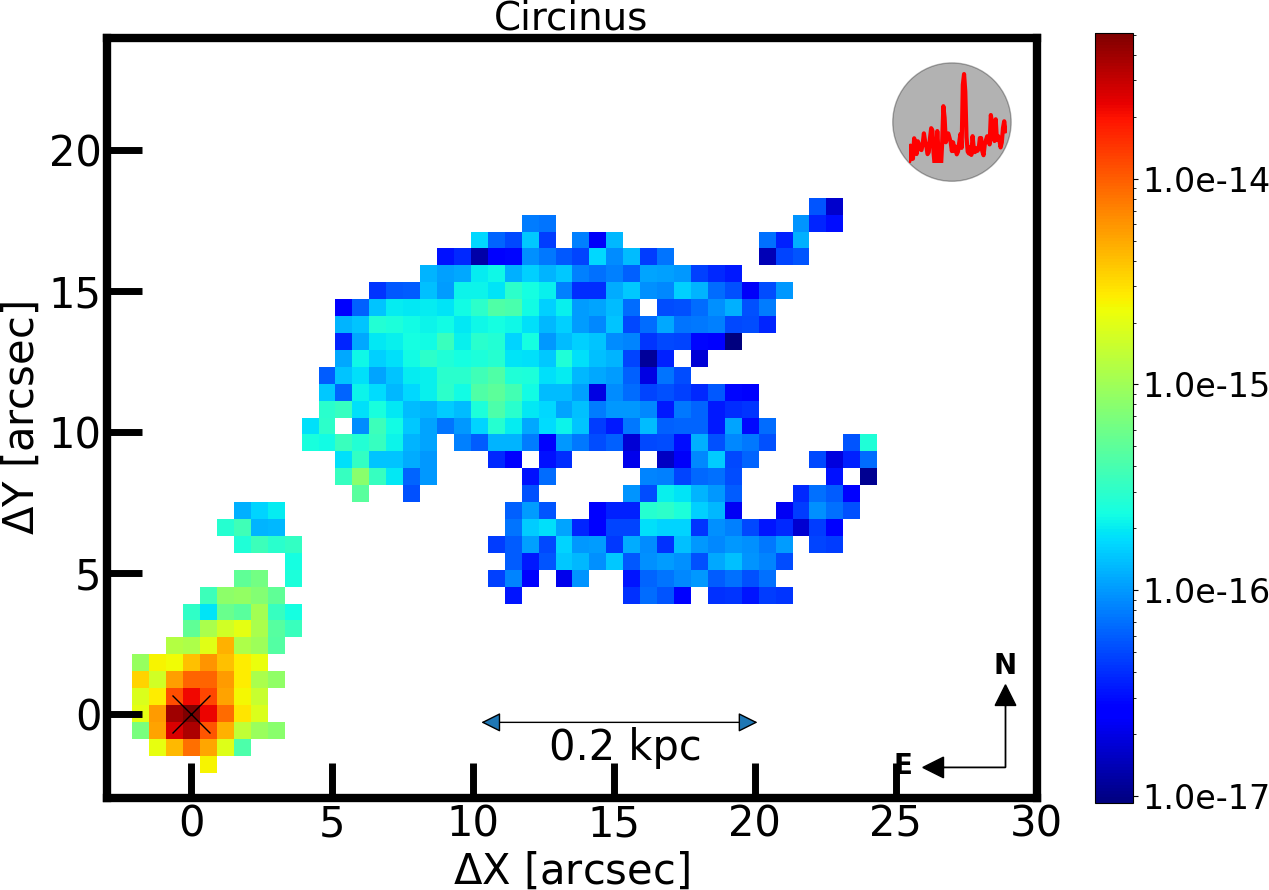}
    \centering
    \includegraphics[width=8.6cm]{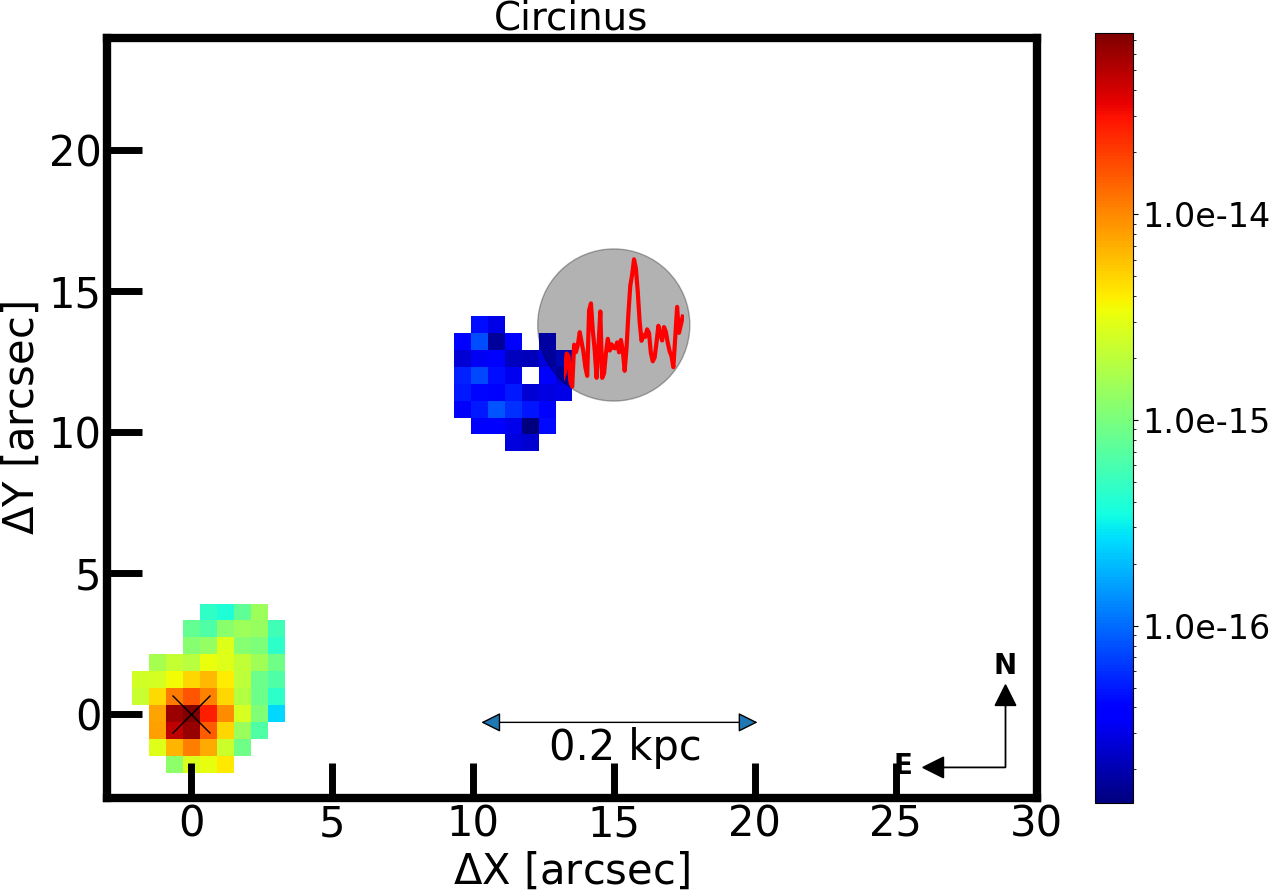}
    \caption{Coronal emission in the Circinus galaxy. In the upper panel,  the flux distribution map of [Fe\,{\sc vii}] (IP = 99~eV) detected with a 3$\times$3 binning is shown along with the spectrum around the [\ion{Fe}{vii}] line at the largest distance from the AGN. The detection of CL emission at the farthest locations was done by scanning the MUSE spectra using a large aperture (see the text at the beginning of section~\ref{sec:extension} for an explanation). In this case, the signal within a circular aperture of  1.8\arcsec\ in radius (gray circle) was employed.  The flux map distribution of [\ion{Fe}{x}] (IP = 235~eV) and the line detection at the farthest distance from the AGN are shown in the bottom panel. For its detection, we integrated a circular aperture of radius 2.4\arcsec. In order to improve the visualization of the [\ion{Fe}{x}] line, the [\ion{O}{i}]~$\lambda$6363 emission, which is blended to the former, was previously removed. In both maps, the colour bar is in units of erg\,cm$^{-2}$\,s$^{-1}$.}
    \label{fig:ext_circinus}
\end{figure}

The optical coronal line emission of Circinus was previously studied by \citet{fonseca-faria_2021}, mainly regarding the emission of [\ion{Fe}{vii}]. In Figure~\ref{fig:ext_circinus} we show the emission of [\ion{Fe}{vii}] in the nuclear region, with flux values of the order of $\sim10^{-13}$~erg\,s$^{-1}$\,cm$^{-2}$. A conspicuous elongated structure is observed in the north-northwest direction from the nucleus and up to a distance of 150~pc. Farther out, we identified extended emission that starts at 250\,pc and ends at 700\,pc elongated to the northwest of AGN. This extended emission has an average flux of the order $10^{-15}$~erg\,s$^{-1}$\,cm$^{-2}$. In the region where the most distant extended [\ion{Fe}{vii}] emission was found, there is a small extension at 600~pc (to the northwest). At the end of this structure, [\ion{Fe}{vii}] was identified at a distance of 684~$\pm$~36~pc. In order to detect it,  we integrated the signal in a circular aperture of 1.8\arcsec\ in radius. We adopted as uncertainty the radius of the integrated region in the circular aperture. This procedure was employed in all the galaxies of the sample. Previously, \citet{oliva_1999}, using optical long-slit spectroscopy, had already
reported extended [Fe\,{\sc vii}] emission up to $\sim$400~pc from the center at
a PA = 318$\degr$.

In the bottom panel of Figure~\ref{fig:ext_circinus} we present  the emission map of the [\ion{Fe}{x}] line. Extended emission around the nucleus is detected, with a morphology and flux very similar to that of [\ion{Fe}{vii}]. In the map produced with the binning of 3$\times$3 spaxels, we identified emission of [\ion{Fe}{x}] distributed in a region size of approximately 14 spaxels, with a maximum flux of $5\times10^{-16} $~erg\,s$^{-1}$\,cm$^{-2}$  at 300~pc from the AGN. However, more extended [\ion{Fe}{x}] emission was identified farther out after integrating a circular region with a radius of 2.4\arcsec at a distance of 408~$\pm$~48\,pc. This value represents an increase of more than an order of magnitude in the size of the [\ion{Fe}{x}] emission as \citet{ardila_2006} reported a maximum extension of 30~pc in that AGN.

We notice that in addition to the above two lines, extended emission of [\ion{Fe}{xi}] and [\ion{S}{ix}] was observed, with maximum extension of $\sim$70~pc from the AGN. These latter two lines have ionisation potential larger than 300~eV, implying the presence of a locally very hard continuum emission at such distances from the central engine. 
 
\subsection{NGC\,1386}

\begin{figure} 
    \centering
    \includegraphics[width=8.6cm]{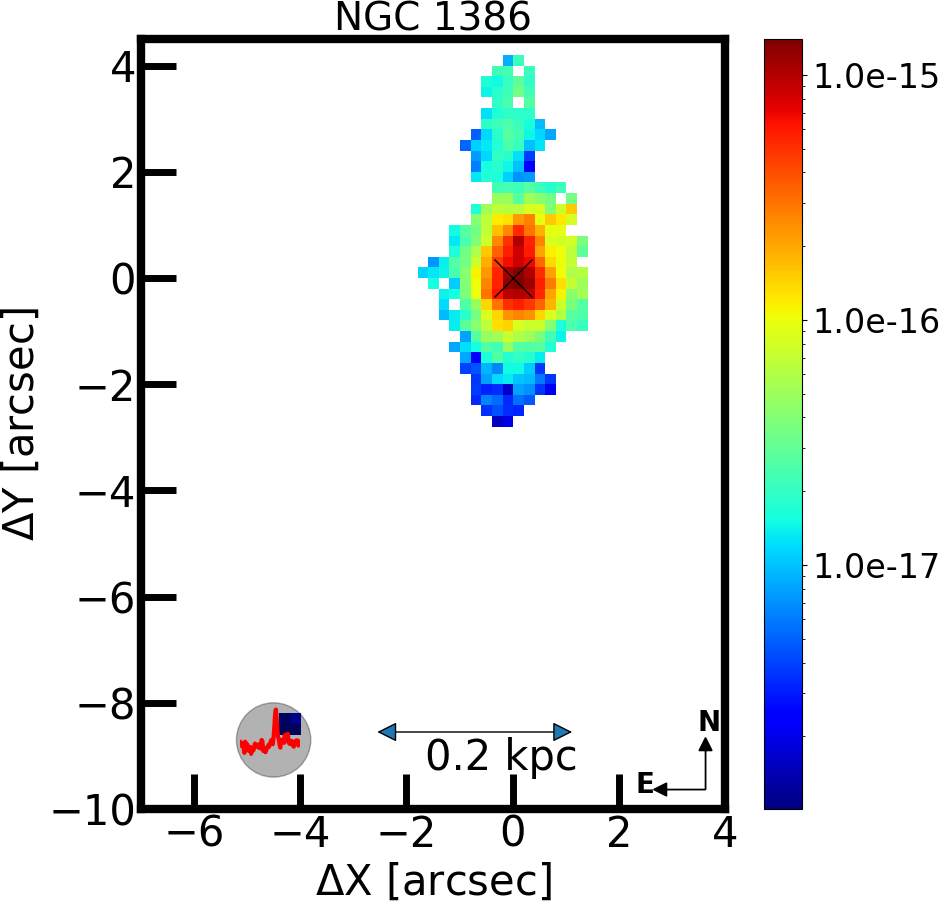} 
    \includegraphics[width=8.6cm]{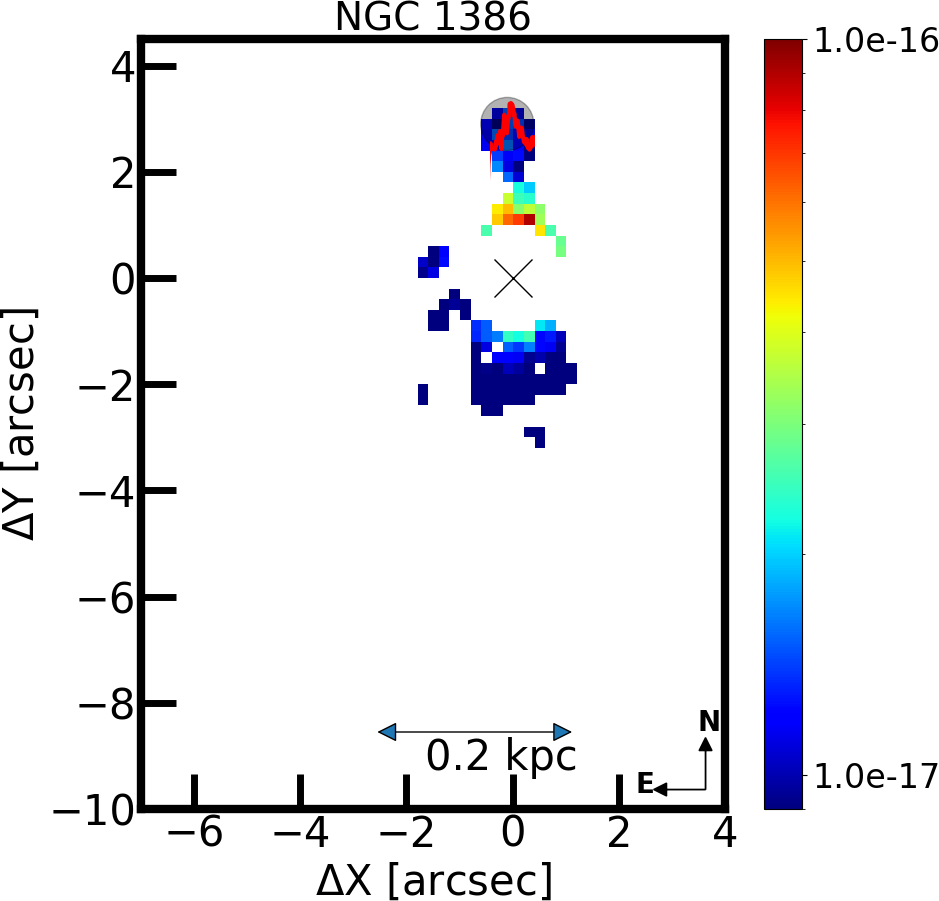} 
    \caption{Similar to Figure~\ref{fig:ext_circinus} for NGC\,1386. The top panel shows the emission map of [\ion{Fe}{vii}] and the spectrum of the most extended emission observed using a circular aperture of radius 0.6$\arcsec$. The bottom panel shows the flux map of [\ion{Fe}{x}]. The nuclear emission was masked because of the complex blend of [\ion{Fe}{x}] and [\ion{O}{i}], making the separation of the individual components very uncertain. The most distant emission was detected after integrating the signal in a circular aperture of 0.4$\arcsec$ in radius. The colour bars are in units of erg\,cm$^{-2}$\,s$^{-1}$.}
    \label{fig:ext_NGC1386}
\end{figure}

NGC\,1386 is one of the galaxies studied by~\citet{ardila_2006}, where extended coronal emission at a scale of $\sim$100~pc from the AGN was first reported. Prominent [\ion{Fe}{vii}] emission was found, with a conspicuous double-peak profile both at the nucleus and in the extended region, suggesting that part of the coronal gas is participating in an outflow. This scenario was later confirmed by~\citet{ardila_2017_ngc_1386} using AO SINFONI/VLT observations. They found extended [\ion{Si}{vi}]~1.963~$\mu$m along the N-S direction to distances reaching 100~pc south of the AGN. The line profiles of the silicon emission were highly complex, with widths implying gas velocities amounting 600~km\,s$^{-1}$. Extended coronal emission
is also marginally reported north of the AGN. \citet{ardila_2006} also reported extended [\ion{Fe}{x}]~$\lambda$6374 and [\ion{Fe}{xi}]~$\lambda$7889 emission up to $\sim$80~pc, north and south of the nucleus.

The MUSE data cube shows the presence of coronal lines of [\ion{Fe}{vii}], [\ion{Fe}{x}], and [\ion{Fe}{xi}] (see Figure~\ref{fig:ext_NGC1386} for the flux distribution of the former two lines). Both [\ion{Fe}{vii}] and [\ion{Fe}{x}] are conspicuous in the inner 1$\arcsec$ radius region centred at the nucleus, where the emission peaks. Moreover, extended emission to the the north and south of the AGN, similar to the gas distribution mapped through the [\ion{Si}{vi}]~1.963~$\mu$m line \citep{ardila_2017_ngc_1386} is observed. Both the north and south extensions appear to be connected to the central, nuclear emission. We notice that the [\ion{Fe}{vii}] emission is more extended than that of [\ion{Si}{vi}] to the north. The [\ion{Fe}{vii}] to the south is detected up to 2.5$\arcsec$ or 150~pc of the AGN. Moreover, we found, for the first time, a blob of [\ion{Fe}{vii}] emission at 519$\pm$32~pc SE of the AGN. This blob seems disconnected to the nucleus or any other prominent structure. The blob is coincident spatially with a much larger blob of emission, already reported by Schimdt et al. (2003) in the [\ion{O}{iii}]~$\lambda$5007 line as well by \citet{jones+21} in both soft and hard X-ray emission bands. 

The [\ion{Fe}{x}] map (bottom panel of Figure~\ref{fig:ext_NGC1386}) confirms that this emission is extended. However, the flux distribution within the inner 1$\arcsec$ is not shown because the  line is strongly blended with [\ion{O}{i}]~$\lambda$6363. Separating both contributions is rather uncertain due to the presence of double-peaks in both lines. It can be seen that the extended emission is more compact than that of [\ion{Fe}{vii}] but still extended at scales of a few hundred parsecs. Indeed, [\ion{Fe}{x}] is detected up to 154$\pm$21~pc north and south of the AGN.

Regarding [\ion{Fe}{xi}]~$\lambda$7889, the emission flux map indicates that the emission is restricted to the cental 1$\arcsec$ region. Thus, it is barely resolved angularly.

\subsection{NGC\,5643}

\begin{figure*} 
    \centering
    \includegraphics[width=8.0cm]{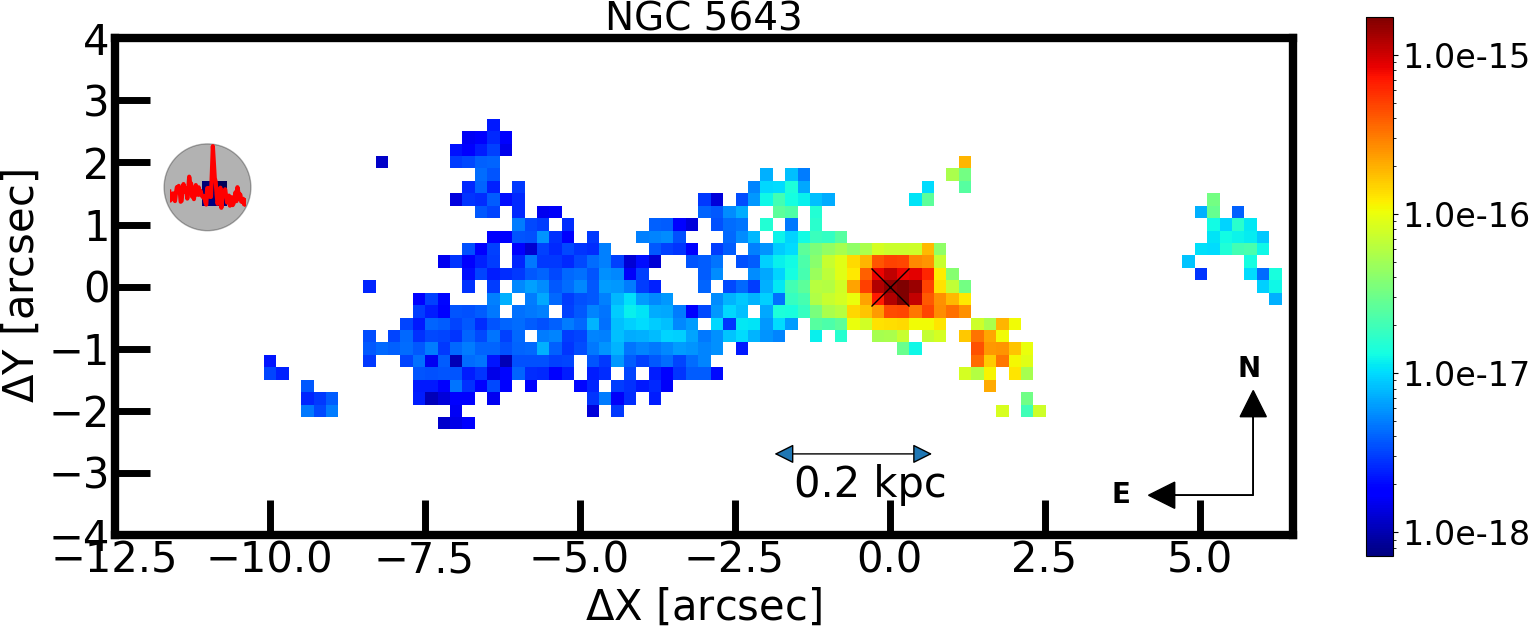} 
    \includegraphics[width=1.0cm]{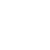}  
    \includegraphics[width=8.0cm]{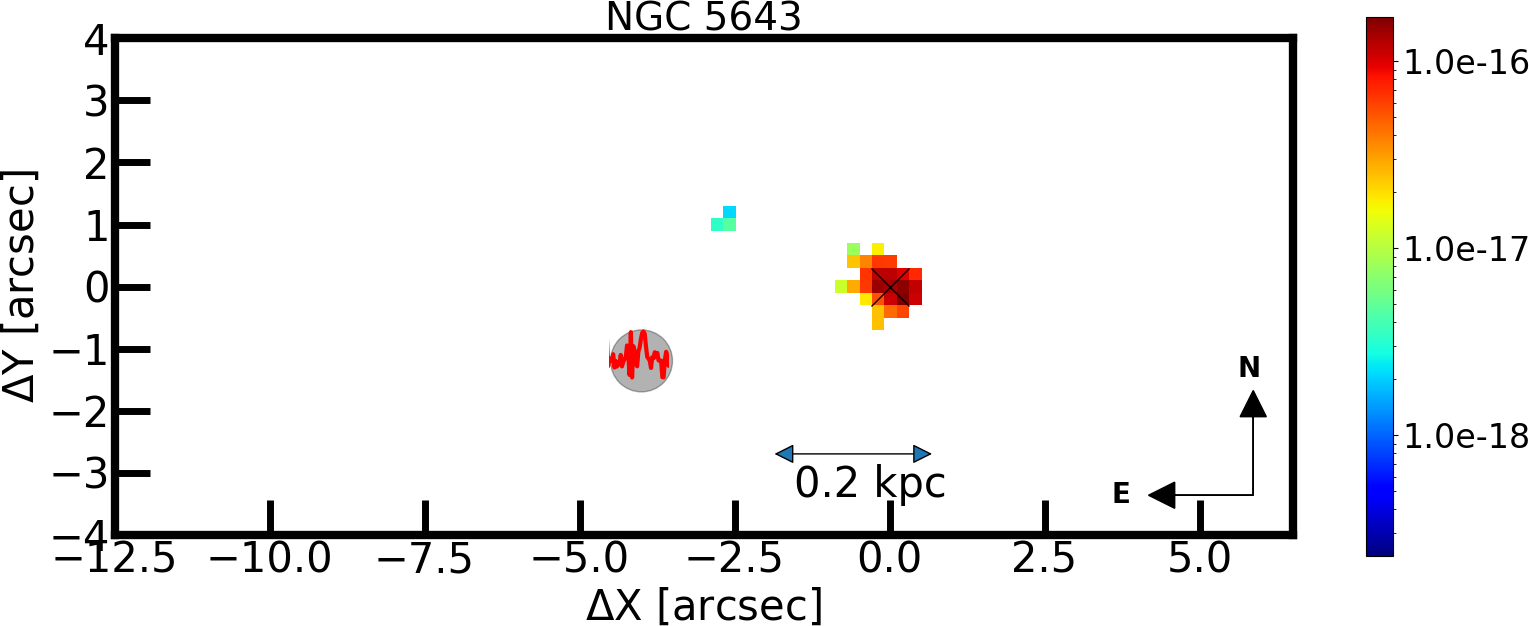} 
    \caption{Similar to Figure~\ref{fig:ext_circinus} for NGC\,5643. The left panel shows the the emission map of [\ion{Fe}{vii}] and the emission line at the point of maximum extension, identified after integration in a circular aperture of radius 0.6$\arcsec$. The right panel shows the emission map of [\ion{Fe}{x}]. The emission line at the maximum extension was found after integration in a circle of radius 0.4$\arcsec$. The colour bars are in units of erg\,cm$^{-2}$\,s$^{-1}$.}
    \label{fig:ext_NGC5643}
\end{figure*}

NGC\,5643 displays extended coronal emission, as seen through the [\ion{Fe}{vii}] line (see the left panel of Figure \ref{fig:ext_NGC5643}). The emission map  is clearly distributed along the east-west direction although the bulk of the off-nuclear emission is concentrated towards the east of the AGN. In that direction, it is detected up to 700\,pc from the centre using the 1$\times$1 binning cube. To the west, extended [\ion{Fe}{vii}] is restricted to a small region of irregular morphology at about 300\,pc from the AGN position. Between the nucleus and the western emission, no other region of significant emission is detected. The maximum extension of [\ion{Fe}{vii}] is identified to the east, at a distance of 845$\pm$46\,pc from the nucleus. It was found after integrating the signal in circular aperture of 0.6$\arcsec$ in radius. 

The intensity of the [\ion{Fe}{vii}] emission peaks at the AGN position. The average flux is $\sim10^{-15}$~erg\,s$^{-1}$\,cm$^{-2}$. In addition, we detect a small bright filament of emission projected to the southwest, with flux values of $\sim10^{-16}$~erg\,s$^{-1}$\,cm$^{-2}$, extending up to 250\,pc from the nucleus. To the west, the average flux of [\ion{Fe}{vii}] is $\sim$2~dex fainter when compared to that of the nucleus. Although the emission to the west of the AGN is significantly less extended when compared to that of the east, on average, the former is about 4 times brighter than that of the latter. The lack of significant extended emission to the west is likely due to the host galaxy that shadows the emission from the counter cone. 

Our data also shows that [\ion{Fe}{x}] (right panel of Figure \ref{fig:ext_NGC5643}) is found only in the nuclear region, constrained to a region size equal to that of the seeing of the observation ($\sim1\arcsec$). However, as in Circinus, we detected [\ion{Fe}{x}] emission  at a maximum distance of 317$\pm$30\,pc SE from the AGN, being identified after integrating the signal in a circular aperture of 0.4\arcsec in radius. It is the first time in the literature that extended coronal emission is reported in this object.  

\subsection{NGC\,1068}
\label{subsec:n1068}

Evidence of extended coronal emission in NGC\,1068 is extensively documented in the literature. Just to name a few, \citet{prieto_2005} reported [\ion{Si}{vii}]~2.48$\mu$m detected at scales of about 100~pc from the AGN. Later, \citet{ardila_2006} found from optical and NIR spectroscopy extended coronal gas at scales of a few hundred parsecs. They reported, for the first time, [\ion{Fe}{vii}]~$\lambda$6087 emission up to $\sim$240~pc north and 180~pc south of the AGN. Moreover, they also found extended [\ion{Fe}{x}]~$\lambda6374$ and [\ion{Fe}{xi}]~$\lambda7892$ at a hundred of parsec scale both North and South from the nucleus. The line profiles displayed a prominent double-peak structure, interpreted in terms of coronal gas outflows. \citet{exposito_2011} found NIR coronal emission associated to nodules observable in the IR broad-band images. They stated that coronal lines observed in the nodules cannot be caused by photoionisation by the central source but are instead caused by a local jet-induced ionising continuum. Moreover, \citet{mazzalay_2010} and \citet{mazzalay_2013} found, by means of optical HST spectroscopy and NIR NIFS/Gemini adaptive optics IFU spectroscopy, respectively, extended CL emission including  [\ion{Fe}{vii}]~$\lambda$6087 in the optical and [\ion{Si}{vii}]~2.48~$\mu$m in the NIR at scales of $\sim$300~pc from the AGN.

\begin{figure} 
    \centering
    \includegraphics[width=8.6cm]{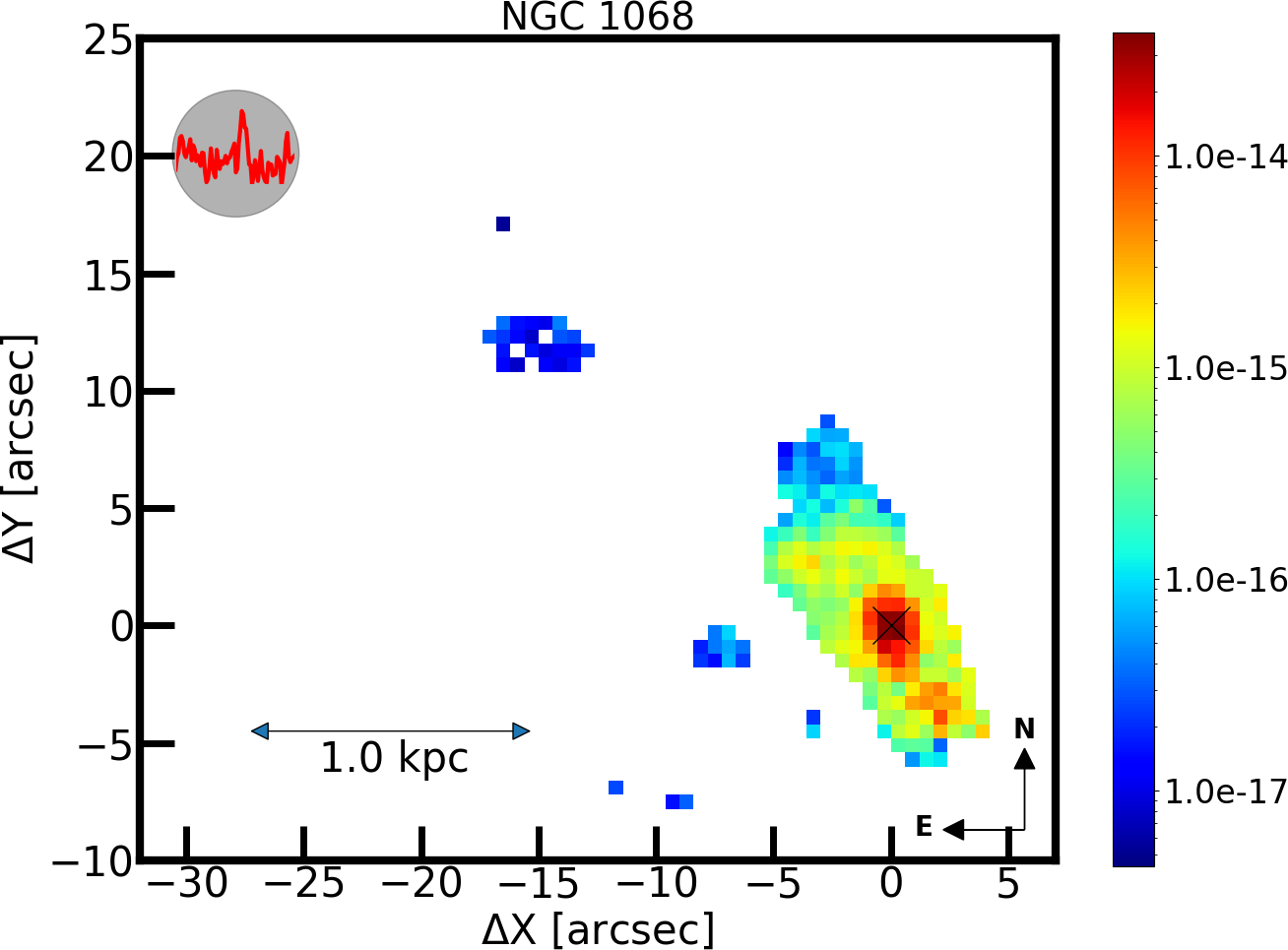} 
    \includegraphics[width=8.6cm]{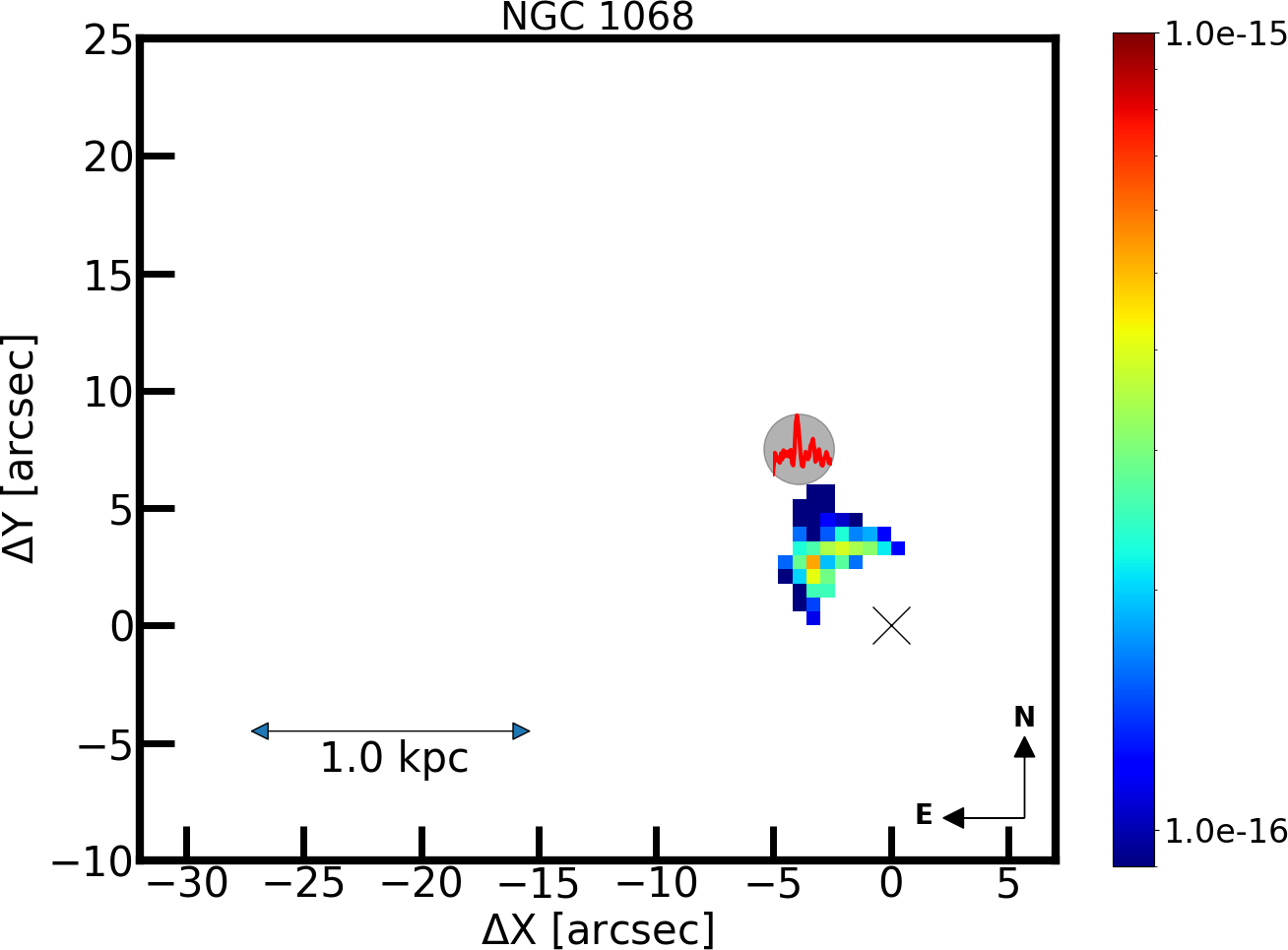} 
    \caption{Similar to Figure~\ref{fig:ext_circinus} for NGC\,1068.
     The top panel shows the emission map of [\ion{Fe}{vii}] along with the integrated emission line at the most extended location from the AGN, observed through a circular aperture of radius 1.2$\arcsec$. The bottom panel is the emission map of [\ion{Fe}{xi}] and the most extended emission integrated in the circular aperture of radius 0.4$\arcsec$. The colour bars are in units of erg\,cm$^{-2}$\,s$^{-1}$.}
    \label{fig:ext_NGC1068}
\end{figure}

Although the above results are impressive, here we found that this AGN presents the most extended coronal emission of our sample, at spatial scales not reported before (see Figure \ref{fig:ext_NGC1068}). In the inner 1~kpc, the [\ion{Fe}{vii}] emission peaks at the nucleus, with a measured flux of $4 \times 10^{-14}$~erg\,s$^{-1}$\,cm$^{-2}$. Moreover, an elongated structure in the same CL is observed in the north-northeast and south-southeast directions from the AGN position.  The average flux of that extended emission amounts $\sim 10^{-15} $~erg\,s$^{-1}$\,cm$^{ -2}$. It reaches extensions of 800\,pc and 400\,pc, respectively, from the centre. Using the data cube with bin 3 $\times$ 3 spaxels, it was possible to identify a cloud detached from the central [\ion{Fe}{vii}] emission at 2~kpc  northeast  of the AGN (see the upper panel of Fig.~\ref{fig:ext_NGC1068}). The average flux of this emission amounts to 5 $\times~10^{-17}$~erg\,s$^{-1}$\,cm$^{-2}$. Further out, also in the northeast direction, we found the maximum extension of [\ion{Fe}{vii}], detected at a distance of 2.82$\pm$0.20\,kpc from the AGN. This emission was detected after integrating the flux in a circular aperture of 1.2$\arcsec$ in radius. This is the largest extension of [\ion{Fe}{vii}] identified in our sample.

It is also important to mention that the extended coronal emission in NGC\,1068 away from the central 5$\arcsec \times 5\arcsec$\, region is well aligned with the direction of the radio-emitting jet. However, the radiating part of the jet  to the NE direction is confined to the central region of the galaxy (i.e., $\sim$500\,pc). Therefore, the emitting clouds of [\ion{Fe}{vii}] observed at $\sim$2\,kpc and $\sim$2.82\,kpc to the NE of the AGN, if originated by the jet, are due to a radio component still not detected. It can also indicate that the jet has a very low radiative efficiency and is depositing a large part of its energy mechanically, along the direction of propagation. We do not discard that these two extended coronal emission regions can be associated to molecular outflows. For instance, the [\ion{Fe}{vii}] emission located at $\sim$2\,kpc from the AGN is spatially very close to the inner radius of the prominent starburst ring displayed by the host galaxy \citep{rico-villas_2020}. Indeed, it is between Super Star Clusters (SSC) 1 and 2 \citep[see Figure~5 of][]{rico-villas_2020}, where susbtantial star formation is taking place. Regarding the even more distant emission at 2.8~kpc, it is indeed in the outer rim of the same star-formation ring, supporting the scenario that these two blobs of emission may be associated to stellar outflows.

Regarding [\ion{Fe}{x}], it has a clearly extended emission (see the bottom panel of Figure~\ref{fig:ext_NGC1068}) although it is considerably more compact than [\ion{Fe}{vii}]. We notice that in Figure~\ref{fig:ext_NGC1068} the [\ion{Fe}{x}] emission at the AGN position and in the central 2$\arcsec \times 2\arcsec$ is masked because of the strong blending with [\ion{O}{i}]~$\lambda6363$, making rather uncertain the characterization of the coronal line profile. Outside this region, the width of the lines decreases, allowing their deblending with confidence.  We identified [\ion{Fe}{x}] in the north-northeast region at a distance of 693$\pm$98~pc from the AGN.  The signal was integrated in a circular aperture of 0.4$\arcsec$ in radius. We also identified an emission cloud to the south at 240\,pc from the central source.

\subsection{ESO 428-G14}

ESO~428-G14 is usually regarded as a showcase of a radio-weak AGN with a jet interacting with the NLR gas. \citet{falcke_1996} had already noticed that the NLR of this object consists of many individual thin strands, which are very closely related to the radio jet producing a highly complex yet ordered structure. First detection of extended coronal emission was reported by \citet{prieto_2005}, who found [\ion{Si}{vii}]~2.48~$\mu$m  at scales of  120$-$160~pc from the nucleus by means of VLT Adaptive Optics imaging. The emission was found to be aligned with the radio jet. Later, \citet{may_2018}  reported [\ion{Si}{vi}]~1.963~$\mu$m  (IP = 168~eV), also co-spatial to the region where the jet propagates.

\begin{figure} 
    \centering
    \includegraphics[width=8.6cm]{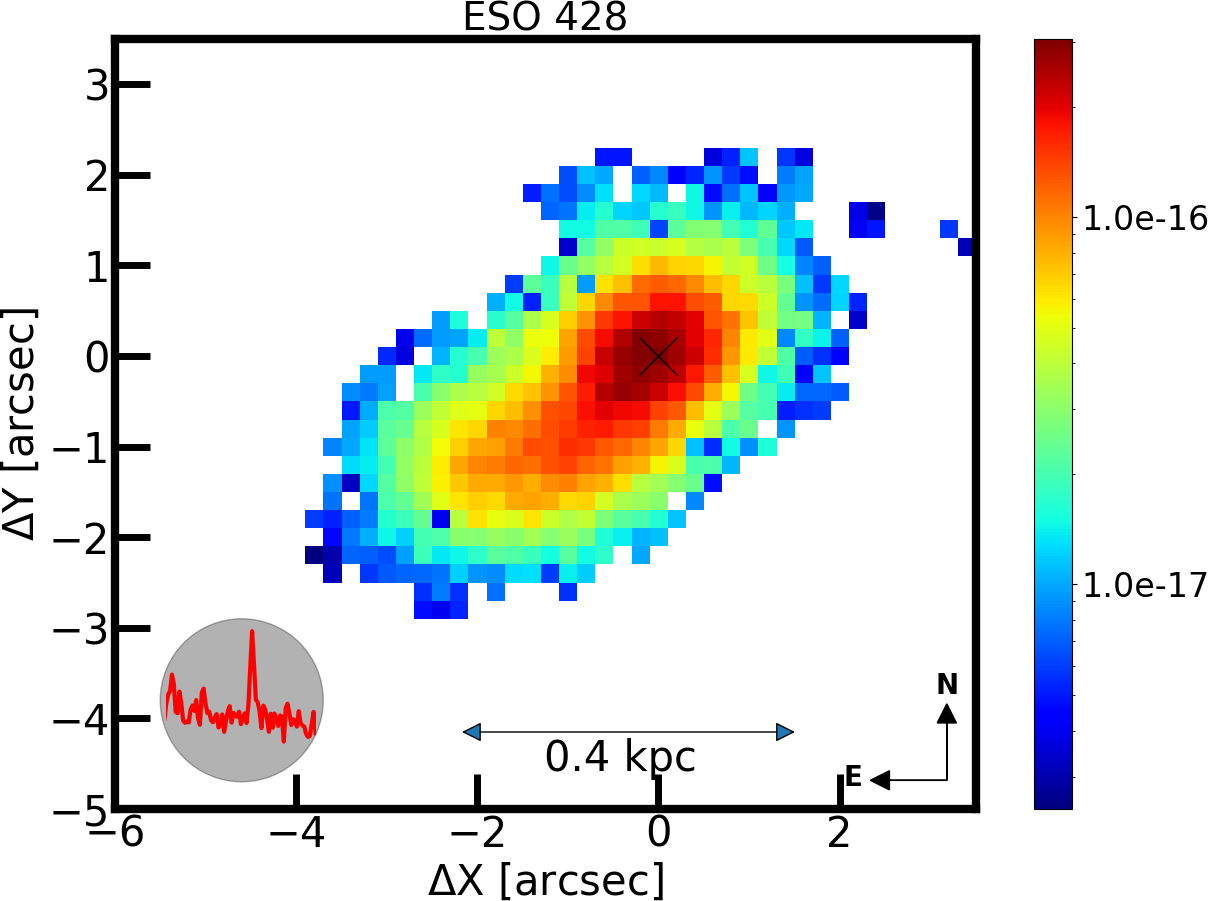} 
    \includegraphics[width=8.6cm]{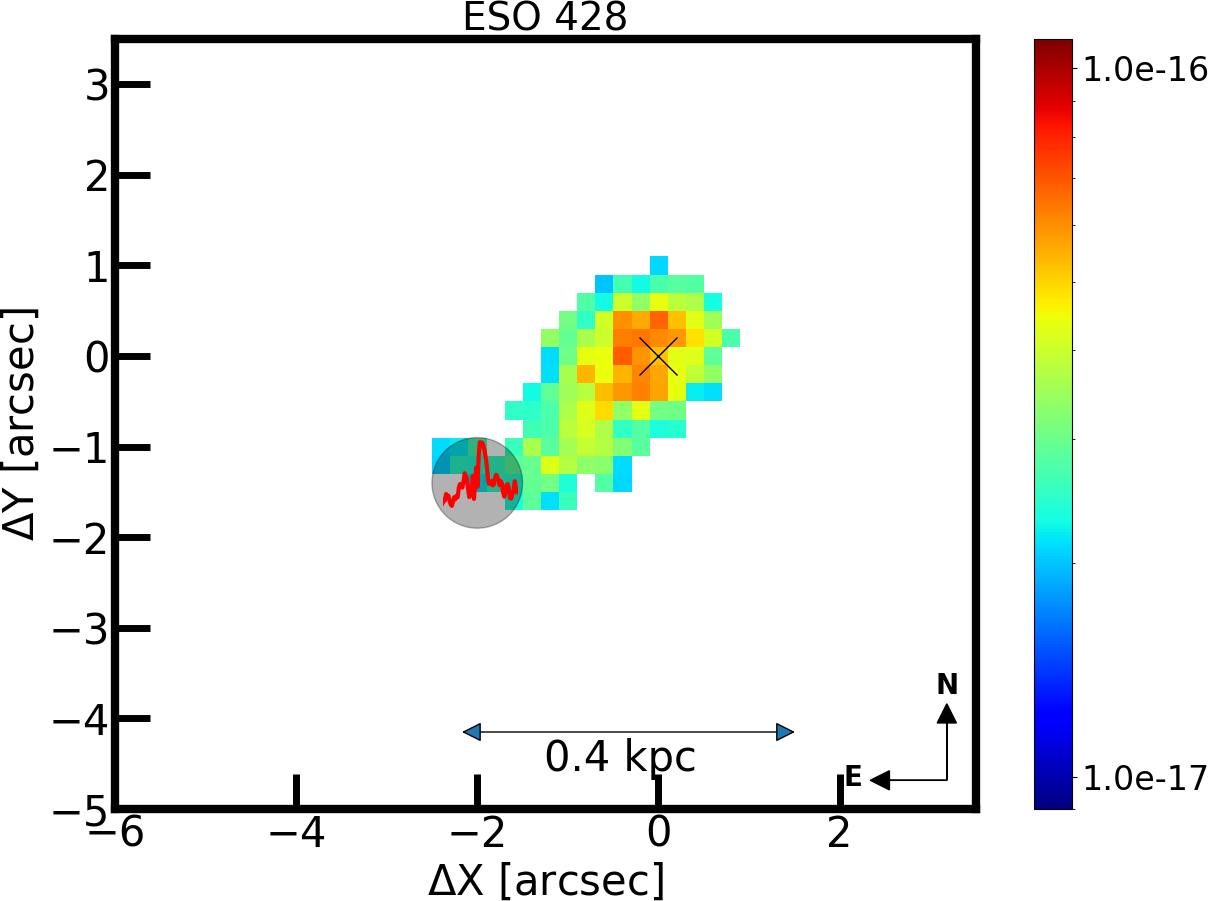} 
    \caption{Similar to Figure~\ref{fig:ext_circinus} for The top panel shows the emission map of ESO\,428 in the [\ion{Fe}{vii}] line and the integrated spectrum of the most extended emission observed after the integration in circular aperture of radius 0.8'' (grey circle). In the bottom panel is the emission map of [\ion{Fe}{X}]   and the most extended emission spectrum obtained by integrating the gray circular aperture of radius 0 .4''. The colour bars are in units of erg\,cm$^{-2}$\,s$^{-1}$.}
    \label{fig:ext_ESO428}
\end{figure}

The upper panel of Figure \ref{fig:ext_ESO428} show the coronal emission map for [\ion{Fe}{vii}] as well as the region of maximum extension of that emission. As in the other cases explored above, this coronal emission is well-resolved spatially. The high-ionisation gas is distributed mostly in the SE-NW direction, peaking at the nucleus, with an average flux of $3 \times 10^{-16}$~erg\,s$^{-1}$\,cm$^{-2}$. The emission flux decreases slowly towards the southeast portion of the AGN, up to $\sim300$ \,pc. The average flux of this extended emission is $10^{-16}$~erg\,s$^{-1}$\,cm$^{-2}$. Then, the flux gradually decreases as we move away from the extended region, reaching flux values of $5\times10^{-18}$~erg\,s$^{-1}$\,cm$^{-2}$ at the edge of the emission region, at about 450~pc from the AGN position. In this same direction, we identified the most distant emission of [\ion{Fe}{vii}] at 638$\pm$86\,pc from the AGN. This result was obtained after integrating the spectrum in a circular aperture of 0.8\arcsec\ in radius. Towards the northwest, [\ion{Fe}{vii}] is detected up to 200~pc of the AGN. In that direction, the emission drops sharply as we move farther out from the peak emission. We found that the orientation angle of the [\ion{Fe}{vii}] emission (PA $\sim$ 135$^\circ$) coincides spatially with that of [\ion{Si}{vi}] (PA = 135$^\circ$) reported in \citet{may_2018}. The coronal emission of the latter line, though, is detected only up to a maximum distance of 170\,pc SE of the AGN. This can be due, in part, to the larger IP of the transition leading to [\ion{Si}{vi}]  and to the fact that this line is usually fainter than that of [\ion{Fe}{vii}] \citep[See Table A1 in][]{prieto+22}. Therefore, in this work, we report a CLR three times larger than the value previously published in the literature, still in the same direction as that of the jet.

Regarding the [\ion{Fe}{x}] emission, the bottom panel of Figure~\ref{fig:ext_ESO428} shows the flux distribution measured for that line and the corresponding emission detected at the location of maximum extension. Evidence of extended emission is identified to the north, northeast and west of the AGN, up to 180~pc from the nucleus. The peak emission coincides with that of [\ion{Fe}{vii}]. We identified the most extended emission of [\ion{Fe}{x}] to the southeast, associated to a small filament of emission. The spectrum of that region, shown in Figure~\ref{fig:ext_ESO428}, results after integration of the signal in a circular aperture of of 1.2\arcsec\ of radius. The coronal line is strongly blended to [\ion{O}{i}]~$\lambda$6363. The maximum extension of [\ion{Fe}{x}] is found at 202$\pm$43\,pc from the AGN.  

\subsection{NGC\,3081}
\begin{figure} 
    \centering
    \includegraphics[width=8.6cm]{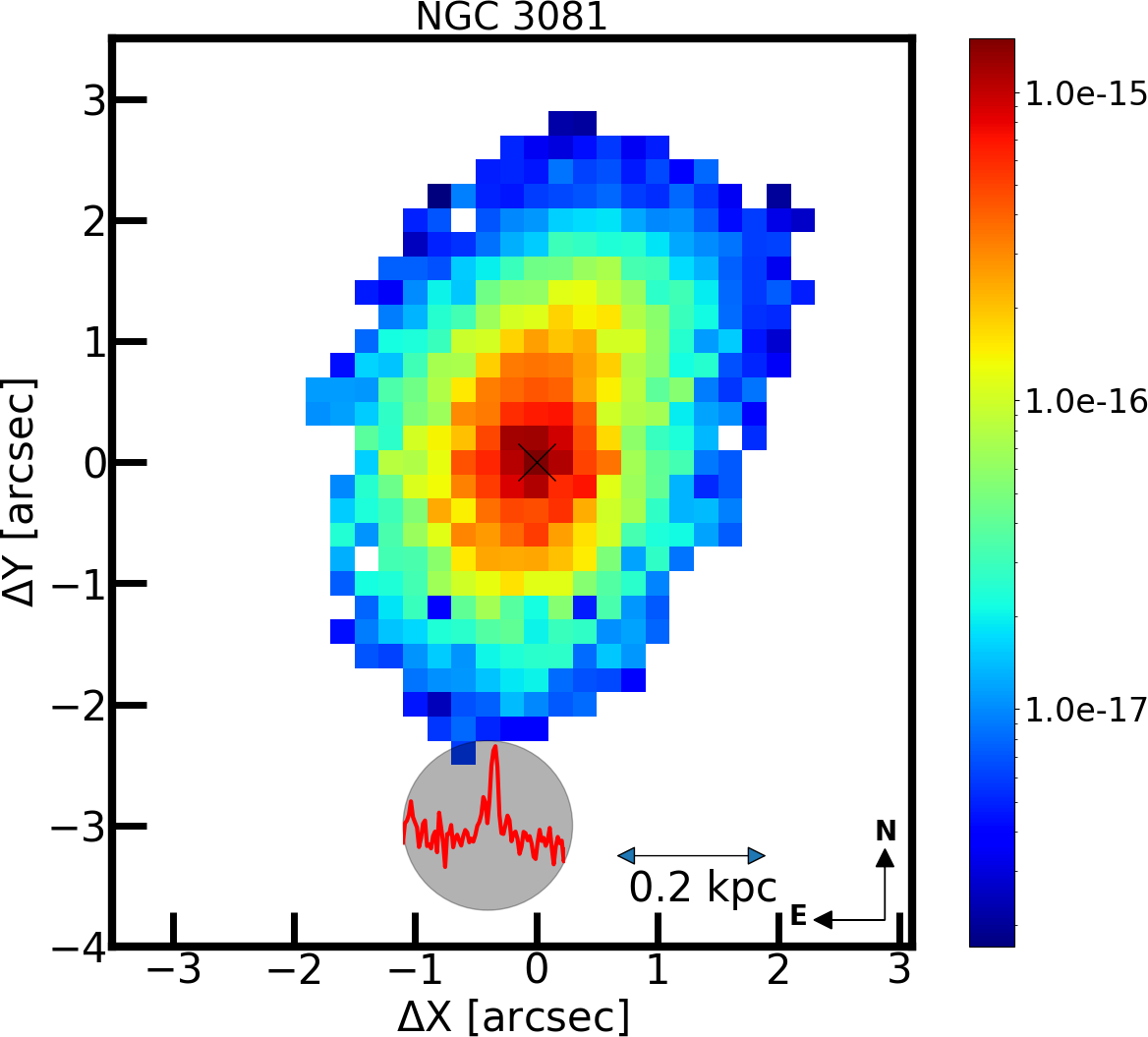}  
    \includegraphics[width=8.6cm]{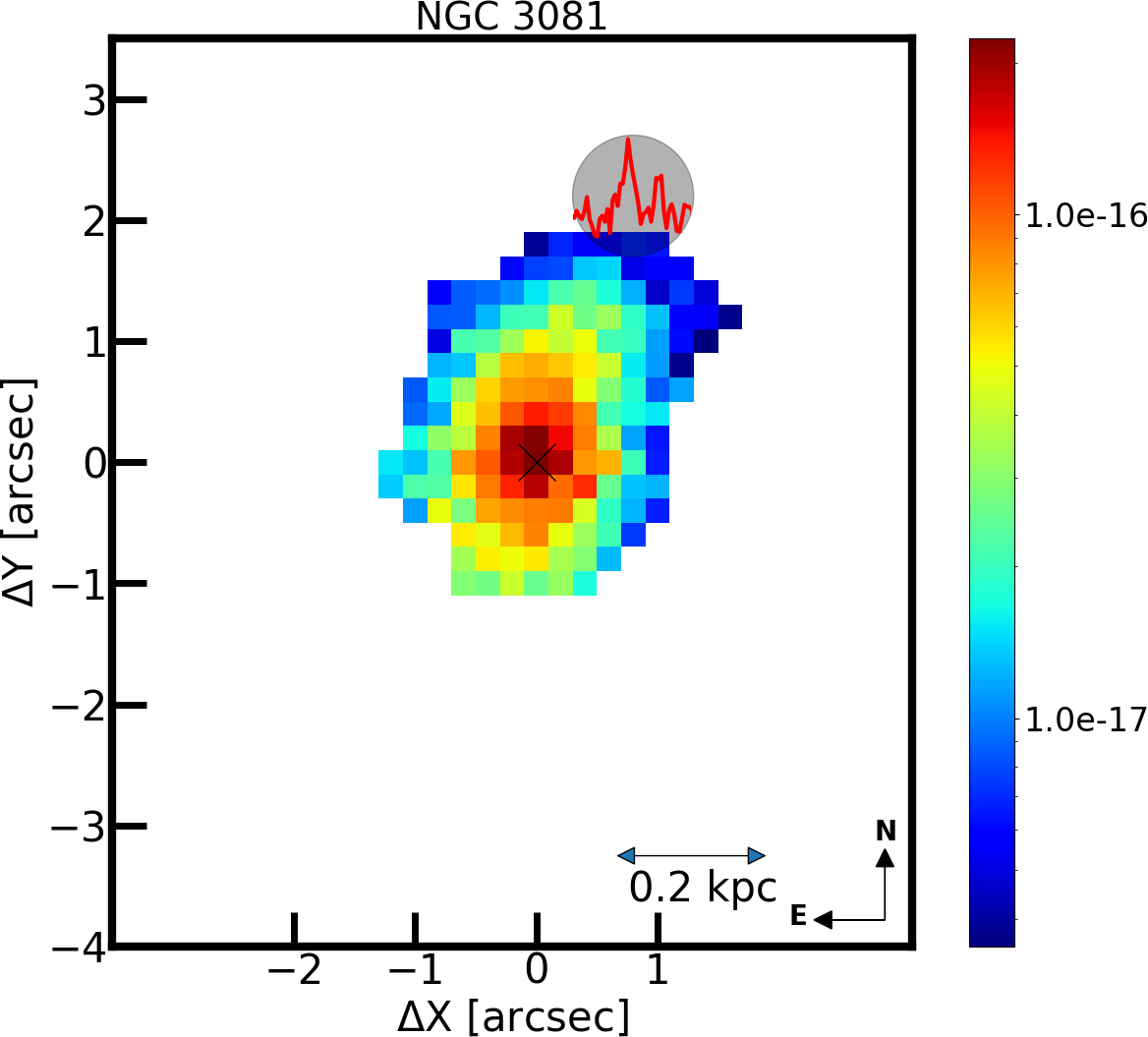}  
    \caption{Similar to Figure~\ref{fig:ext_circinus} for NGC\,3081. The top panel shows the [\ion{Fe}{vii}] map and the corresponding emission line of the most extended emission observed by the integration of the signal in a circular aperture of radius 0.6$\arcsec$. The bottom panel is the [\ion{Fe}{x}] map with the most distant emission, found after the integration of the signal in a circular aperture of radius 0.6$\arcsec$. The colour bars are in units of erg\,cm$^{-2}$\,s$^{-1}$.}
    \label{fig:ext_NGC3081}
\end{figure}

NGC 3081 is a galaxy with intense star formation located in a series of nested ringlike features \citep{buta+04}. A spectropolarimetric study by \citet{moran+00} established that it is a Seyfert~1, obscured by dense material at a few parsecs from the nucleus. However, no other previous or latter works have confirmed its Seyfert~1 nature and in all publications it is classified as a Seyfert~2 source. Here, we consider NGC\,3081 as a Type~II AGN based on its spectroscopic characteristics.
It has a bright resolved compact nucleus, with an ionisation cone extending to the
north. Previous works have already pointed out that NGC\,3081 is a strong coronal line emitter AGN. 
\citet{appoes88} reported the detection of  [\ion{Fe}{vii}],  [\ion{Fe}{x}], [\ion{Fe}{xi}] and [\ion{Fe}{xiv}], although this latter feature is strongly blended with [\ion{Ca}{v}]. \citet{mazzalay_2010} found by means of STIS/HST observations extended optical coronal emission in [\ion{Ne}{v}]~$\lambda$3425 at scales of $\sim$50~pc. In the near-infrared, \citet{Reunanen+03} reported the detection of the coronal lines [\ion{Si}{vi}]~1.964~\mum\ and [\ion{Si}{vii}]~2.484~\mum. Moreover,
coronal lines of [\ion{Ca}{viii}] 2.321~\mum\ and [\ion{Al}{xi}] 2.043~\mum\ are also detected. \citet{prieto_2005} resolved the [\ion{Si}{vii}]~2.484~\mum\ emission region using AO+VLT. They found that it was extended up to $\sim$120~pc from the AGN, coinciding with the UV continuum emission on the same region \citep[See Fig. 2f in][]{prieto_2005}. In the mid-infrared, \citet{lutz+03} report the presence of the [\ion{Si}{ix}]~3.94~\mum\ in its nuclear spectrum. 
 
The MUSE cube (see Figure~\ref{fig:ext_NGC3081}) shows strong coronal emission of [\ion{Fe}{vii}] (upper panel) and [\ion{Fe}{x}] (bottom panel). Both emissions peak at the nucleus and display a conspicuous extended region. The former line reaches a maximum intensity of $\sim 1.03 \times 10^{-15}$~erg\,cm$^{-2}$\,s$^{-1}$. Outwards, the emission steadily decreases, reaching a flux of $\sim 1.0 \times 10^{-17}$~erg\,cm$^{-2}$\,s$^{-1}$ at 380~pc both N and S of the AGN. The location where the [\ion{Fe}{vii}] gas reaches its maximum extension is found at 472$\pm$94~pc S of the AGN, at the outer tip of a prominent plume of coronal gas that seems to be connected to the AGN.  

The [\ion{Fe}{x}] emission is more compact than that of [\ion{Fe}{vii}] but spatially resolved and extended predominantly to the NW of the AGN. Similarly to  [\ion{Fe}{vii}], a plume of high ionisation gas bends slightly to the NW at 1$\arcsec$ N of the AGN. At the nucleus, the strength of the line reaches 1.5~$\times 10^{-16}$~erg\,cm$^{-2}$\,s$^{-1}$ and decreases by a factor of $\sim$20 at 365$\pm$62~pc to the N-NW of the AGN, where it reaches the maximum extension. Towards the south, the [\ion{Fe}{x}] gas is detected at a maximum distance of 180~pc from the nucleus.  

\subsection{NGC\,5728}

\begin{figure} 
    \centering
    \includegraphics[width=8.6cm]{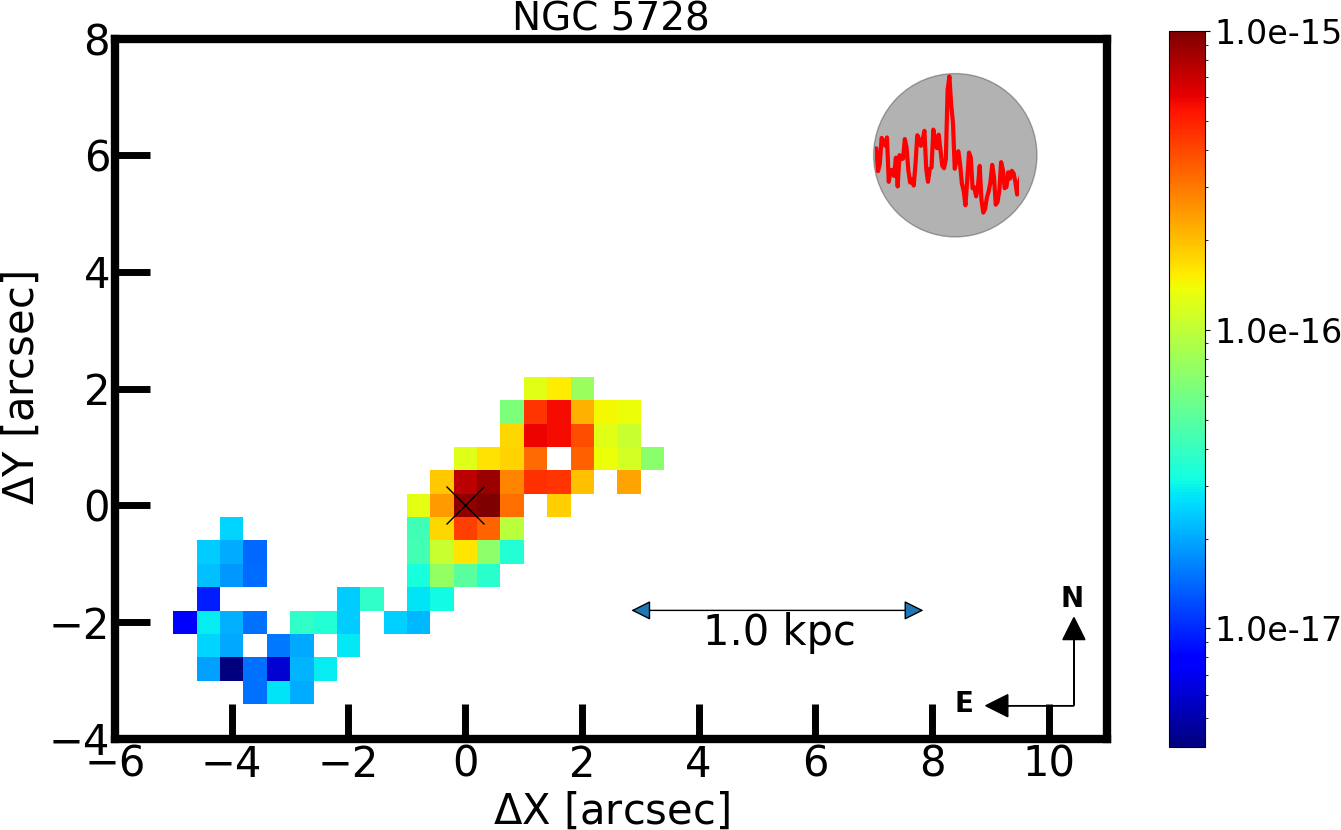} 
    \includegraphics[width=8.6cm]{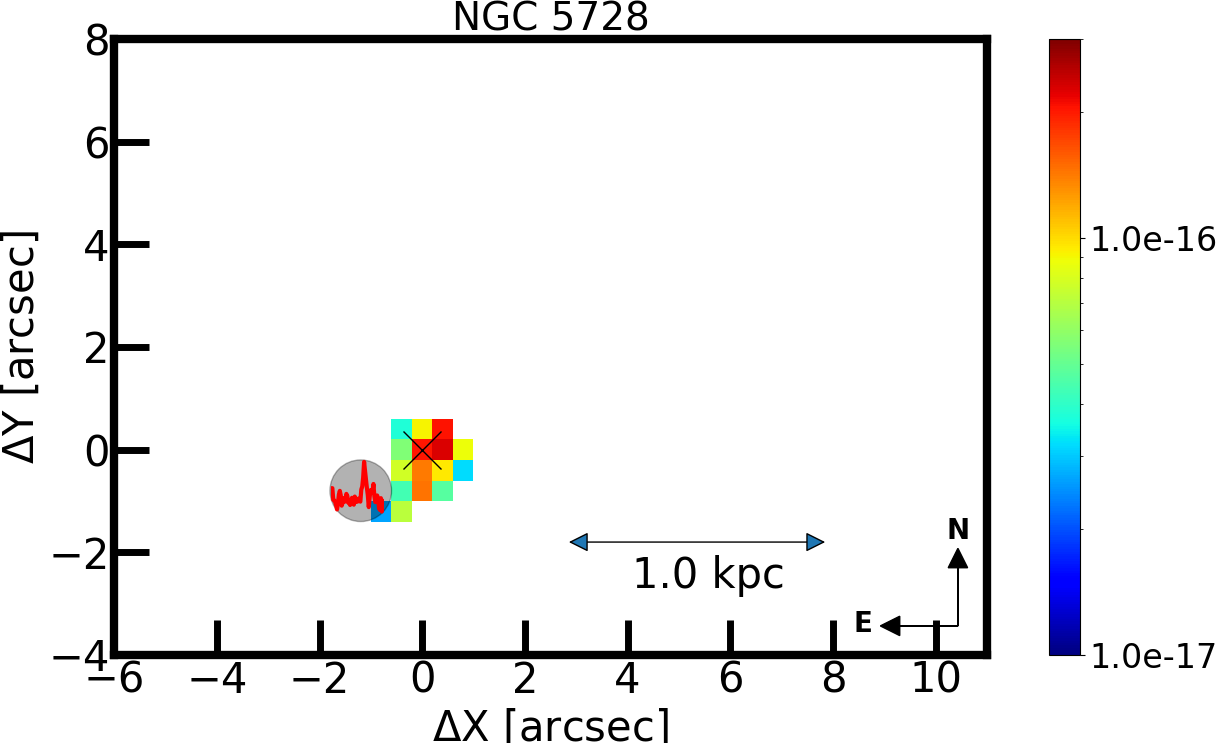} 
    \caption{Similar to Figure~\ref{fig:ext_circinus} for NGC\,5728.
     The most distant [\ion{Fe}{vii}] emission (upper panel) is detected after integration in a circular aperture of radius 1.2$\arcsec$. For [\ion{Fe}{x}], the most extended emission is obtained after integration in a circular aperture of radius 1.2$\arcsec$. The colour bars are in units of erg\,cm$^{-2}$\,s$^{-1}$.}
    \label{fig:ext_NGC5728}
\end{figure}

Figure~\ref{fig:ext_NGC5728} displays the coronal emission maps for NGC\,5728. In the top panel we show the flux distribution of the [\ion{Fe}{vii}] line. In the nuclear region, we identified two bright peaks of emission, aligned to the SE-NW direction, along which the extended emission is also distributed. The brightest of the two peaks coincides with the position of the nucleus. It carries a flux of $10^{-15}$~erg\,s$^{-1}$\,cm$^{-2}$. The second maximum is located at 1.84\arcsec\ ($\sim$600\,pc) to the NW of the first, with a flux of 2$\times 10^{-16}$~erg\,s$^{-1}$\,cm$^{-2}$. Moreover, extended emission is clearly observed along the southeast direction up to 1.2~kpc from the AGN. The emission is restricted to a narrow arm that ends in a structure resembling a hook. The average flux along that component is $2\times10^{-17}$~erg\,s$^{-1}$\,cm$^{-2}$.  Northwest of the AGN, we identify the most distant emission of [\ion{Fe}{vii}] in this object, at 2023$\pm$235\,pc. This result was obtained after integrating the signal in a circular aperture of 1.2\arcsec\ in radius. The [\ion{Fe}{vii}] gas seems to be aligned along the same position angle as that of the extended coronal emission to the SE. The emission to the NW is rather faint, at flux levels of $<5 \times 10^{-17}$~erg\,cm$^{-2}$\,s$^{-1}$. For this reason, in the map shown in Fig.~\ref{fig:ext_NGC5728}, no detectable emission is observed. It is necessary to integrate a significant number of spaxels to unveil it. 

In contrast to the [\ion{Fe}{vii}] emission, the [\ion{Fe}{x}] map (bottom panel) is limited to the nuclear region, being constrained to the size of the seeing of the observations. On the original map, each spaxel equals 0.4$\arcsec$. Using a larger spaxel bin, we detect [\ion{Fe}{x}] to the southeast, outside the unresolved nuclear region. The maximum distance from the AGN where [\ion{Fe}{x}] is  identified is at 282~$\pm$~78\,pc. For its detection, the signal was integrated into a circular aperture of a 1.2$\arcsec$ in radius. This is the galaxy of the sample with the smallest extended [\ion{Fe}{x}] emission region. The flux at the AGN position is $\sim2 \times 10^{-16}$~erg\,cm$^{-2}$\,s$^{-1}$.

\subsection{IC 5063}

The most important aspects of the coronal emission in IC\,5063 were detailed in \citet{fonseca-faria+23}. Here we summarize the main findings regarding the coronal emission found by them. 

\begin{figure}
    \centering
    \includegraphics[width =8cm]{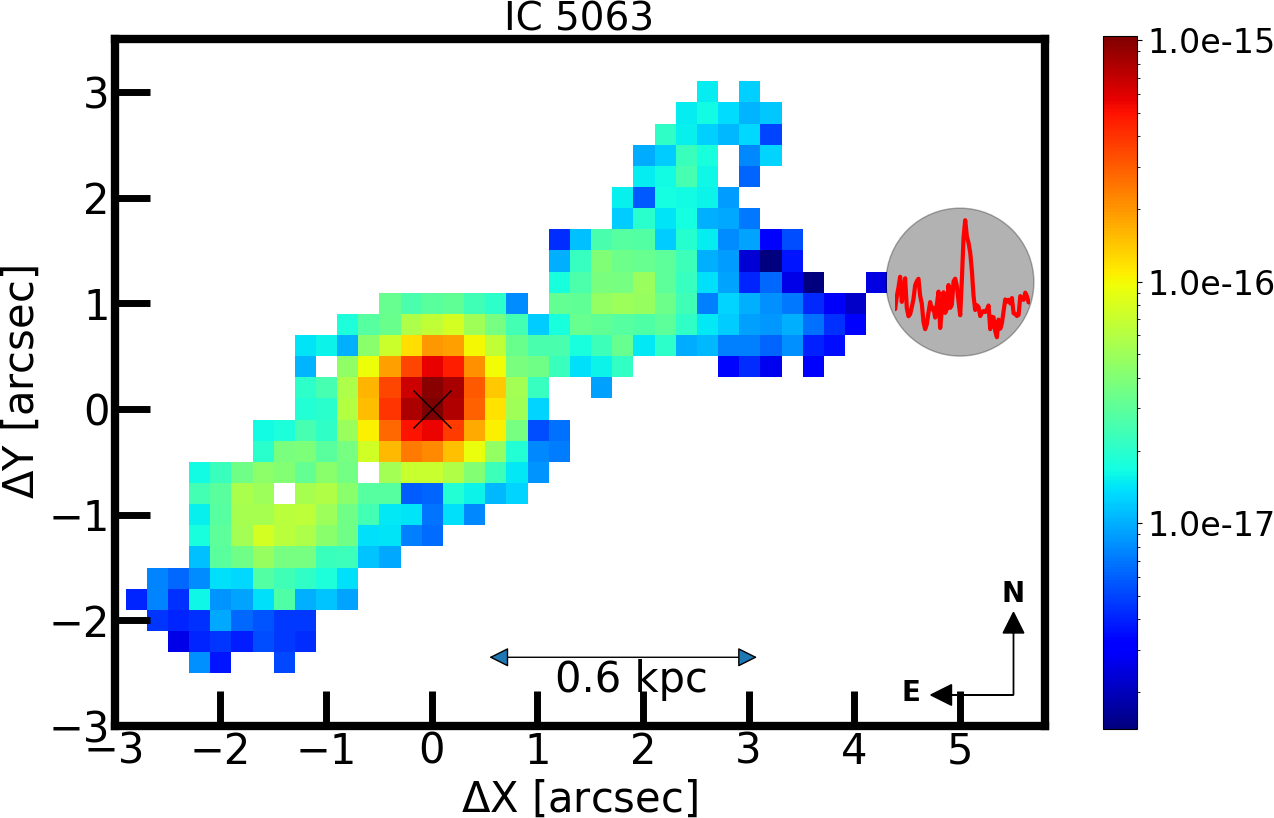}
    \includegraphics[width=1cm]{F1.png}  
    \includegraphics [width=8cm]{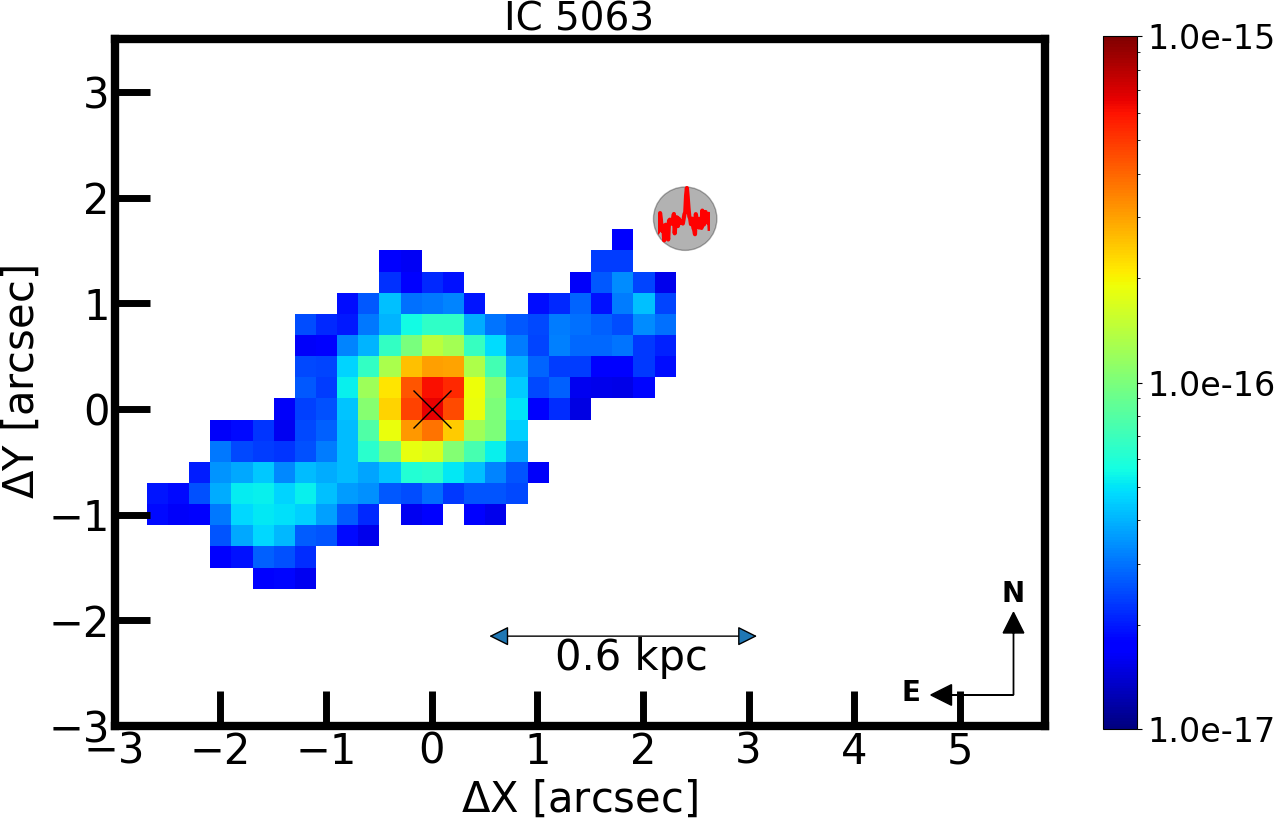}
    \caption{Same as Figure~\ref{fig:ext_circinus} for IC\,5063.   In the upper panel, the maximum extension of the [\ion{Fe}{vii}] emission was identified after integration of the signal in a circular aperture of radius 0.6\arcsec. In the bottom panel the location of the most extended [\ion{Fe}{x}] emission was obtained after integrating the signal in circular aperture of radius 0.2\arcsec. In both cases, the corresponding emission line at the position of maximum extension (red curve) is shown within the circular aperture drawn. The colour bars are in units of erg\,cm$^{-2}$\,s$^{-1}$.}
    \label{fig:ext_IC5063}
\end{figure}

The [\ion{Fe}{vii}] flux distribution (see the upper panel of Figure~\ref{fig:ext_IC5063}) presents three distinctive peaks, with the brightest one coinciding with the AGN position. In addition, two secondary off-nuclear peaks of emission are identified, one at 500\,pc to the northwest of the AGN and the other at 400\,pc to the southeast of the central engine. The most distant [\ion{Fe}{vii}] emission is identified at 1193~$\pm$~39~pc NW of the centre. In the opposite direction, to the southeast, [\ion{Fe}{vii}] emission is detected up to $\sim$680~pc from the galaxy nucleus.

[\ion{Fe}{x}] shares a similar flux distribution in the central 4$\times$4 arcseconds as [\ion{Fe}{vii}] (see the bottom panel of Figure~\ref{fig:ext_IC5063}). The most intense emission is predominantly concentrated around the nucleus and at the two secondary peaks of emission, observed at the same spatial location in the two coronal lines. We identified [\ion{Fe}{x}] emission at a maximum distance of 696~$\pm$~46~pc from the AGN, after integrating the signal in a circular aperture of 0.2\arcsec\ in radius. That region coincides with the base of the NW  extension seen in [\ion{Fe}{vii}]. To the SE, [\ion{Fe}{x}] is detected at 450~pc from the nucleus. 

[\ion{Fe}{xi}]$\lambda$7889, [\ion{S}{vii}]$\lambda$7611 and [\ion{Fe}{xiv}]$\lambda$5311 are detected at the nucleus. In all cases, the emission regions are unresolved under seeing-limited conditions.

\subsection{NGC\,3393}

\begin{figure*} 
    \centering
    \includegraphics[height=6cm]{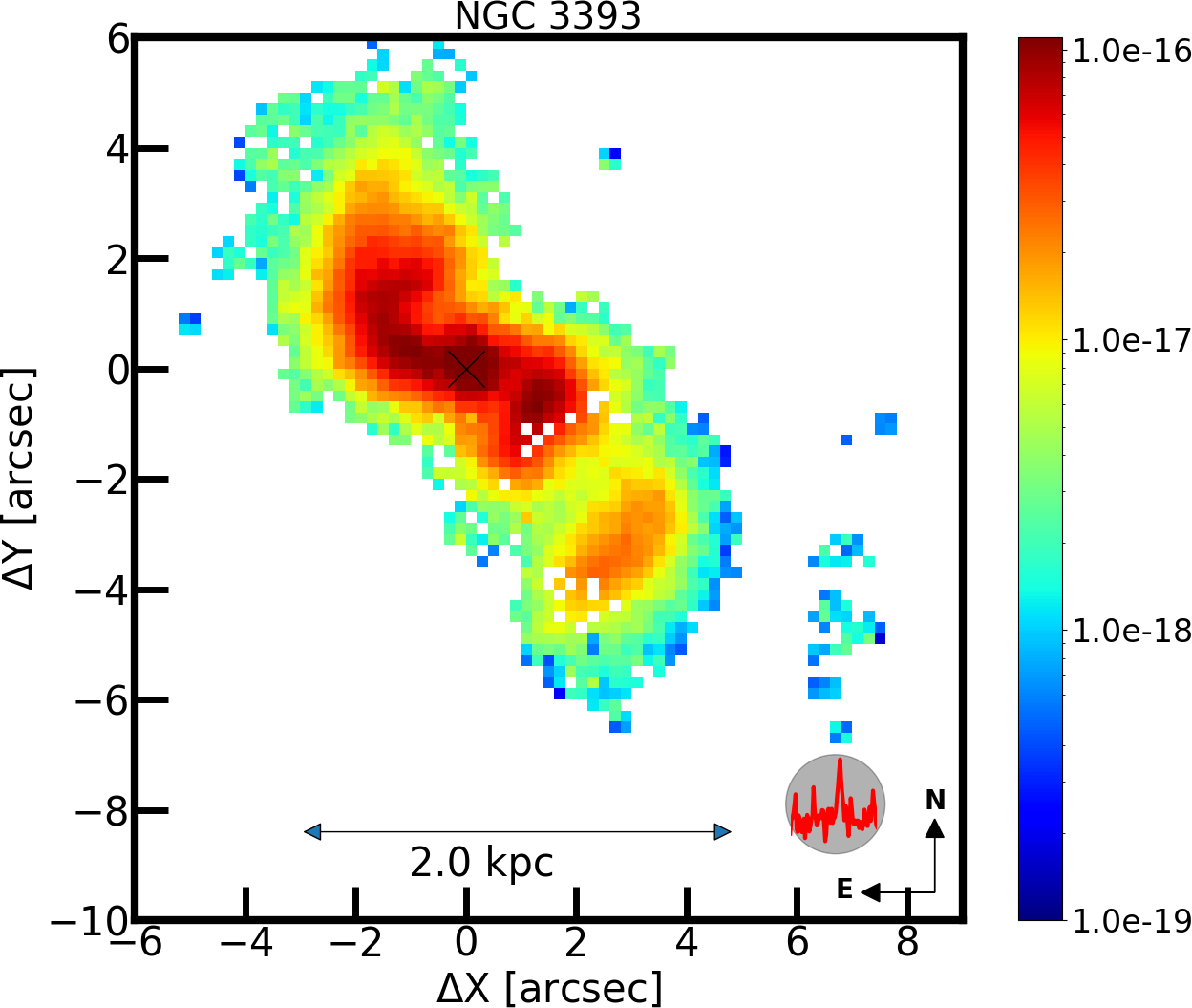} 
    \includegraphics[width=1.0cm]{F1.png}  
    \includegraphics[height=6cm]{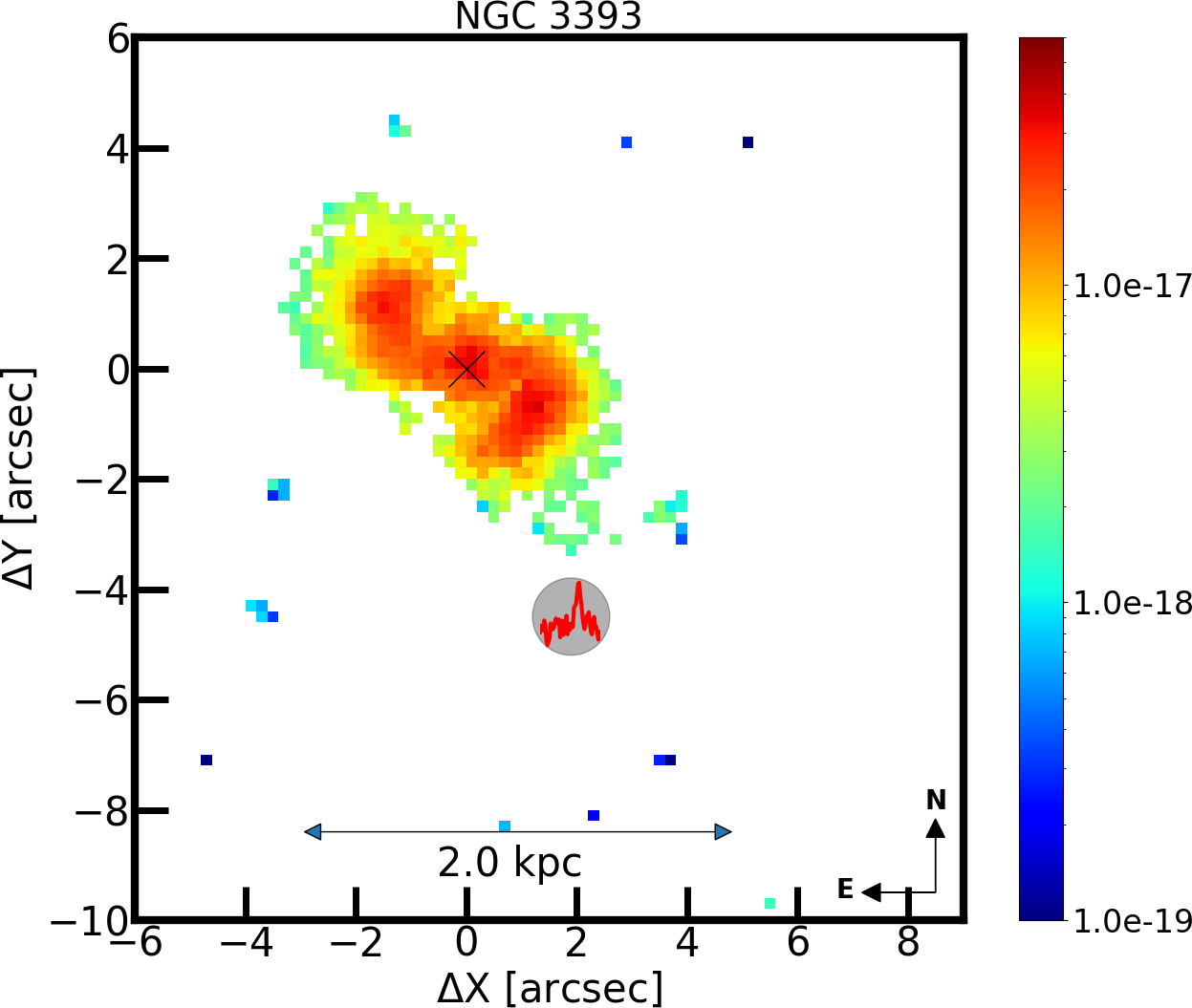} 
    \caption{Similar to Figure~\ref{fig:ext_circinus} for NGC\,3393.
     The most distant [\ion{Fe}{vii}] emission (left panel) is detected after integration in an circular aperture of radius 0.8$\arcsec$ (grey circle). [\ion{Fe}{x}] emission  is found up to $\sim$1.2~kpc SW of the AGN after integrating the signal in a circular aperture of radius 0.4$\arcsec$. The colour bars are in units of erg\,cm$^{-2}$\,s$^{-1}$.}
    \label{fig:ext_NGC3393}
\end{figure*}

NGC\,3393 (see Figure \ref{fig:ext_NGC3393}) presents one of the most striking high-ionisation gas distribution of the sample. The coronal emission ([\ion{Fe}{vii}] and [\ion{Fe}{x}]) is oriented along the northeast-southwest direction, and in the innermost kiloparsec it follows a clear "S" shape, previously observed in [\ion{O}{iii}]~\citep{maksym+19}. The `S' structure has several knots of emission in the inner 2.5$\arcsec \times 2.5\arcsec$ region, with flux values in the [\ion{Fe}{vii}] line larger than $\sim10^{-16}$~erg\,s$^{-1}$\,cm$^{-2}$. At the outer edges of the `S' structure, the flux gradually decreases to values of $\sim2\times10^{-18}$~erg\,s$^{-1}$\,cm$^{-2}$.

To the southwestern portion of the `S' structure, along the axis where the extended emission is distributed (PA $\sim$ 45$^\circ$ E of N), the [\ion{Fe}{vii}] flux decreases and then increases again, reaching a secondary peak at $\sim$1~kpc of the AGN. The emission flux in that region reaches $2\times 10^{-17}$erg\,s$^{-1}$\,cm$^{-2}$. The extended structure is strongly elongated in the direction perpendicular to that of the main emission axis of [\ion{Fe}{vii}]. The brightest part has a bean-shaped morphology. Farther out, in the SW direction, at 1.8~kpc from AGN and away from the `S' structure, we find a faint extended region emitting [\ion{Fe}{vii}] with flux values of $\sim10^{-18}$~erg\,cm$^{-2}$\,s$^{-1}$. We identified [\ion{Fe}{vii}] emission up to a distance of 2.63\,$\pm$\,0.20\,kpc from the AGN. This result was obtained after integrating the observed flux in a circular aperture of 0.8\arcsec\ in radius.
In the opposite direction, to the northeast of the AGN, [\ion{Fe}{vii}] emission gas is observed up to 1.2\,kpc of the AGN.  

The [\ion{Fe}{x}] flux distribution (right panels of Figure~\ref{fig:ext_NGC3393}) reveals spatially resolved emission within the "S" feature, coinciding with the most intense region emitting [\ion{Fe}{vii}]. The maximum extension of [\ion{Fe}{x}]  coincides with the  brightest portion of the bean-shaped structure detected in [\ion{Fe}{vii}], at a distance of 1.24\,$\pm$0.15\,kpc SW from the AGN. It was identified after integrating the flux in a circular aperture of 0.4\arcsec\ in radius. We notice that the central 'S' feature is fully detected in [\ion{Fe}{x}]. This emission line peaks at the position of the nucleus. 

It is important to mention here that first report of the S-shaped structure was made by \citet{cooke_2000} by means of pre-COSTAR and WFPC2 Hubble Space Telescope (HST) images. They show that it occupies the central few arseconds of the nucleus. By combining the HST with VLA radio data to investigate the gas kinematics and ionisation of the line-emitting gas, they found that the radio lobes are responsible for creating the S-shape arms and that the observed emission is likely due to photoionisation by the central source. Thereafter, \citet{maksym_2017} and \citet{maksym+19} added {\sc chandra}, mid-UV imaging, and {\sc spitzer} data to the existing WFC3 HST and VLA observations to carry out an exquisite analysis of the central few hundred parsecs of NGC\,3393 aiming at investigating the spatially resolved relative contributions of photoionization and shocks. They found  that the circumnuclear ISM of NGC\,3393 is a complex multiphase medium, with the likely contributions of shocks in regions where radio outflows from the AGN most directly influence the ISM. Finally, they also report the presence of the Ne\,{\sc ix}~ 0.905~keV transition, which points to strong shocks driven by AGN feedback.

As seen from above, although the 'S' structure has been widely studied, to the best of our knowledge, this is the first time that extended optical coronal emission at kiloparsec scales is reported in that source.
The preferred direction of this emission coincides with that of the radio-emitting jet widely reported in the literature \citep{maksym_2017,finlez_2018,venturi_2021}.

\section{The Size of the CLR in AGN}

\label{CapSizeCoronal}

As discussed in Sect.~\ref{sec:extension}, most galaxies of this work had previous reports of extended emission in the [\ion{Fe}{vii}] line at scales of a few hundreds of parsecs at the most. In Table \ref{tab:clr_size} we summarize the maximum coronal line extension found here for the [\ion{Fe}{vii}] (column 2) and [\ion{Fe}{x}] (column 3) emission by means of MUSE IFU spectroscopy for the 9 AGN of the sample. It can be seen that the size of the coronal emission measured by means of the [\ion{Fe}{vii}] in some sources has increased by about an order of magnitude with respect to previous values reported in the literature. Similar results were also found for [\ion{Fe}{x}].  

\begin{table}
    \centering
    \begin{tabular}{lcc}
     \hline \hline
     Galaxy &  {[}\ion{Fe}{vii}{]} & {[}\ion{Fe}{x}]\\
     & (pc) & (pc) \\ \hline

 Circinus  &  684$\pm$36  & 408$\pm$48   \\
 NGC\,1386 &  519$\pm$32  &  154$\pm$21  \\ 
 NGC\,5643 &  845$\pm$46  &  317$\pm$30  \\
 NGC\,1068 &  2820$\pm$197  & 693$\pm$98  \\
 ESO\,428  &  638$\pm$86  &  202$\pm$43  \\
 NGC\,3081 &  472$\pm$94 &  365$\pm$62  \\
 NGC\,5728 &  2023$\pm$235  & 282$\pm$78  \\
 IC\,5063  &  1193$\pm$139  & 696$\pm$46   \\
 NGC\,3393 &  2631$\pm$203  & 1241$\pm$152   \\
  \hline
    \end{tabular}
    \caption{Maximum distance from the AGN where the coronal emission of [\ion{Fe}{vii}] (column 2) and [\ion{Fe}{x}] (column 3) was detected.}
    \label{tab:clr_size}
\end{table}

\begin{figure*}
    \centering
    \includegraphics[width=8.2cm]{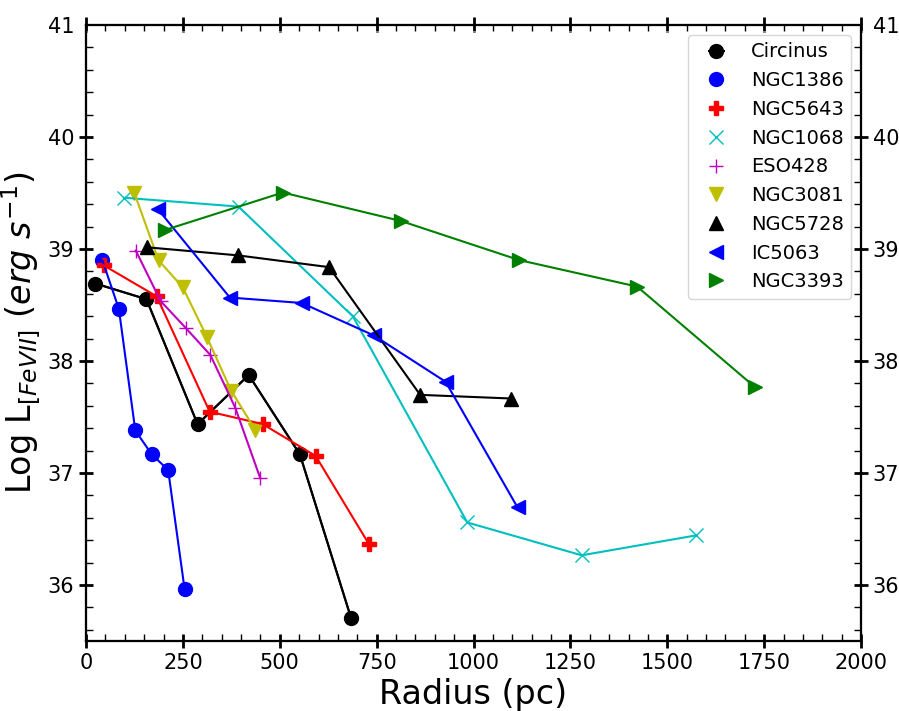}  
    \includegraphics[width=8.2cm]{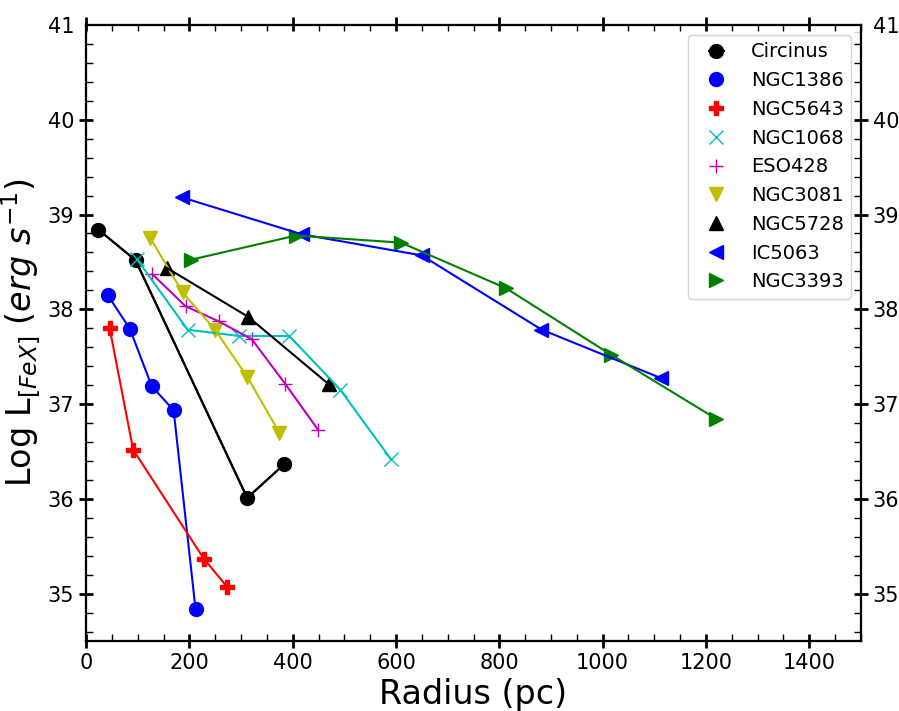}  
    \caption{Luminosity vs radial distance of the [Fe\,{\sc vii}] (left panel) and \fe10b\ (right panel) coronal emission found for the galaxy sample. Each object is identified with a different symbol or colour combination, as marked in the upper right corner of each panel.}
    \label{fig:radial_profile}
\end{figure*}

\begin{figure*}
    \centering
    \includegraphics[width=17cm]{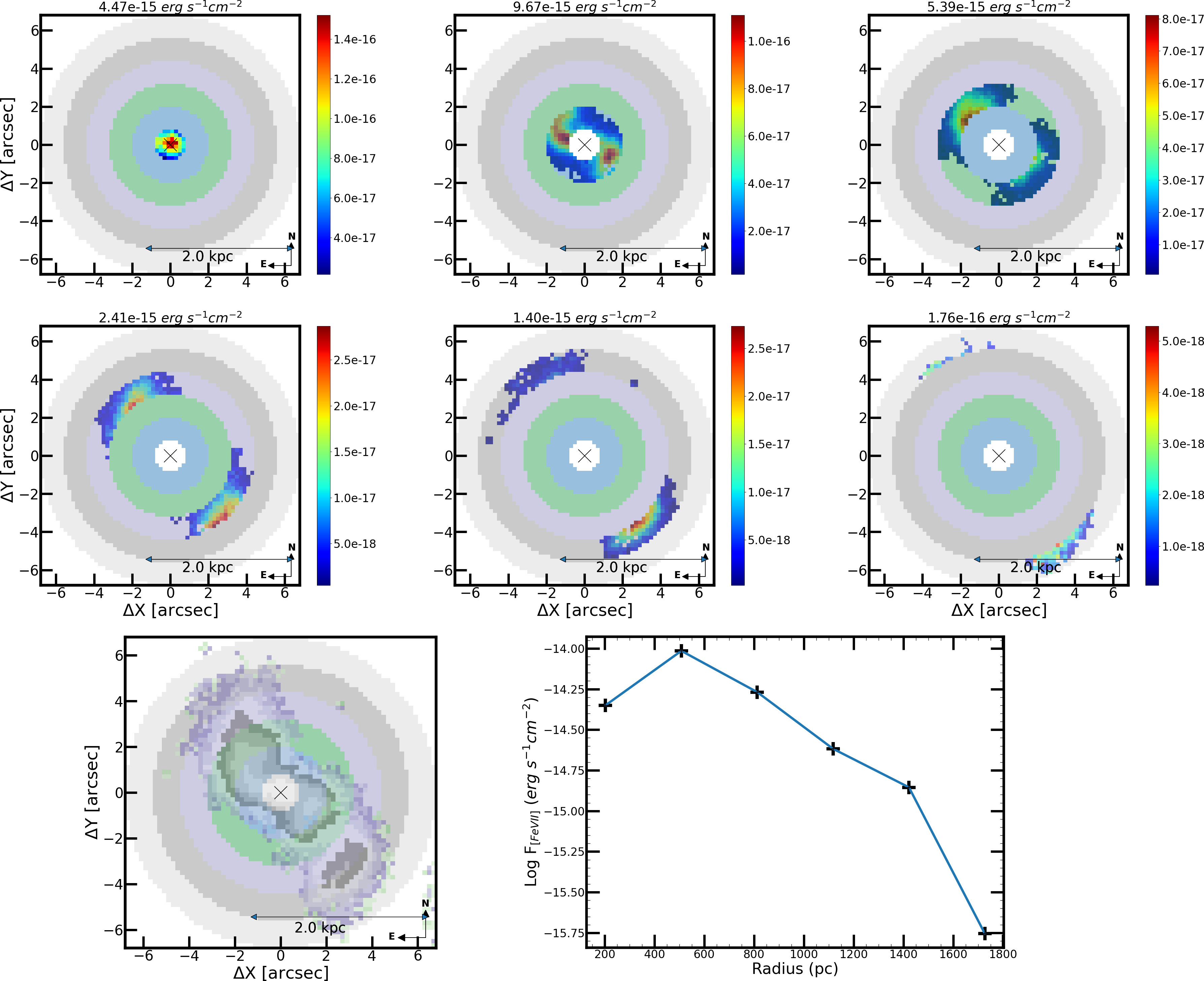}  
    \caption{Example of the method employed to construct the radial flux distributions shown in Figure~\ref{fig:radial_profile} and applied to the [\ion{Fe}{vii}] emission in NGC\,3393. We first divided the emission region into a nuclear region and five concentric rings. The nuclear region is centred at the AGN position in the $V-$band and has a radius close to that of the seeing.  Column 2 of Table~\ref{tab:relation_contrib_nuc_ext} lists the radius of the nuclear region in each AGN. We then divided the emission region from the nuclear radius to the maximum emission in the resolved spaxels into 5 concentring rings of equal radius. The nuclear aperture and the rings, overlaid to the [Fe\,{\sc vii}] emission is plotted in the left panel of the bottom row. Then, we integrated the flux within the nuclear region and each ring to produce a plot of flux vs radius $R$, shown in the right panel of the bottom row. The individual panels in the first two rows show the flux at the nuclear region and within each concentric ring. The number at the top of each panel shows the corresponding integrated flux emission.}
    \label{fig:rings_N3393}
\end{figure*}

Figure~\ref{fig:radial_profile} shows the variation of the integrated flux of the [Fe\,{\sc vii}] (left panel) and \fe10b\ (right panel) emission with the distance to the AGN. It was constructed by dividing the corresponding coronal emission map in a circular aperture of radius similar to that of the seeing of the observation (see second column of Table~\ref{tab:relation_contrib_nuc_ext}), centred at the AGN position, and five outer concentric rings. The innermost circular region is named "nuclear emission region" hereafter. The flux inside the nuclear emission region and that of each ring was integrated to build the plot of flux vs distance.  The procedure described above is illustrated in Figure~\ref{fig:rings_N3393}, which shows the results obtained for the [Fe\,{\sc vii}] emission in NGC\,3393. In the top of each individual panel we annotated the integrated flux measured in the corresponding region. All other radial profiles in Figure~\ref{fig:radial_profile} were constructed similarly and are shown in Figures~\ref{fig:rings_Circinus} to~\ref{fig:rings_N3393}. 

It is important to mention that the maximum radial extension of the coronal lines in Figure~\ref{fig:radial_profile} takes into account the most extended resolved spaxel in  Figures~\ref{fig:ext_circinus} to~\ref{fig:ext_NGC3393}. For that reason, they are smaller than the values reported in Figure~\ref{fig:radial_profile}. Moreover, the extension quoted is the projection of that distance in the plane of the sky. They were not deprojected into the galaxy plane because of the large uncertainties in determining the inclination of that emission relative to the line-of-sight. Therefore, the values found could represent a lower limit to the size of the CLR. However, we believe that it does not affect the results considerably due to the following reasons. Two sources have a very compact CLR and radio emission (NGC\,1386 and NGC\,3081). In the Circinus Galaxy the ionisation cone, where the CLR is located, is very close to the plane of the sky, and NGC\,5643 and NGC\,3393 are face-on spirals. Thus, in more than half of the sample, the deprojection would not be an issue.

From Figure~\ref{fig:radial_profile} we see that the size of the CL emission varies considerably from source to source, from a few hundred parsecs in NGC\,1386 to nearly 2~kpc in NGC\,1068 and NGC\,3393. In NGC\,1068 we did not take into account the emission region found at 2.8~kpc because it likely is due to a molecular outflow (see Sect.~\ref{subsec:n1068}) instead of AGN-driven. Our analysis confirms that Circinus is not the only object with extended coronal emission to scales of several hundreds parsecs \citep{appenzeller_1991,ardila_2020}. In fact, most galaxies display a CL region extended up to kiloparsecs scales. Such is the case of NGC\,3393, IC\,5063, NGC\,5728, and NGC\,1068. Previous efforts at resolving the [\ion{Fe}{vii}] emission by \citet{negus_2021,negus+23} suggest CL emission at kiloparsec-scales from the central source.  Their results show CLR reaching 1.3-23~kpc from the galactic centre, with an average distance of 6.6~kpc. We notice that their sample is dominated by faint sources and that the spatial resolution of MANGA is considerably poorer than that of MUSE. Therefore, \citet{negus_2021} and \citet{Negus2023} results should be treated with caution and as upper limits.

The results for [\ion{Fe}{x}] are yet more surprising. With an ionisation potential of 235~eV, the emission region of that line has usually been found to be constrained to the central few tens of parsecs from the AGN. Here, to the best of our knowledge, we detected for the first time that emission in  all objects of the sample at scales larger than 150~pc. In both IC~5063 and NGC\,3393, the emission region of that line reaches $\sim$1.2~kpc from the AGN, with most sources showing emission region sizes between $\sim$300 -- 500~pc. Moreover, in NGC\,3393, the emission peaks outside the nucleus, at $\sim$500~pc from the AGN.

Figure ~\ref{fig:radial_profile} also reveals very interesting trends regarding the radial distribution of the coronal emission. Except in NGC\,1386, in most cases the decrease in flux with distance is not as sharp as expected in a central source ionisation scenario. Secondary peaks of emission along the radial direction are observed, with the decline in emission flux being $<$1 dex in the inner 500~pc. NGC\,3393 is yet more extreme as the emission flux peaks at $\sim$500~pc from the centre and, at 1~kpc away from the AGN, the flux measured is half the one at the nucleus. This result is indicative of either strongly varying physical conditions with distance (i.e., the gas density decreases sharply outside the nuclear region) or there are additional sources of gas ionisation like shocks. In the first case, \citet{ferguson1997} showed that gas density of 1000~cm$^{-3}$ can produce [\ion{Fe}{vii}] emission at a maximum distance of 102~pc from the AGN. At larger distances that line could not be formed because of the strong dilution of the AGN radiation. Also, a decreasing density will hamper the detection of  coronal emission  as its flux goes proportional to the density, which will make their detection very hard for such rather low abundance metallic ion. For [\ion{Fe}{x}] the physical conditions for developing an extended emission due solely to the radiation from the central source are even more stringent than for [\ion{Fe}{vii}]. At AGN luminosities similar to that of our sample ($L_{\rm bol} \sim 10^{43.5}$~erg\,s$^{-1}$) that line cannot be formed at distances larger than 20~pc from the AGN regardless of the gas density \citep[][and references therein]{ferguson1997,Contini_2001,contini+02}; see Fig 13 in \citet{ardila_2006}. Therefore, the pure central AGN ionisation scenario hardly explain the observed size of the CLR.

The second hypothesis involves shocks, which have been probed to be an efficient mechanism to enhance coronal line emission in regions located at distances larger than 200~pc from the AGN \citep{fonseca-faria_2021, fonseca-faria+23}. These shocks can be powered either by a radio-jet or by a starburst region. Indeed, the first possibility was found to be at work in Circinus and IC\,5063. The above authors support that claim by means of photoionisation modelling that employs the combined effect of radiation from the central source and shocks. This approach indeed reproduces the observed emission line ratios, including those involving CLs.

At this point, it is worth comparing our results to those derived for Seyfert~1 galaxies. To this purpose we employ the results of \citet{muller_2011}, who used SINFONI/VLT and OSIRIS/Keck adaptive optics IFU data to study the CLR of 7 local {\it bona-fide} AGN by means of the [\ion{Si}{vi}]~1.963$\mu$m line (IP = 168eV) in the near-infrared region. Their sample was composed of four Seyfert~1 (NGC\,3783, NGC\,4151, NGC\,6814, and NGC\,7469), one Seyfert~1.9 (NGC\,2992), and two Seyfert~2 (Circinus and NGC\,1068).  For the Type~I sub-sample, they report very compact emission regions, with the most extended one amounting 140~pc in radius (NGC\,3783). The other sources  displayed CLR radius equal or smaller than 100~pc. These numbers are about an order of magnitude smaller than the results found here.  It is possible that the CLR in most sources of \citet{muller_2011} extends further out. However, due to the small field-of-view (FoV) of the detectors employed (a few hundred of parsecs by side; see Table~1 in \cite{muller_2011}), the full extent of the CLR could not be fully assessed. For example, Circinus and NGC\,1068, the two AGN common to theirs and our work, display a [\ion{Si}{vi}]1.963~$\mu$m emission radius of 8~pc and 150~pc, respectively. Here, we found for [\ion{Fe}{x}] (a CL with still larger IP than that of [\ion{Si}{vi}]) that the their emission region extends to $\sim$400~pc and $\sim$700~pc from the AGN, respectively. 

It is also possible that projection effects may play a very important role in the case of Type I AGNs. If we assume an extended CLR aligned along the radio-jet in Type~Is, as in the Type~II AGN studied here, the emission would be considerably more beamed towards the observer in the former Type. As a result, the nuclear component should dominate over the extended component. This indeed seems to be the case. \citet{muller_2011} report a CLR consisting of a bright compact nucleus and an extended diffuse emission along a preferred position angle. In whaterver scenario, for a fair comparison with Type~I AGN, a sample observed with IFUs with larger FoV should be employed.

\subsection{Relative contribution of the nuclear and extended coronal emission}

\begin{figure*}
\includegraphics[width=8cm]{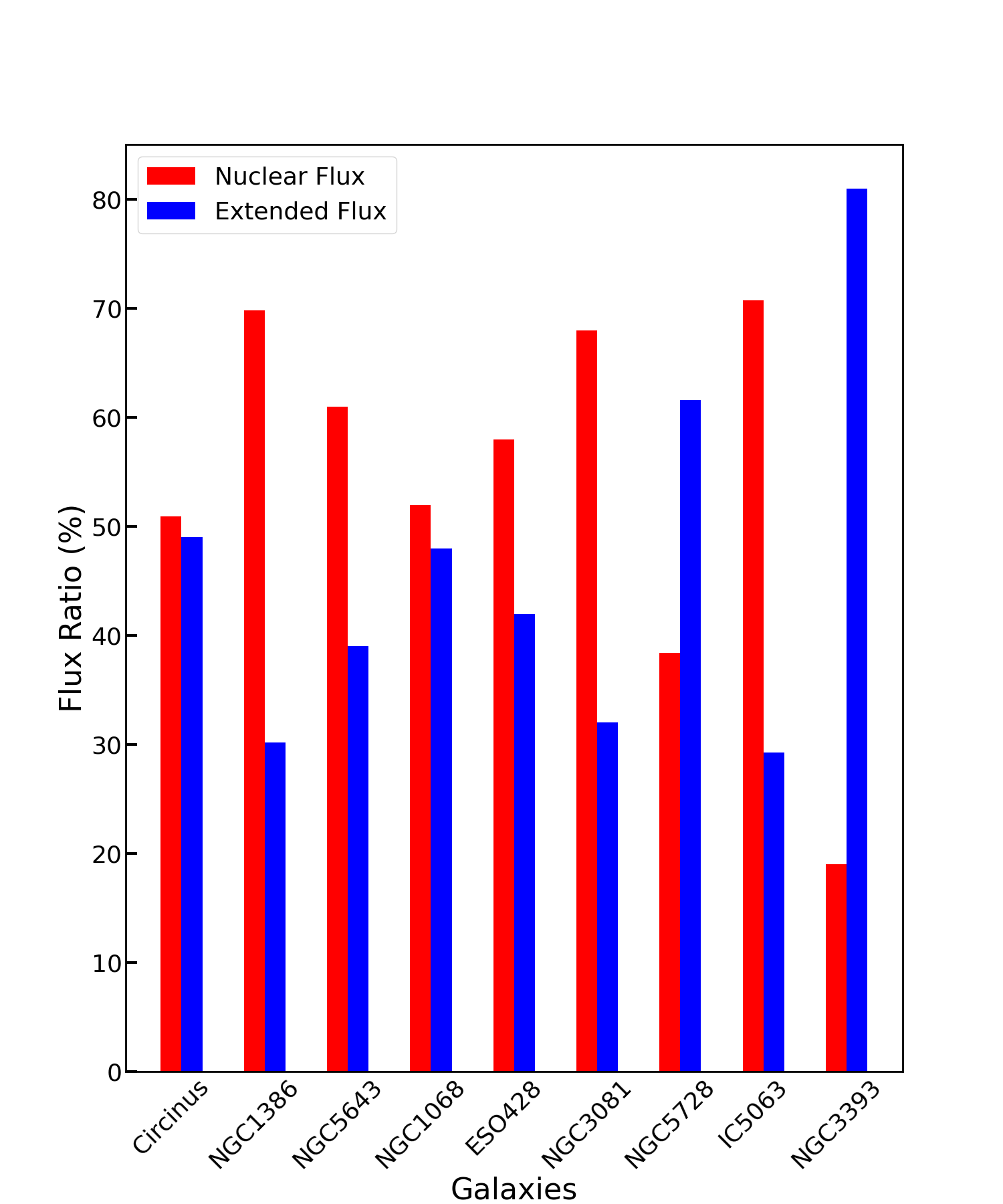}
\includegraphics[width=8cm]{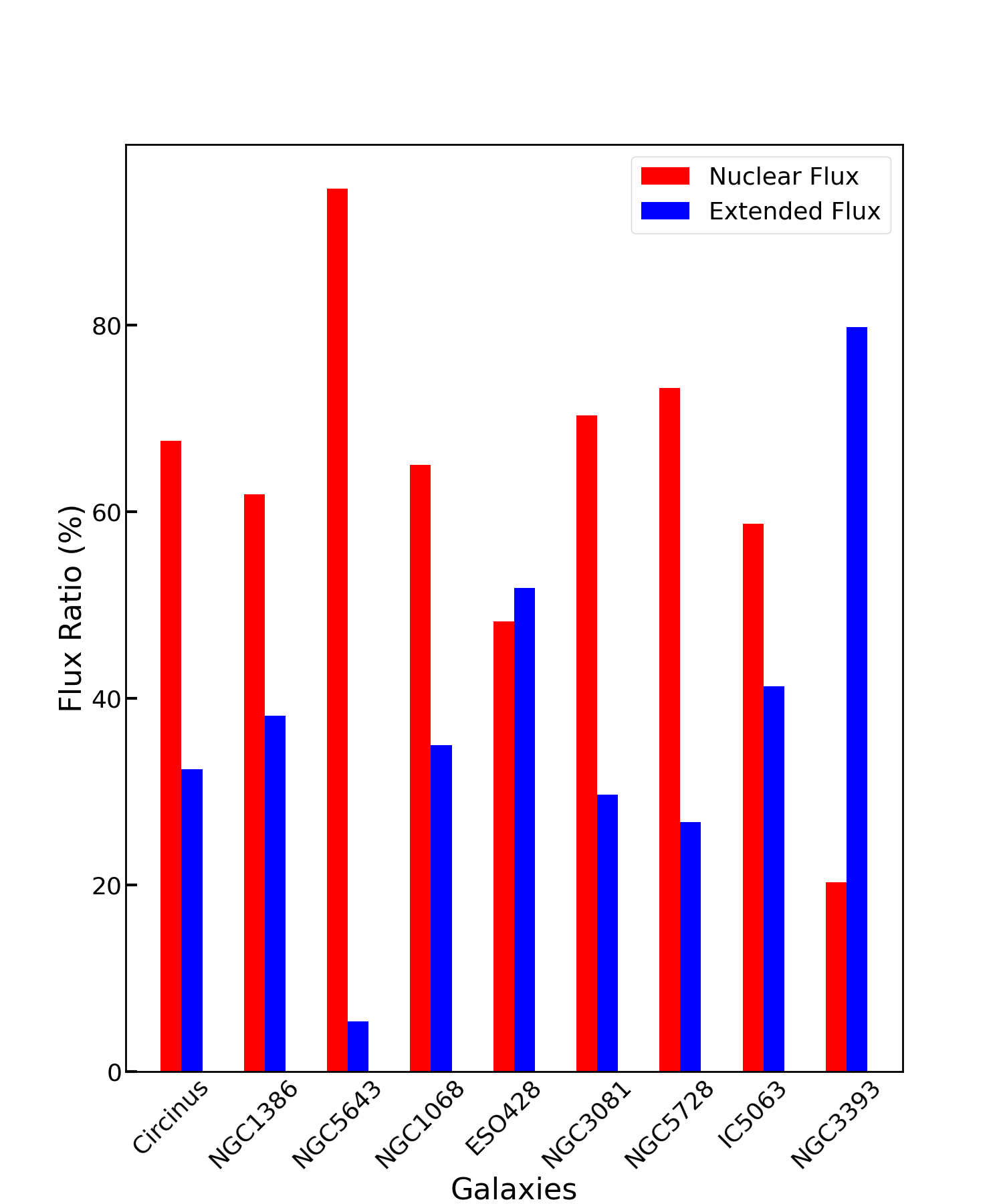}  
    \centering
    \caption{Relative contribution of the nuclear (red bar) and the extended (blue bar) emission to the total coronal emission flux for [\ion{Fe}{vii}] (left panel) and [\ion{Fe}{x}] (right panel) in the galaxy sample. The nuclear flux is  the integrated emission in the resolution beam of the observations, $R_{\rm nuc}$ in Table~\ref{tab:relation_contrib_nuc_ext}. The extended flux is the emission outside the nuclear one.} 
    \label{fig:barplot_fe7and10}
\end{figure*}

We investigate the relative contribution of the nuclear and extended CL emission to the total CL flux observed. To this purpose we obtained the total integrated CL flux in each AGN and divided it into two contributions: that of the nucleus (nuclear emission) and the one detected out of it (extended emission). The results are plotted in Figure~\ref{fig:barplot_fe7and10} while the individual values are listed in Table~\ref{tab:relation_contrib_nuc_ext}. Columns 4 and 5 of that table correspond to the nuclear and extended [\ion{Fe}{vii}] emission, respectivey, while columns 7 and 8 show the same for [\ion{Fe}{x}].

\begin{table*}
    \centering
    \caption{Total flux and Relative contribution of the nuclear and extended [\ion{Fe}{vii}] and [\ion{Fe}{x}] emission.}
    \label{tab:relation_contrib_nuc_ext}
    \begin{tabular}{lccccccc}
     \hline \hline
     Galaxy & $R_{\rm nuc}$ & \multicolumn{3}{|c|}{{[}\ion{Fe}{vii}{]} } & \multicolumn{3}{|c|}{{[}\ion{Fe}{x}]}\\
     \hline
     & (pc)  & $F_{\rm \left[\ion{Fe}{vii}\right]Tot}^*$ & nuclear & extended & $F_{\rm \left[\ion{Fe}{x}\right]Tot}^{**}$ & nuclear & extended \\ \hline

 Circinus  &  24 & 45.5$\pm$7.7 & 49$\%$    & 51$\%$ & 477$\pm$81    & 68$\%$  & 32$\%$   \\
 NGC\,1386 &  42 & 3.70$\pm$0.16 & 70$\%$   & 30$\%$ & 7.42$\pm$0.74 & 62$\%$  & 38$\%$   \\ 
 NGC\,5643 &  46 & 3.42$\pm$0.34 & 62$\%$   & 38$\%$ & 1.96$\pm$0.27 & 95$\%$  &  5$\%$  \\
 NGC\,1068 &  100 & 44.6$\pm$3.6 & 53$\%$   & 47$\%$ & 42.0$\pm$9.14 & 65$\%$  & 35$\%$   \\
 ESO\,428  &  129 & 2.54$\pm$0.21 & 58$\%$  & 42$\%$ & 7.55$\pm$1.34 & 48$\%$  & 52$\%$   \\
 NGC\,3081 &  125 & 3.63$\pm$0.31 & 68$\%$  & 38$\%$ & 6.22$\pm$0.68 & 70$\%$  & 30$\%$   \\
 NGC\,5728 &  157 & 1.40$\pm$0.23 & 38$\%$  & 62$\%$ & 1.94$\pm$0.42 & 74$\%$  & 26$\%$   \\
 IC\,5063  &  186 & 1.9$\pm$0.17  & 71$\%$  & 29$\%$ & 15.2$\pm$2.73 & 59$\%$  & 41$\%$    \\
 NGC\,3393 &  203 & 2.37$\pm$0.11 & 19$\%$  & 81$\%$ & 5.11$\pm$0.56 & 20$\%$  & 80$\%$    \\
   \hline
 \multicolumn{8}{l}{$^*$ Total {[}\ion{Fe}{vii}{]} flux in units of 10$^{-14}$ erg\,cm$^{-2}$\,s$^{-1}$} \\
 \multicolumn{8}{l}{$^{**}$ Total {[}\ion{Fe}{x}{]} flux in units of 10$^{-15}$ erg\,cm$^{-2}$\,s$^{-1}$} \\
    \end{tabular}
   \end{table*}


From Figure~\ref{fig:barplot_fe7and10}, we found a very large variation of the contribution of the extended emission to the total flux in the galaxy sample. For the  [\ion{Fe}{vii}] line, it varies from 29\% up to 81\%, while for [\ion{Fe}{x}] the extended emission accounts from $<5\%$ to 80\% of the total flux.  IC\,5063 and NGC\,1386 are the AGN with the smallest contribution of the extended [\ion{Fe}{vii}] emission ($\sim$30\%), followed by NGC\,5643 and NGC\,3081 (both with 38\%). In the former two AGN, the nuclear emission accounts for about 70\% ~of the total emission even though in IC\,5063 the extended CL gas is observed up to distances of 1.1~kpc from the AGN. In contrast, in NGC\,3393 and NGC\,5728 the extended emission dominates over the nuclear one. We have already found that in the former AGN, the peak intensity of the [\ion{Fe}{vii}] emission gas was located at $\sim$500~pc from the central engine. Now we see that the extended component of that line contributes with 81\%~of the total flux. A similar result in that same AGN was observed for the [\ion{Fe}{x}] emission, where the extended component carries 80\% of the total flux. In NGC\,5728, although the [\ion{Fe}{vii}] emission peaks at the nuclear region, the extended component carries a large fraction of the total flux, accounting to 62\%. For [\ion{Fe}{x}], though, the nuclear region dominates (74\% of the total flux). We notice that both in Circinus and NGC\,1068 the nuclear and extended emission of [\ion{Fe}{vii}] contributes with approximately 50\% to the total flux while for [\ion{Fe}{x}] the nuclear component carries $\sim$65 of the total flux. 

The above results strongly contrast to the ones reported by \citet{muller_2011}. From the nuclear and total flux of the [\ion{Si}{vi}] line listed in their Table~3 (columns~2 and~3, respectively) for the four Seyfert~1 of their sample, the nuclear component is responsible for nearly 70\% of the total flux emission. Thus, the CLR is basically nuclear-dominated in Type I AGN. Here, we found that some Type~II AGN display a CLR that is clearly extended-dominated. For instance, in NGC\,3393 the nuclear contribution to the total flux in the [\ion{Fe}{vii}] and [\ion{Fe}{x}] lines amounts to only 20\% in both cases (see Table~\ref{tab:relation_contrib_nuc_ext}). ESO\,428 is slightly extended-dominated, particularly in the [\ion{Fe}{x}]. It is interesting to note that our results for NGC\,1068, one of the two Seyfert~2s of \citet{muller_2011} common to both works, are in very good agreement despite the lines studied are different. For this AGN about half of the nuclear coronal emission contributes to the total emission. In Circinus, the other Seyfert~2 in common to both works, a fair comparison is not possible because of the small FoV of the SINFONI dectector employed by \citet{muller_2011}. The bulk of the extended emission found here is out of sight of the field explored by them. 

\begin{figure*}
    \includegraphics[width=18.cm]{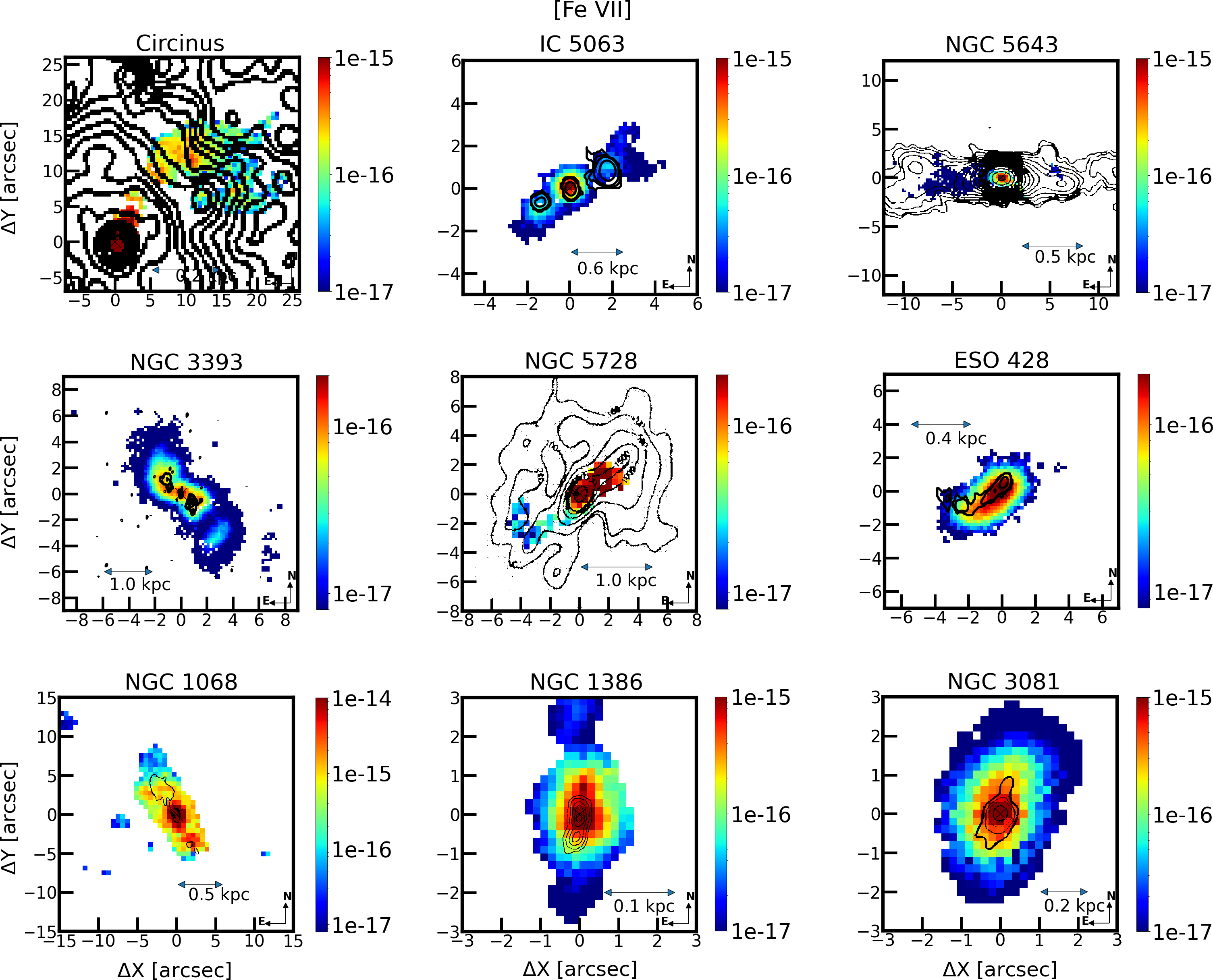}
    \caption{[\ion{Fe}{vii}] flux distribution maps overlaid with the corresponding radio emission (black contours). For Circinus, the ATCA 13\,cm contours are plotted \citep{elmouttie+98}. For IC\,5063, we plot the 1.7\,cm map \citep{morganti+07} while for NGC\,5728 and ESO\,428, the 6\,cm radio map \citep{leipski+06} is shown. For the remainder of the objects, the 3.6~cm VLA contours \citep{mundell+09} are drawn. The colour bar is in units of erg\,cm$^{-2}$\,s$^{-1}$}.
    \label{fig:Fevii_com_radio}
\end{figure*}

At this point, we notice that all objects of our sample have small-scale jets that, although not as powerful as those found in radio-loud AGN, may still deposit a significant amount of energy in the off-nuclear region. If, for instance, the jet-driven shocks are strong at some locations in the medium, it will produce [\ion{Fe}{x}] first (or even higher IP species) at the shock front, and then further out extended coronal gas of lower IP (i.e., [\ion{Fe}{vii}]). This scenario is possible as the CL emission in all cases is strongly aligned with the axis of the jet propagation and co-spatial to the radio emission, as can be seen in Figure~\ref{fig:Fevii_com_radio}. Therefore, it is tempting to consider the effect of jet-driven shocks as the source of local gas excitation, leading to the enhancement of the CL emission. This was already probed in IC\,5063 \citep{fonseca-faria+23}, where such an effect is noticed. In NGC\,3393 \citet{maksym+19}, using $Chandra$ X-ray observations, had already shown an increase of the gas excitation in the regions coinciding with the radio peaks, as well as at $\sim 5\arcsec$ to the SW, where the bean-like feature is located. Moreover, in these locations, an increase of X-ray Ne\,{\sc ix} emission is observed, coincident with the places where [\ion{Fe}{vii}] and [\ion{Fe}{x}] peak. X-ray modelling carried out by \citet{maksym+19} in NGC\,3393 suggests that the off-nuclear regions of enhanced Ne\,{\sc ix} emission is consistent with having a collisional component with flux comparable to that of the reprocessed photoionizing emission. 

We also notice that in NGC\,5728, \citet{TrindadeFalcao2023} found from $Chandra$ soft and hard X-ray analysis that the bicone spectra (where the CL is located) is dominated by a mix of photoionization and shocked gas emission, with shock velocities of $\sim$1100 - 1200~\kms\ and  likely explained by jet-ISM interactions. We will examine, in the next section, that there is a very good coincidence between the radio-jet and the extended region with the largest values of coronal emission. We notice also that in all the objects mentioned above, extended X-ray emission has also been detected, with models pointing out to a collisional gas component, likely attributed to shocks. 

\subsection{Assessing the role of the radio-jet in shaping the CLR}
\label{sec:radio}

The results gathered in the previous sections prompted us to investigate the role of the radio-jet in shaping the CLR in the AGN of the sample. This hypothesis is further supported by Figure~\ref{fig:Fevii_com_radio}, where it can be seen that the CLR is strongly shaped by the radio-jet. To this purpose, we collected  information in the literature about the jet size and the corresponding flux at 1.4~GHz for the galaxy sample. The selection of that frequency is motivated by the fact that it is the most common radio frequency where data from our sample are available. If the flux at that specific frequency was not available, we employed the one reported at another frequency (8.4~GHz or 10~GHz) and assumed a power-law spectrum ($S_\nu \propto \nu^{-\alpha}$) with $\alpha$ = 1 to rescale to the 1.4 GHz. If $\alpha$ were available, we employed the one published instead. Table~\ref{tab:radio_data} lists the values found and the corresponding references. We notice that in some cases, error bars were not reported in the literature. For this reason, they are missing in that Table. The [\ion{Fe}{vii}] and [\ion{Fe}{vii}] total luminosities were derived using the total flux listed in columns 2 and 5 of Table~\ref{tab:relation_contrib_nuc_ext}. The distance adopted, $D$, is listed in the last column of Table~\ref{tab:radio_data} and was taken from NED.

\begin{table*} \centering  
\caption{Radio data and coronal line luminosity} \label{tab:radio_data}
\begin{tabular}{llccccc} \hline \hline

ID & Galaxy & $R_{\rm jet}$ & $S_{\rm 1.4 GHz}$ & $L_{\rm \left[FeVII\right]}^*$  &  $L_{\rm \left[FeX\right]} ^*$ & $D^{**}$  \\
   &         &  [kpc]        &   [mJy] & [$\times10^{38}$~erg\,s$^{-1}$] &  [$\times10^{37}$ erg\,s$^{-1}$] & Mpc\\
\hline
1  & Circinus     & 1.2$^{\rm a}$  &  2571$^{\rm a}$  &  7.89$\pm$1.00  & 82.6$\pm$14 & 3.8 \\ 
2  & NGC\,1386    & 0.08$^{\rm b}$ &  61.8$^{\rm f}$  &  5.67$\pm$0.25   & 11.4$\pm$1.2 & 11.3 \\
3  & NGC\,5643    & 2.0$^{\rm b}$  & 70.2$^{\rm h}$   &  9.61$\pm$1.10  & 5.50$\pm$0.75 & 15.3 \\
4  & NGC\,1068    & 0.8$^{\rm b}$  & 840$^{\rm j}$    &   144$\pm$12   & 138$\pm$30 & 16.4 \\
5  & ESO\,428-G14 & 0.6$^{\rm e}$  & 113.7$^{\rm h}$  & 14.0$\pm$1.20   & 41.1$\pm$7.3 & 21.3 \\
6  & NGC\,3081    & 0.14$^{\rm f}$ &  5.7$^{\rm f}$   & 44.6$\pm$3.9   & 75.8$\pm$8.7 & 32 \\
7  & NGC\,5728    & 1.75$^{\rm d}$ &  70$^{\rm d}$    & 25.1$\pm$4.2   & 33.9$\pm$7.6 & 38.5 \\
8  & IC\,5063     &  1.0$^{\rm b}$ & 1260  $^{\rm g}$ &  52.0$\pm$4.7  & 414$\pm$72 &  47.7 \\
9  & NGC\,3393    & 2.12$^{\rm c}$ &  54.4$^{\rm i}$  & 75.8$\pm$3.4   & 162$\pm$18 & 51.9 \\
  \hline
\multicolumn{7}{l}{$^*$ This work; see text for further information. $^{**}$ Taken from NED.} \\
\multicolumn{7}{l}{References: a - \citet{elmouttie_1998}; b - \citet{venturi_2021}; c - \citet{morganti+99}} \\
\multicolumn{7}{l}{d - \citet{singh+13}; e - \citet{condon+98}; f - \citet{mundell+09}; g - \citet{morganti_1998};} \\
\multicolumn{7}{l}{h - \citet{leipski+06}; i - \citet{koss+15}; j - \citet{garciaburillo+14}.}
\end{tabular}
\end{table*}

\begin{figure}
    \includegraphics[width=8.8cm]{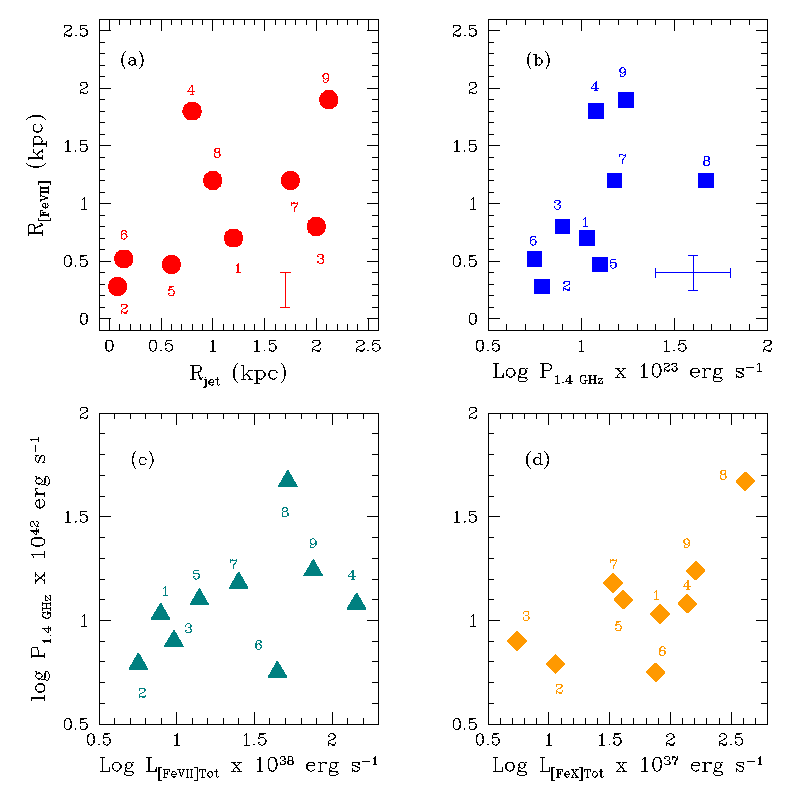}
    \caption{(a) Radius of the [\ion{Fe}{vii}] emission region (R$_{\rm \left[FeVII\right]}$, in kpc) as given in Figure~\ref{fig:radial_profile} vs jet radius (R$_{\rm jet}$, in kpc, see Table~\ref{tab:radio_data}). The numbers close to each data point refers to the ID of the individual sources as indicated in the first column of Table~\ref{tab:radio_data}. The mean error bar in R$_{\rm \left[FeVII\right]}$ is indicated.   Values of R$_{\rm jet}$ were taken from the literature and lack of uncertainties; (b) Same as (a) but now vs the Logarithm of the jet power at 1.4 GHz (P$_{\rm 1.4 GHz}$, see Table~\ref{tab:radio_data}). The error bar in P$_{\rm 1.4 GHz}$ is the maximum value in this quantity for NGC\,1386 (see Table~\ref{tab:radio_data}). All other sources have smaller uncertainties; (c) Logarithm of P$_{\rm 1.4 GHz}$ vs the total [\ion{Fe}{vii}] luminosity, L$_{\rm \left[FeVII\right]}$. Uncertainties in this quantity amount to 10\% in log at the most; (d) Logarithm of the jet power at 1.4 GHz vs the total [\ion{Fe}{x}] luminosity, L$_{\rm \left[FeVII\right]}$. Uncertainties in this quantity amounts to 20\% at the most.}
    \label{fig:jetvsclr}
\end{figure}

Figure~\ref{fig:jetvsclr} displays plots relating the jet radius and the jet power at 1.4 GHz to various CLR properties such as the radius of the emission region found here and the total luminosity of the coronal lines. In panel (a) we plot the the radius of the [\ion{Fe}{vii}] emission region ($R_{\rm \left[FeVII\right]}$) versus the radius of the jet ($R_{\rm jet}$). The number close to each point identifies the corresponding AGN as listed in the first column of Table~\ref{tab:radio_data}. A clear positive trend is seen in the data, with the radius of the CLR scaling linearly with the jet radius. This result strongly suggest that the size of coronal emission is dependent of the size of the jet. 

Panel (b) relates the jet power at 1.4 GHz ($P_{\rm 1.4GHz}$) with $R_{\rm \left[FeVII\right]}$. $P_{\rm 1.4GHz}$ was determined by applying Eq. (16) from \citet{birzan+08}, which relates the cavity (jet) power and the 1.4~GHz
radio luminosity of the source in objects showing cavities in
their X-ray haloes filled by radio emission. In this process, the radio flux at 1.4~GHz, $S_{\rm 1.4~GHz}$, listed in column 4 of Table~\ref{tab:radio_data} was employed. It can be seen that a positive trend is also observed. Overall, the more powerful the jet the more extended the CLR. 
The bottom two panels relates $P_{\rm 1.4GHz}$ to the total [\ion{Fe}{vii}] luminosity ($L_{\rm \left[FeVII\right]Tot}$, panel c) and the total [\ion{Fe}{x}] luminosity ($L_{\rm \left[FeX\right]Tot}$, panel d). The positive trend in both plots is evident.

All plots shown in Figure~\ref{fig:jetvsclr} support our hypothesis that in radio weak AGN the extended coronal emission is strongly jet-driven. The jet would deposit part of its power as kinetic energy to drive shocks into the ISM and produce coronal emission. The most powerful the jet, the most extended and luminous the CLR. We ruled out any potential luminosity vs luminosity relationship effect, as the increased in CL luminosity do not correlate with the distance to the source.

\section{Ionisation degree of the emission gas}
\label{sec:degree_gas}

\begin{figure*}
    \includegraphics[width=18cm]{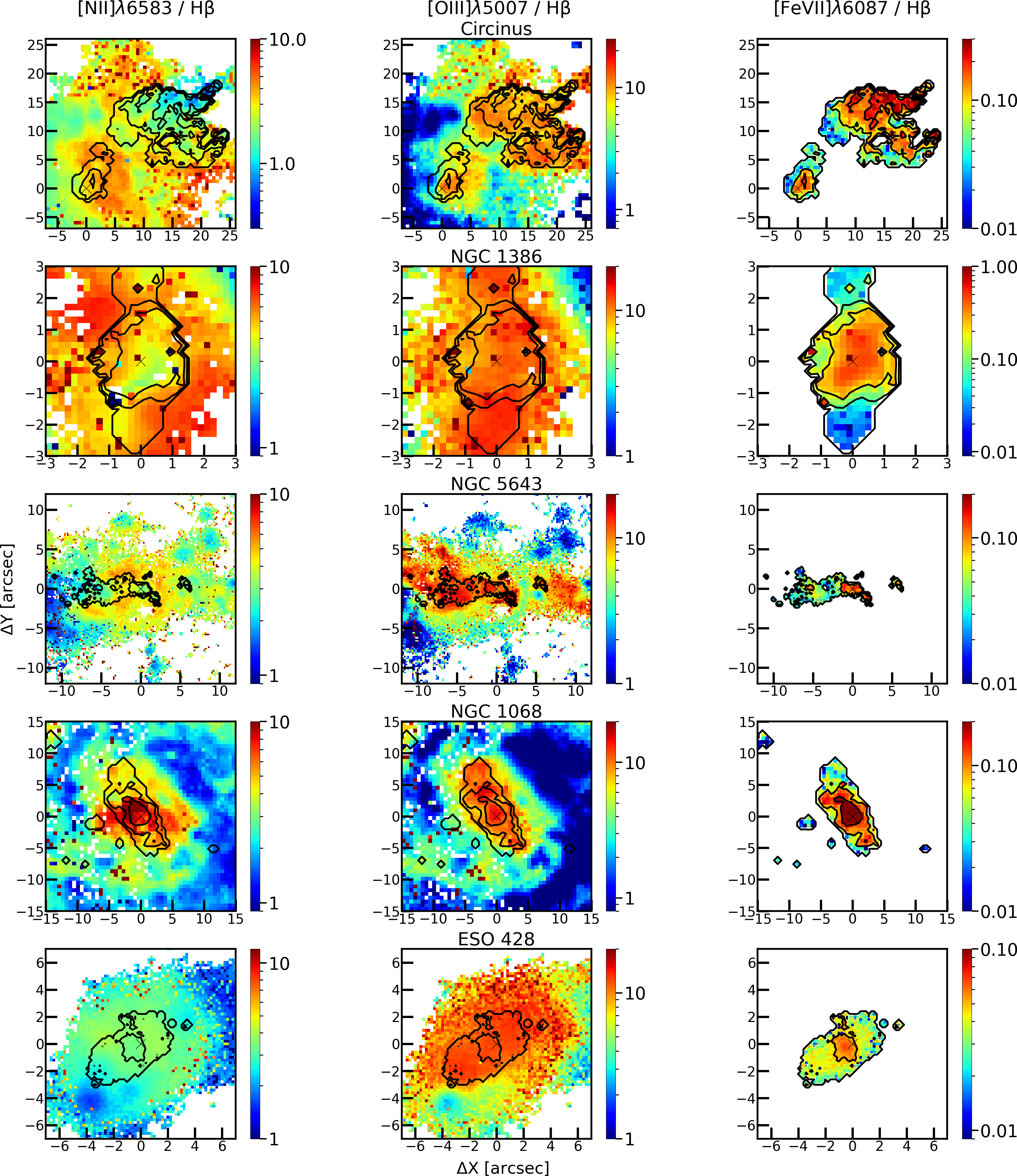}
    \caption{Extinction corrected emission line flux ratios [\ion{N}{ii}]/H$\beta$ (left column), [\ion{O}{iii}]/H$\beta$ (middle column), and [\ion{Fe}{vii}]/H$\beta$ (right column) for Circinus, IC\,5063, NGC\,5643, NGC\,3393, and NGC\,5728 (top to bottom). The contours in black represents the extent of the [\ion{Fe}{vii}] emission.}
    \label{fig:gals_radio1}
\end{figure*}

\begin{figure*}
    \includegraphics[width=18.cm]{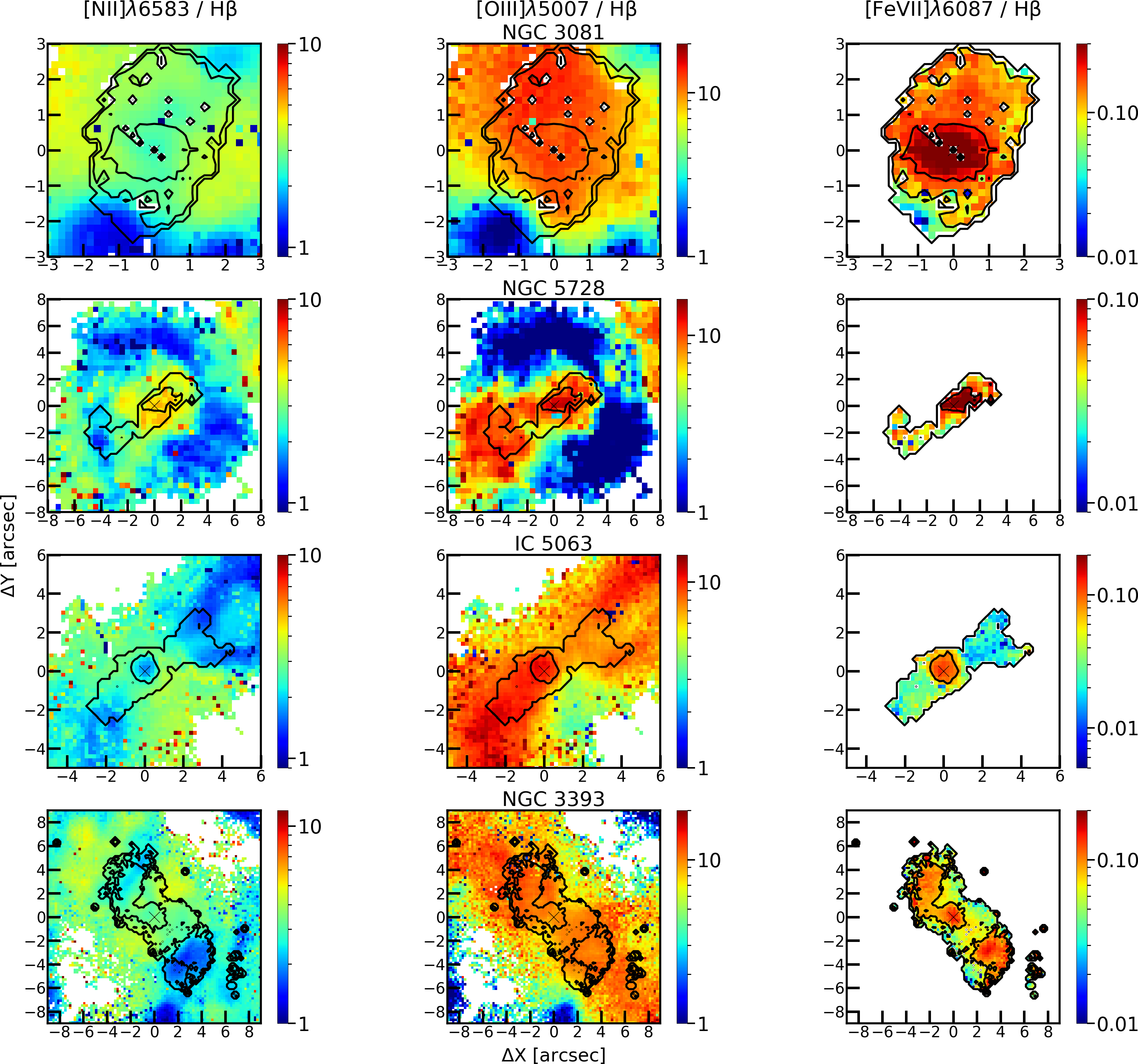}
    \caption{Extinction corrected emission line flux ratios [\ion{N}{ii}]/H$\beta$ (left column), [\ion{O}{iii}]/H$\beta$ (middle column), and [\ion{Fe}{vii}]/H$\beta$ (right column) for ESO\,428, NGC\,1068, NGC\,1386, and NGC\,3081 (top to bottom). The contours in black represents the extent of the [\ion{Fe}{vii}] emission.}
    \label{fig:gals_radio2}
\end{figure*}

The study of the ionisation degree of a gas cloud allows the determination of the nature of the radiation source that illuminates it, that is, if the source is of stellar origin or a harder source like an AGN or a LINER. In this process, line ratios between ions are used, usually of the same element or relative to hydrogen, to avoid potential biases introduced by metallicity or differences in critical density of the transitions involved. For this purpose, we will use in this Section the lines of [N\,{\sc ii}]\,$\lambda6583$, [O\,{\sc iii}]\,$\lambda5007$ and [Fe\,{\sc vii}]\,$\lambda6087$ as representatives of low, medium, and high ionisation gas, respectively. We will also employ the H$\beta$ line as representative of a gas that is not affected by density or abundance variations across the regions examined.

In Figures~\ref{fig:gals_radio1} and ~\ref{fig:gals_radio2}, the line flux ratios [N\,{\sc ii}]\,$\lambda6583$\,/\,H$\beta$ (left column), [O \,{\sc iii}]\,$\lambda5007$\,/\,H$\beta$ (middle column) and [Fe\,{\sc vii}]\,$\lambda6087$\,/\,H$\beta$ (right column) are displayed for the 9 galaxies studied here. For comparison purposes, black contours were superimposed, representing values of the ratio [\ion{Fe}{vii}]\,$\lambda6087$\,/\,H$\beta$ in the regions where both lines are detected simultaneously. The aim is to compare the extension and variation of the values of each ratio within the extended CLR of the AGN. Each galaxy is identified at the top of the central column panels. For simplicity, throughout this section we will refer to the above three line flux ratios as [N\,{\sc ii}]\,/\,H$\beta$, [O\,{\sc iii}]\,/\,H$\beta$ and [Fe\,{\sc vii}]\,/\,H$\beta$, respectively.

Figure~\ref{fig:gals_radio1} to~\ref{fig:gals_radio2} show that in all objects [O\,{\sc iii}]\,/\,H$\beta$ displays values larger than 10 in the region that is co-spatial to [\ion{Fe}{vii}] and even outside it but along the same direction where the coronal emission is observed.  Values of up to 20 in that line ratio are detected a few kiloparsecs away from the AGN. In all the galaxy sample but NGC\,1068, the emission of both [O\,{\sc iii}] and [\ion{Fe}{vii}] is mostly confined to a structure associated with to the ionisation cone. In the cross-cone region, [O\,{\sc iii}]\,/\,H$\beta$ decreases considerably, down to values between 1-2. In contrast, [N\,{\sc ii}]\,/\,H$\beta$ is usually smaller along the the region where [\ion{Fe}{vii}] is detected and reaches higher values in the direction perpendicular to the axis where the oxygen-emitting gas and ionised iron is detected. This result was previously noticed by \citet{venturi_2021} in their study of the galaxies of the MAGNUM project. They found for IC\,5063, NGC\,5643, NGC\,1068, and NGC\,1386  a strong (up to 800–1000~\kms) and extended ($\sim$1~kpc) emission-line velocity spread perpendicular to the direction of the AGN ionisation cones and jets in all four targets. They interpret these results as due to the action of the jets perturbing the gas in the galaxy disc,  impacting on and strongly affecting its material. Ultimately, their results demonstrate that low-power jets are indeed capable of affecting the host galaxy. However, the lack of extended CL emission in the direction perpendicular to the radio jet suggests that these shocks are not energetic enough to produce highly ionised gas.
 
In IC\,5063, NGC\,5643, NGC\,5728 and ESO\,428, [Fe\,{\sc vii}]\,/\,H$\beta$ peaks at the AGN position, with values between 0.06 and 0.1. In the off-nuclear region, the value of the ratio decreases by less than one dex, remaining quite homogeneous, with values between 0.02 and 0.05, regardless of the distance to the nucleus. 
On the other hand, in NGC\,3393, [Fe\,{\sc vii}]\,/\,H$\beta$ displays mean values of 0.1 in three regions: at the AGN position, at the northeast of the AGN, along the upper part of the ``S'' shaped structure,   and in the cloud located $\sim$1.5\,kpc southwest of the central source. At other locations, [Fe\,{\sc vii}]\,/\,H$\beta$ has values between 0.01 and 0.06. We note that in the southwest region, at 1.2\,kpc of the AGN, [Fe\,{\sc vii}]\,/\,H$\beta$ reaches the highest value identified in this object (0.11). In this same region, we also found extended emission of [\ion{Fe}{x}]. These results reflect the sudden increase of gas ionisation at such distances from the central source.
 
NGC\,1068 presents relatively constant values of [O\,{\sc iii}]\,/\,H$\beta$ ($\sim$\,10) across the region that is co-spatial to [Fe\, {\sc vii}]. Furthermore, two extended regions to the north and south, at $\sim$2.5~kpc from the AGN, display high values ($\sim$\,6) of [O\,{\sc iii}]\,/\,H$\beta$. These locations are associated to the starburst ring that surrounds the AGN.  Similar to what is observed in most galaxies, [N\,{\sc ii}]\,/\,H$\beta$ has the largest values in the nucleus and outside of the region where [\ion{ Fe}{vii}] is detected.
It is also important to highlight that in NGC\,1068, [Fe\,{\sc vii}]\,/\,H$\beta$ has the highest values measured in the sample, reaching values greater than 0.2 in the central 300~pc. Moreover, we detect a second structure, which starts at 150\,pc and extends up to 500\,pc in the NE direction of the AGN, with values of $\sim$\,0.1. At the edges of this region, the value of the ratio decreases to 0.05. South of the AGN, [Fe\,{\sc vii}]\,/\,H$\beta$ has mean values of 0.03. This region is co-spatial with the maximum extent of [\ion{Fe}{x}] shown in Figure \ref{fig:ext_NGC1068}.

In ESO\,428, the highest values of [O\,{\sc iii}]\,/\,H$\beta$ ($\sim$\,20) are observed at distances greater than 400\,pc NE of the AGN. This very high-ionisation region begins at the NE boundary where [Fe\,{\sc vii}] is no longer detected. Moreover, the [Fe\,{\sc vii}]\,/\,H$\beta$ peak does not coincide with the AGN position but at a $\sim$50~pc SE of it. In the extended region, the values of that ratio is quite homogeneous, between 0.04 and 0.06, accros a region that has a total extension of $\sim$700~pc.

Overall, from the line ratios studied here, we found that regions with a high degree of ionisation are associated with the ionisation cones previously detected in the sample. The comparison of Figures~ \ref{fig:gals_radio1} and~\ref{fig:gals_radio2} with the radio maps shown in Figure~\ref{fig:Fevii_com_radio} allows us to conclude that the highly ionised gas is aligned with the radio jet. Maximum values of [O\,{\sc iii}]\,/\,H$\beta$ and [Fe\,{\sc vii}]\,/\,H$\beta$ are found in the nucleus and along the radio-emitting jet. In contrast, [N\,{\sc ii}]\,/\,H$\beta$ has maximum values in the cross-cone region, perpendicular to the jet propagation direction. Low values of this latter ratio are identified where the [Fe\,{\sc vii}] emission is observed.

The emission of [Fe\,{\sc vii}], in addition to being more compact than that of [O\,{\sc iii}], is considerably more collimated than the latter. Because the coronal lines are not produced by stellar processes, the results show that this emission unambiguously traces the region of highest ionisation within the cone. It is possible that more extended coronal gas could exist. However, deeper observations would be needed to confirm this possibility.

\begin{figure*}
    \includegraphics[width=8.2cm]{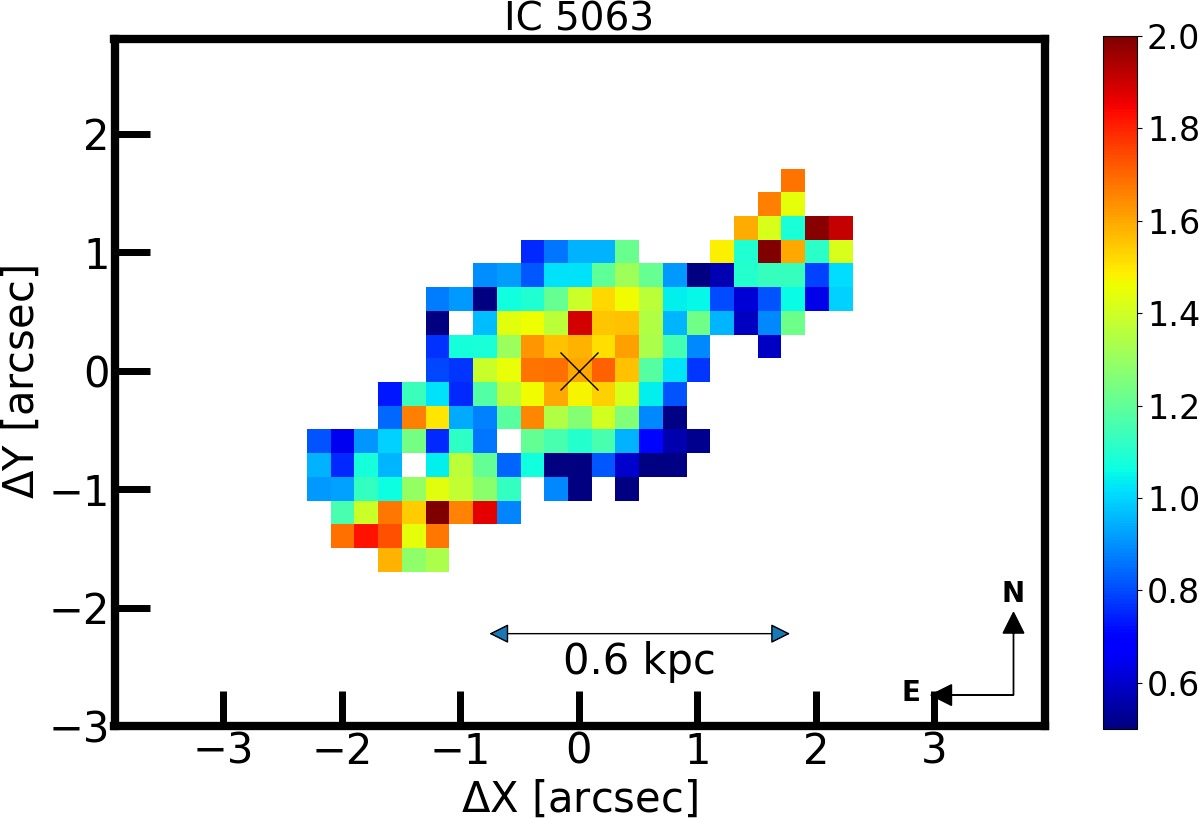}
    \includegraphics[width=8.2cm]{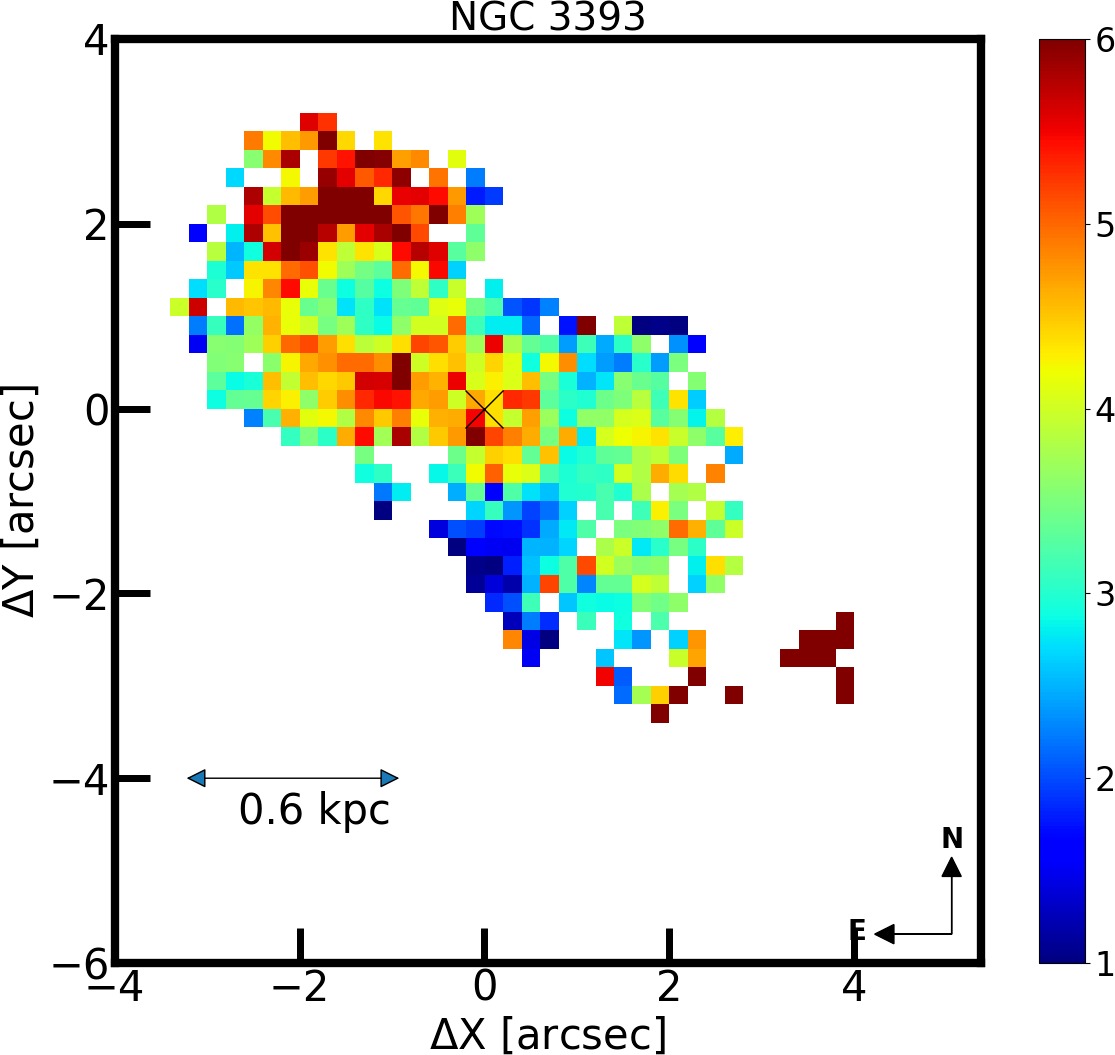}
    \includegraphics[width=8.2cm]{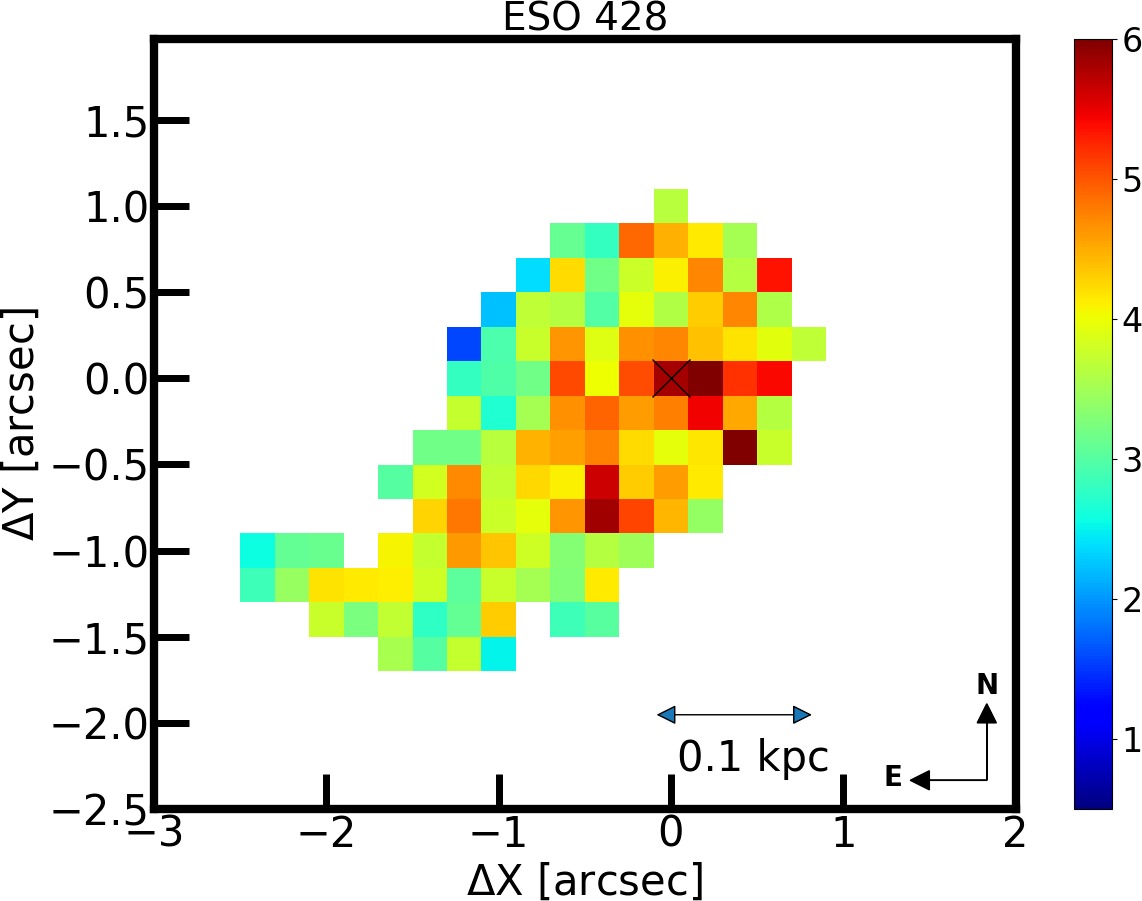}
    \includegraphics[width=8.2cm]{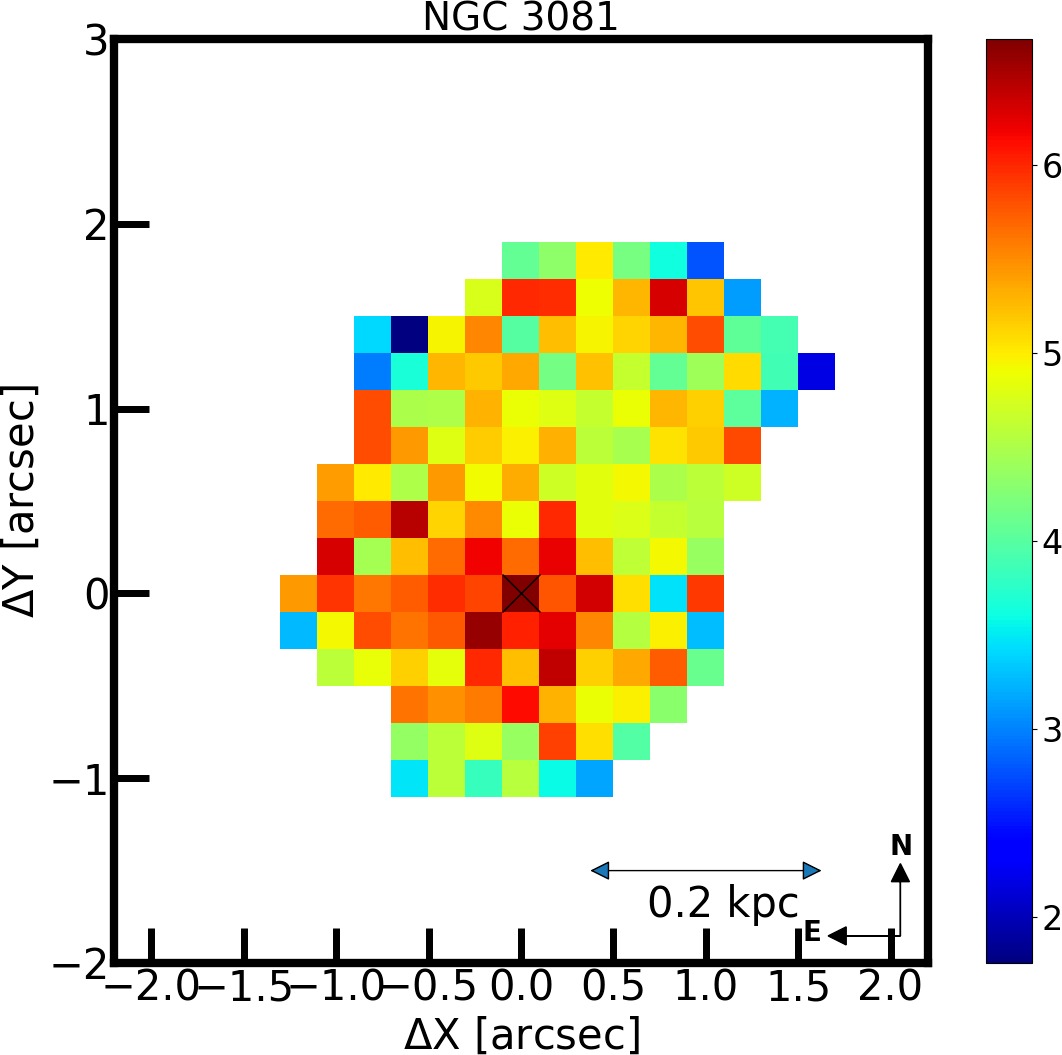}
    \caption{Emission line flux ratio map [\ion{Fe}{vii}/[\ion{Fe}{x}] for the galaxies where both lines are considerably extended. The black cross marks the position of the AGN as determined by the peak of the [\ion{Fe}{x}] emission.} 
    \label{fig:fe7pfe10}
\end{figure*}

We also constructed ionisation maps by means of the line flux ratio [\ion{Fe}{vii}]/[\ion{Fe}{x}] for the galaxies where both lines are considerably extended (IC\,5063, NGC\,3393, ESO\,428, and  NGC\,3081). The results are shown in Figure~\ref{fig:fe7pfe10}. It can be seen the increase of the ionization state of the gas when moving outwards from the active centre. For instance, in NGC\,3393 (top right panel), at the regions where the radio-jet is impacting the ISM gas, the line ratio drops to values between 2 $-$ 4 while at the nucleus and the regions surrounding the jet, values between 4 and 6 are measured. Similarly, in IC\,5063 (top left panel) the regions close the off-nuclear peaks in the extended radio emission that ratio reaches values $<$1, implying [\ion{Fe}{x}] stronger than [\ion{Fe}{vii}]. Finally, In ESO\,428, the relative contribution of the extended emission goes from 42\% in [\ion{Fe}{vii}] to 52\% in [\ion{Fe}{x}], given futher support to the scenario proposed.  

\section{Conclusions}
\label{sec:final}

We have studied the corona line region of nine local Seyfert 2 galaxies with previous evidence of displaying bright and extended (up to a few hundred of parsecs scale) coronal emission, aimed at assessing its size, morphology and mechanisms contributing to its formation. To this purpose, deep MUSE/VLT data is employed. The coronal lines of [\ion{Fe}{vii}]~$\lambda$6087 (IP = 99 eV) and [\ion{Fe}{x}]~$\lambda$ (IP=235 eV) are taken as proxy of that emission. The main results are as follows.

\begin{itemize}
  
\item We have clearly resolved the coronal line region in all objects of the sample. The [\ion{Fe}{vii}] emission region reaches linear extensions ranging from several hundred parsecs to a few kiloparsecs. The [\ion{Fe}{x}] emission is extended to several hundred of parsecs to 1.3 kpc from the AGN. These results represent a major step forward in the determination of the true size of the CLR as in some cases the extension of the coronal gas was found larger by an order of magnitude relative to previous results. In all objects studied, the coronal emission is strongly aligned along the direction of the radio jet. No evidence of coronal gas extended in the direction perperdicular to that of the jet propagation was found, while low-ionisation gas indeed show strong emission in that latter direction.

\item We found that the decrease of CL flux with distance is not as sharp as expected in a central source ionisation scenario. Secondary peaks of emission along the radial direction are observed, with the decline in emission flux being $<$1 dex in the inner 500 pc. NGC 3393 is yet more extreme as the emission flux peaks at $\sim$500 pc from the centre. At 1~kpc away from the AGN the flux measured is half the one at the nucleus. This result is indicative of either strongly varying physical conditions with distance (i.e., the gas density decreases sharply outside the nuclear region) or there are additional sources of gas ionisation.

\item The excellent spatial coincidence of the CL emission and the radio-jet suggests that jet-driven shocks enhances the observed coronal line emission. These shocks can be originated in enhanced radio regions associated with knots in the jet, where shock re-acceleration of relativistic particles transported by the jet occur as a consequence of density changes through the  medium as the jet propagates. 

\item We investigate the relative contribution of the nuclear and the extended CL emission flux to the total flux observed.  For [\ion{Fe}{vii}], we found that the extended emission varies from 29\% to 80\%, while for [\ion{Fe}{x}] it ranges from 5\% to 80\%. Overall, we notice that the coronal line region is extended-dominated. A comparion with a small sample of Type~I sources shows the opposite, being all sources nuclear-dominated. Overall, we interpret this result in terms of jets that may deposit a
significant amount of energy in the off-nuclear region. We notice that in the objects where this effect is the largest, extended X-ray emission has also been detected, with models pointing out to a collisional gas component, likely attributed to shocks.  

\item We studied, for the first time in the literature, the relationship between the radio-jet size and the jet-power with coronal line properties such as the size and luminosity of the coronal emission. In all cases, a positive trend is observed, suggesting a strong role of the radio-jet in shaping the extended coronal line emission in AGN. Compact jets are strongly associated to compact CLR while extended, more powerful jets, are associated to more luminous and extended CLR.   

\end{itemize}

Our work demonstrates the importance of the CLR in any study that seeks to explain the physical conditios of the NLR gas. The high-excitation emission gas is a fundamental ingredient that allows the study of additional sources of gas excitation not always detectable by means of low-ionisation lines. The inclusion of more coronal lines in other wavelength intervals like the ones in the NIR or the MIR is a must in order to understandand that emission.

\section*{Acknowledgements}

ARA acknowledges Conselho Nacional de Desenvolvimento Científico e Tecnológico (CNPq) for partial support to this work under grant 313739/2023-4. LGDH acknowledges support by National Key R\&D Program of China No.2022YFF0503402, and National Natural Science Foundation of China (NSFC) project number E345251001. AP acknowledges support from  Spain I+D+i   PID2020-114092GB-I00. RAR acknowledges the support from Conselho Nacional de Desenvolvimento Cient\'ifico e Tecnol\'ogico (CNPq; Proj. 303450/2022-3, 403398/2023-1, \& 441722/2023-7), Funda\c c\~ao de Amparo \`a pesquisa do Estado do Rio Grande do Sul (FAPERGS; Proj. 21/2551-0002018-0), and Coordena\c c\~ao de Aperfei\c coamento de Pessoal de N\'ivel Superior (CAPES;  Proj. 88887.894973/2023-00). This research has made use of the NASA/IPAC Extragalactic Database, which is funded by the National Aeronautics and Space Administration and operated by the California Institute of Technology. We thank to the anonymous referee for his/her useful comments and suggestions to improve this manuscript.

\section*{Data Availability}

The  data  underlying  this  article  are  available  in  the  European Southern Observatory  archive.  The  data  can  be  obtained  in  raw quality through the MUSE raw data query form (\url{http://archive.eso.org/wdb/wdb/eso/muse/form}). Science quality data can be obtained from the Spectral data products query form (\url{http://archive.eso.org/wdb/wdb/adp/phase3spectral/form?collectionname=MUSE}).



\bibliographystyle{mnras}
\bibliography{referencia} 




\appendix

\section{On-line material}

\begin{figure*}
    \includegraphics[width=17cm]{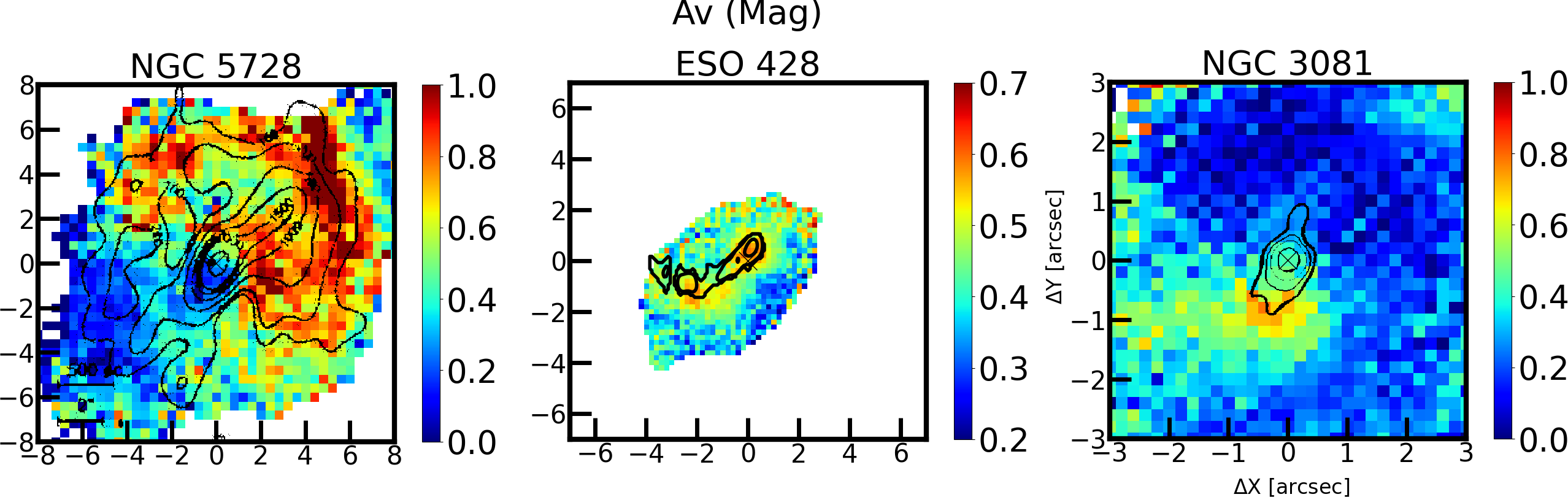}
    \caption{Extinction map for NGC\,5728, ESO\,428 and NGC\,3081. The black contours is the observed radio emission. }
    \label{figAped:Av_NGC5728_ESO428_NGC3081}
\end{figure*}

\begin{figure*}
    \centering
    \includegraphics[width=17cm]{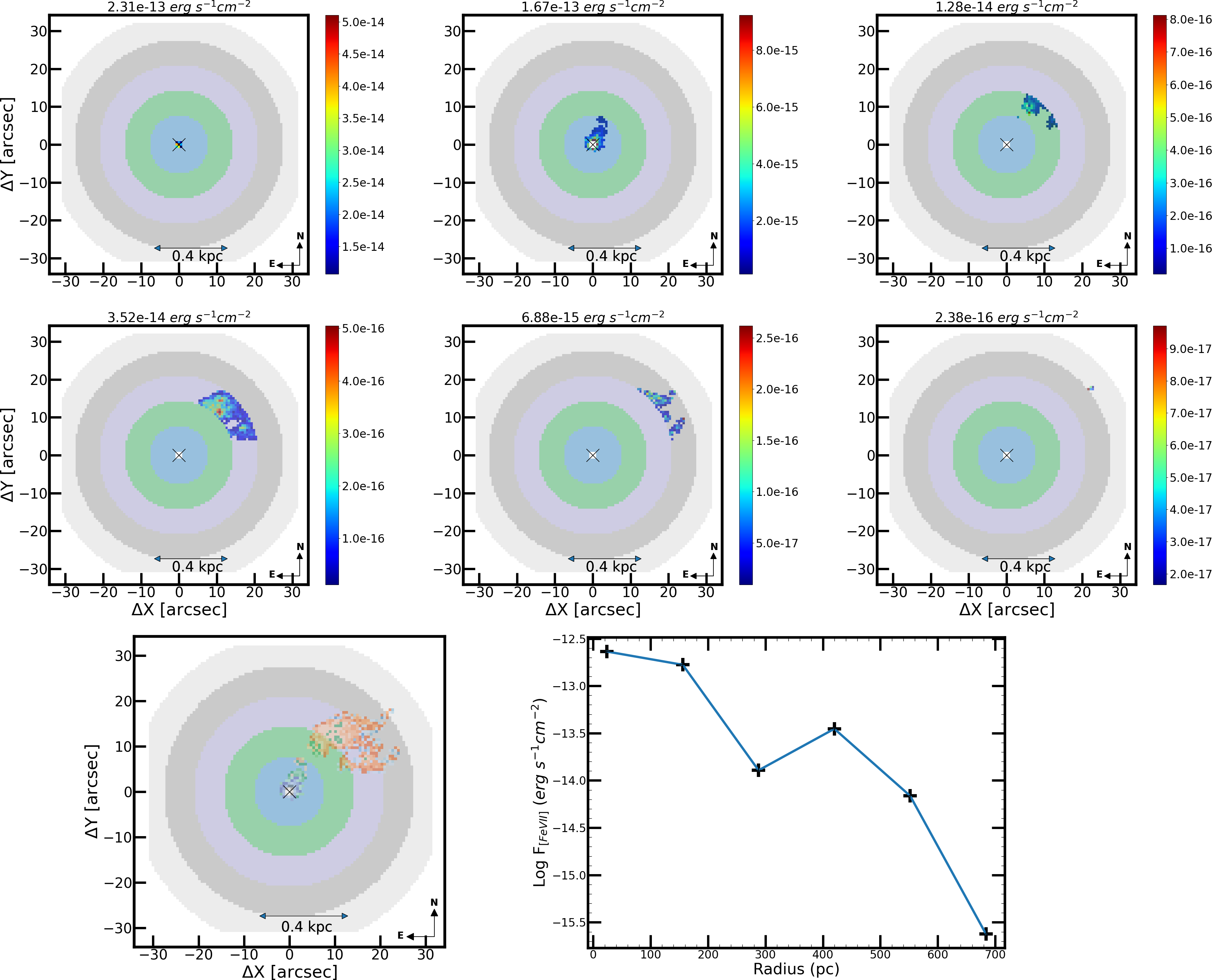}  
    \caption{Radial flux distribution similar to that of  Figure~\ref{fig:rings_N3393} and applied to the [\ion{Fe}{vii}] emission in Circinus. The individual panels in the first two rows show the flux within the nuclear region and each concentric ring. The number at the top of each panel shows the integrated emission. The left bottom panel is the total [\ion{Fe}{vii}] emission overlaid to the nuclear region and rings. The bottom right panel is the resulting flux vs radius profile.}
    \label{fig:rings_Circinus}
\end{figure*}

\begin{figure*}
    \centering
    \includegraphics[width=17cm]{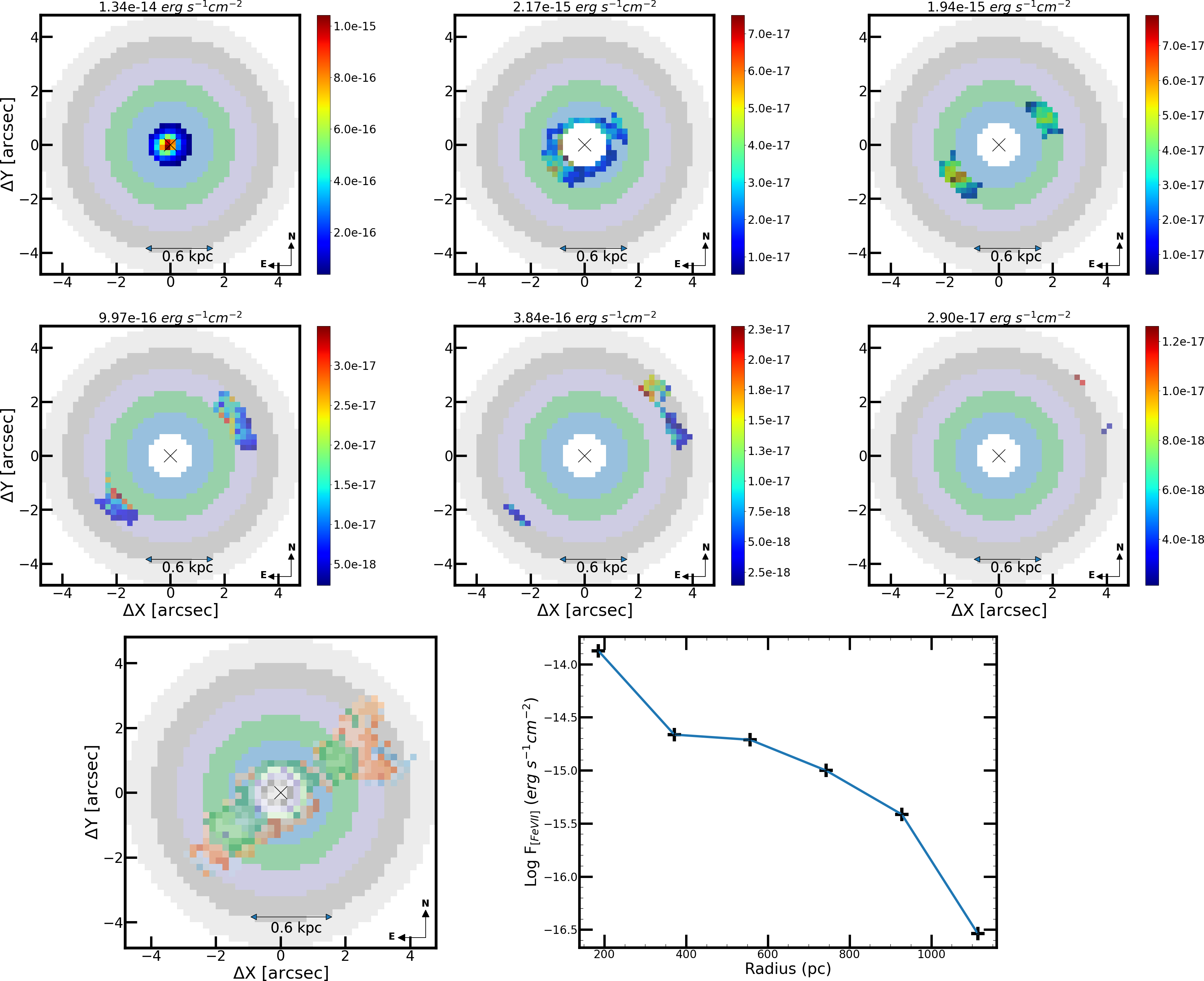}  
    \caption{Same as Figure~\ref{fig:rings_Circinus} but now for the IC~5063.}
    \label{fig:rings_IC5063}
\end{figure*}

\begin{figure*}
    \centering
    \includegraphics[width=17cm]{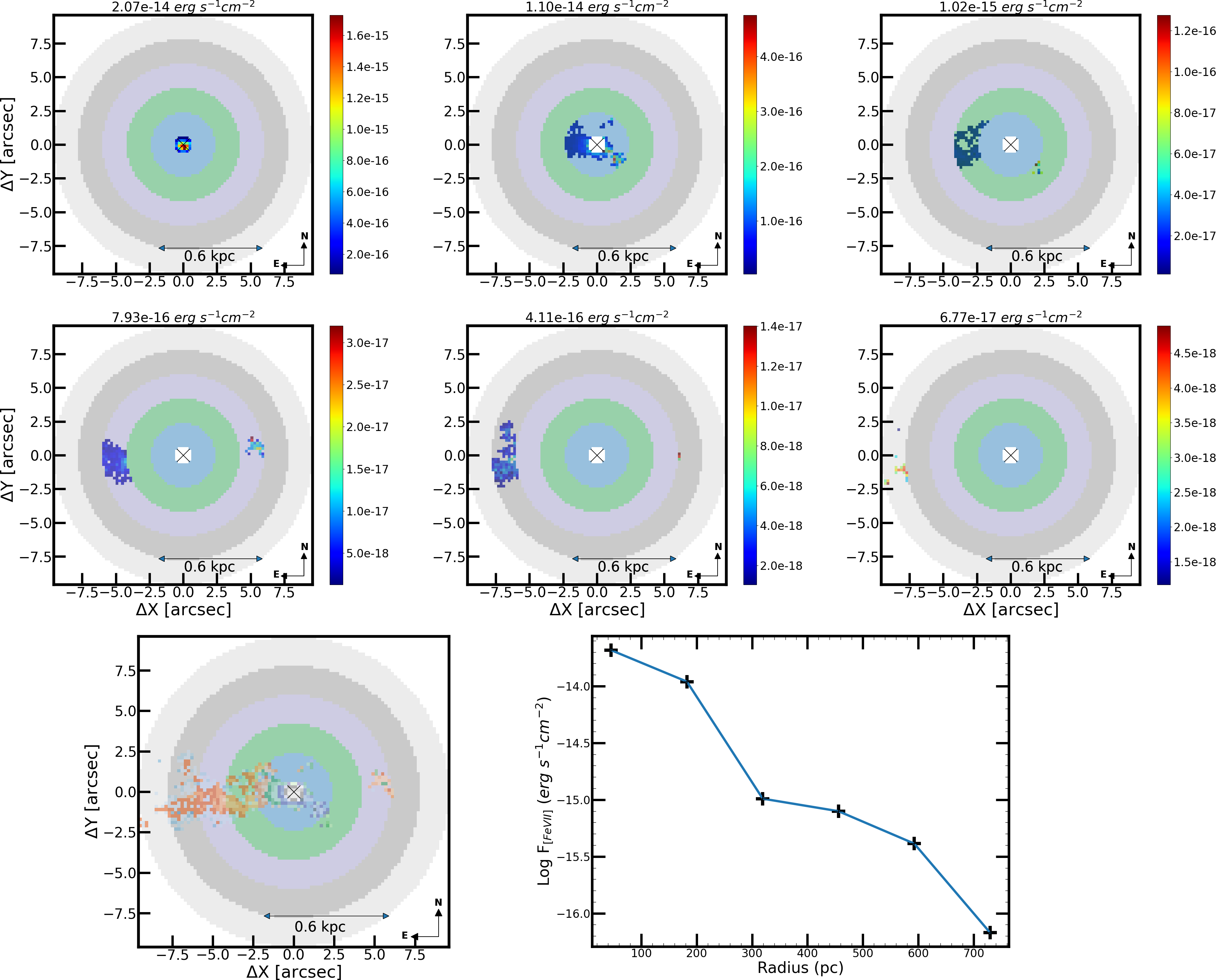}  
    \caption{Same as Figure~\ref{fig:rings_Circinus} but now for the NGC~5643.}
    \label{fig:rings_NGC5643}
\end{figure*}

\begin{figure*}
    \centering
    \includegraphics[width=17cm]{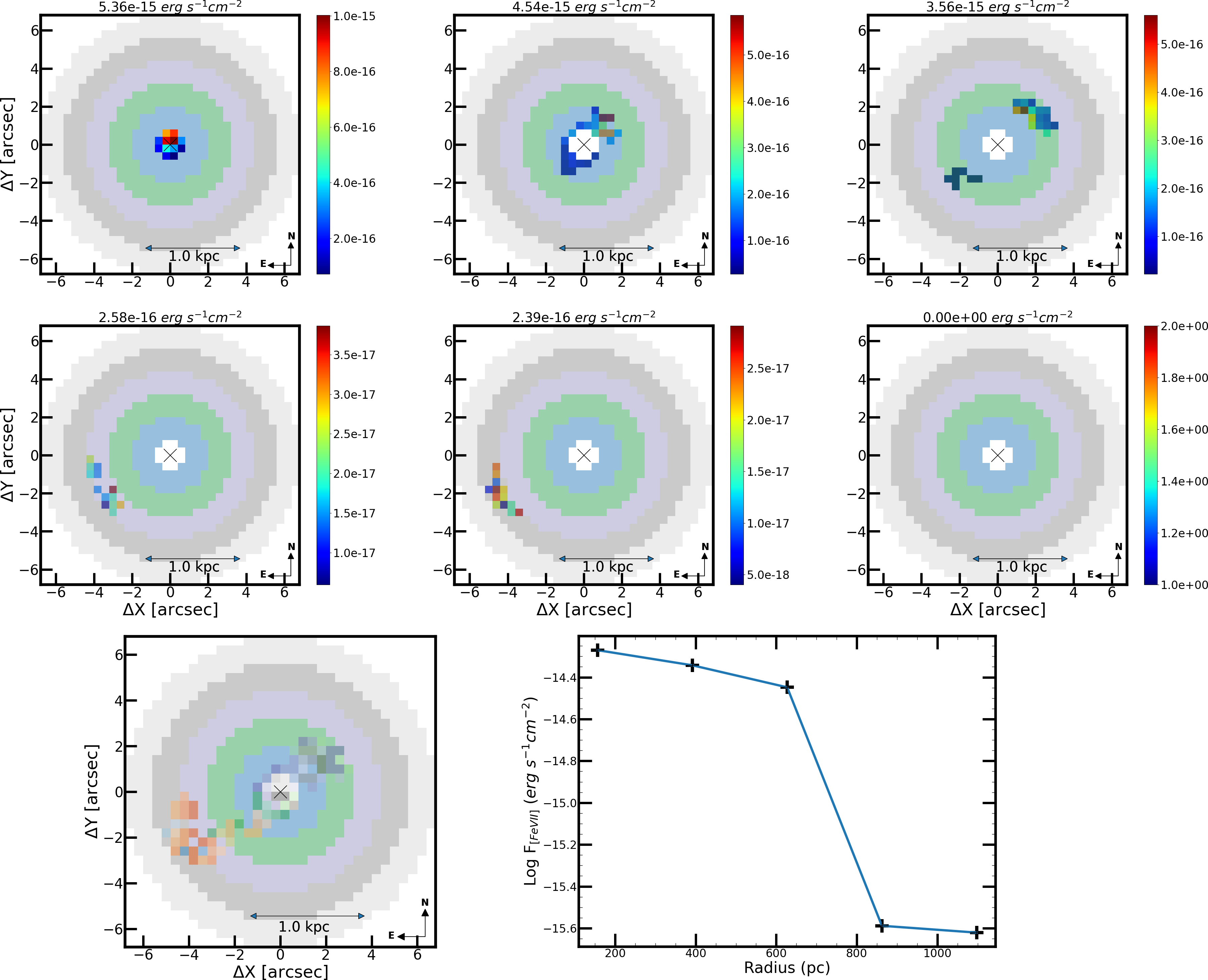}  
    \caption{Same as Figure~\ref{fig:rings_Circinus} but now for the NGC~5728.}
    \label{fig:rings_NGC5728}
\end{figure*}

\begin{figure*}
    \centering
    \includegraphics[width=17cm]{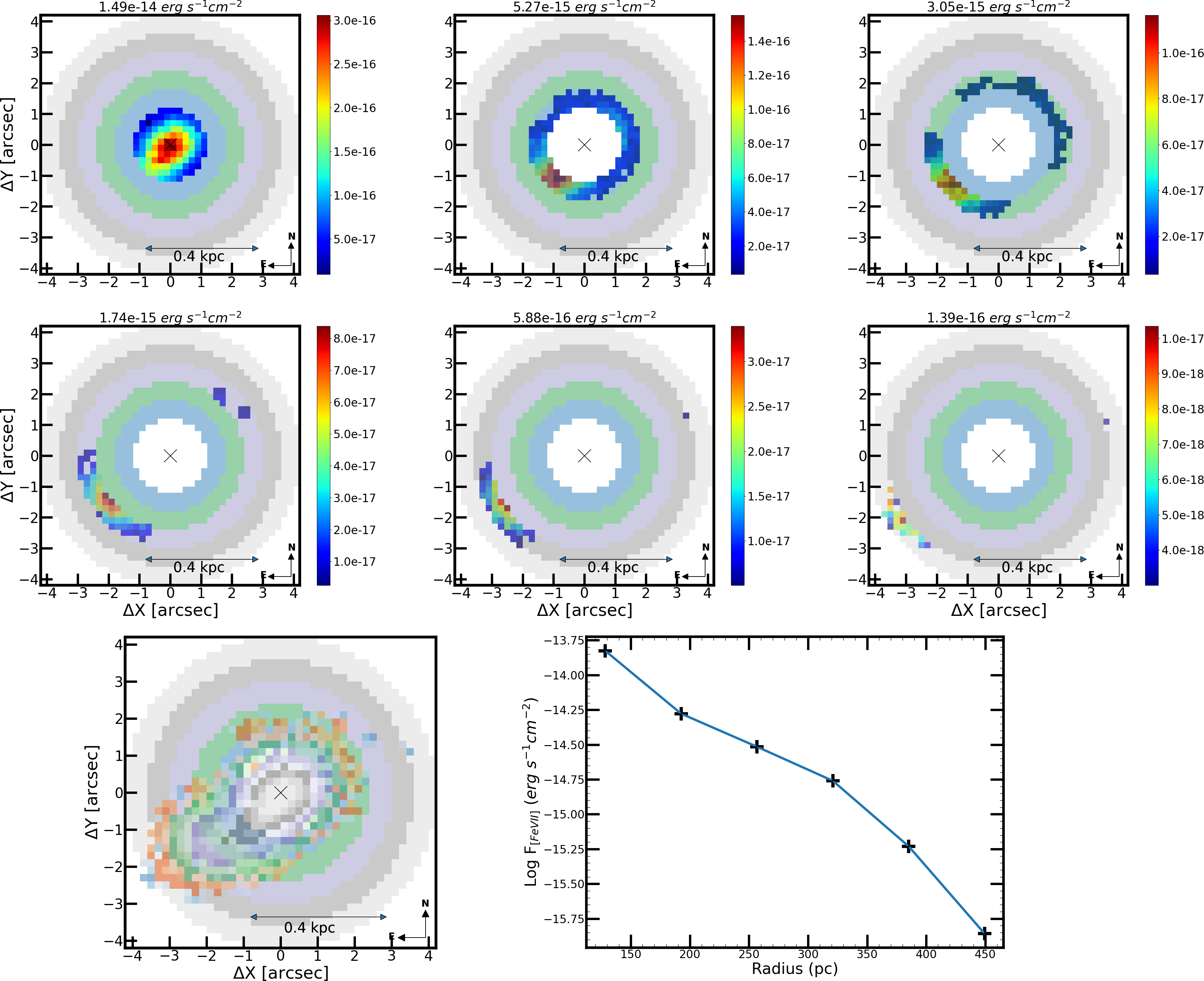}  
    \caption{Same as Figure~\ref{fig:rings_Circinus} but now for the ESO~428.}
    \label{fig:rings_ESO428}
\end{figure*}

\begin{figure*}
    \centering
    \includegraphics[width=17cm]{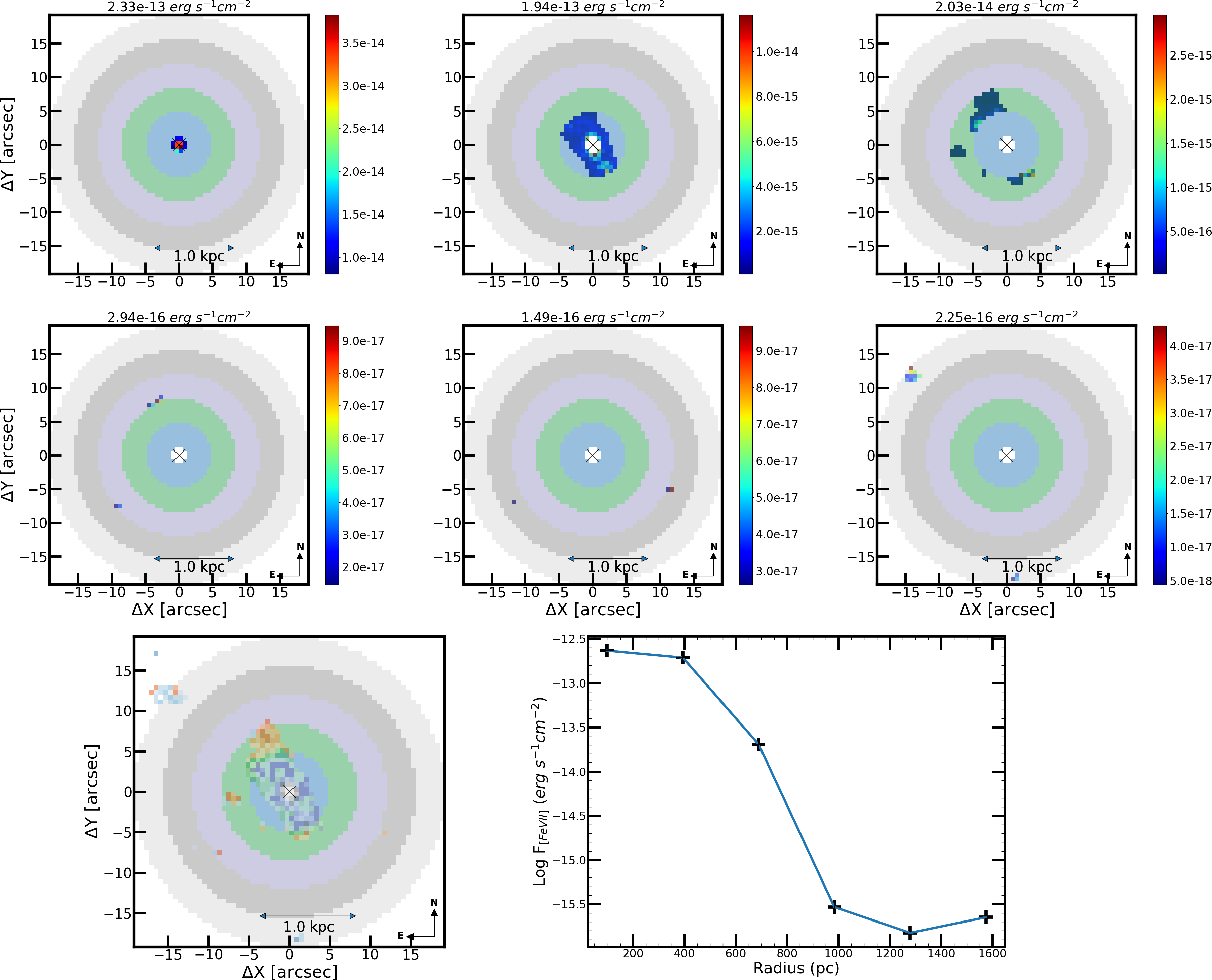}  
    \caption{Same as Figure~\ref{fig:rings_Circinus} but now for the NGC~1068.}
    \label{fig:rings_NGC1068}
\end{figure*}

\begin{figure*}
    \centering
    \includegraphics[width=17cm]{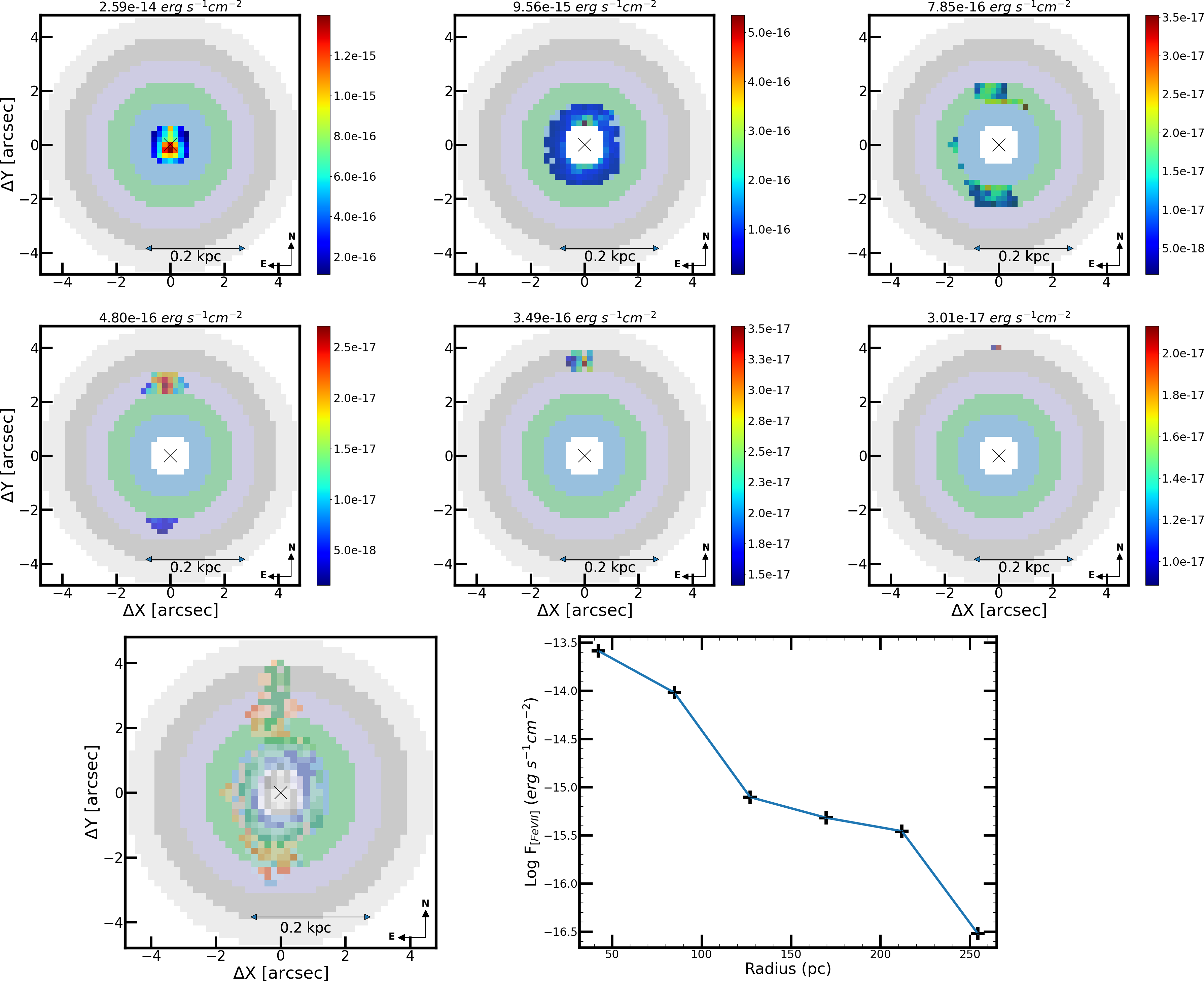}  
    \caption{Same as Figure~\ref{fig:rings_Circinus} but now for the NGC~1386.}
    \label{fig:rings_NGC1386}
\end{figure*}

\begin{figure*}
    \centering
    \includegraphics[width=17cm]{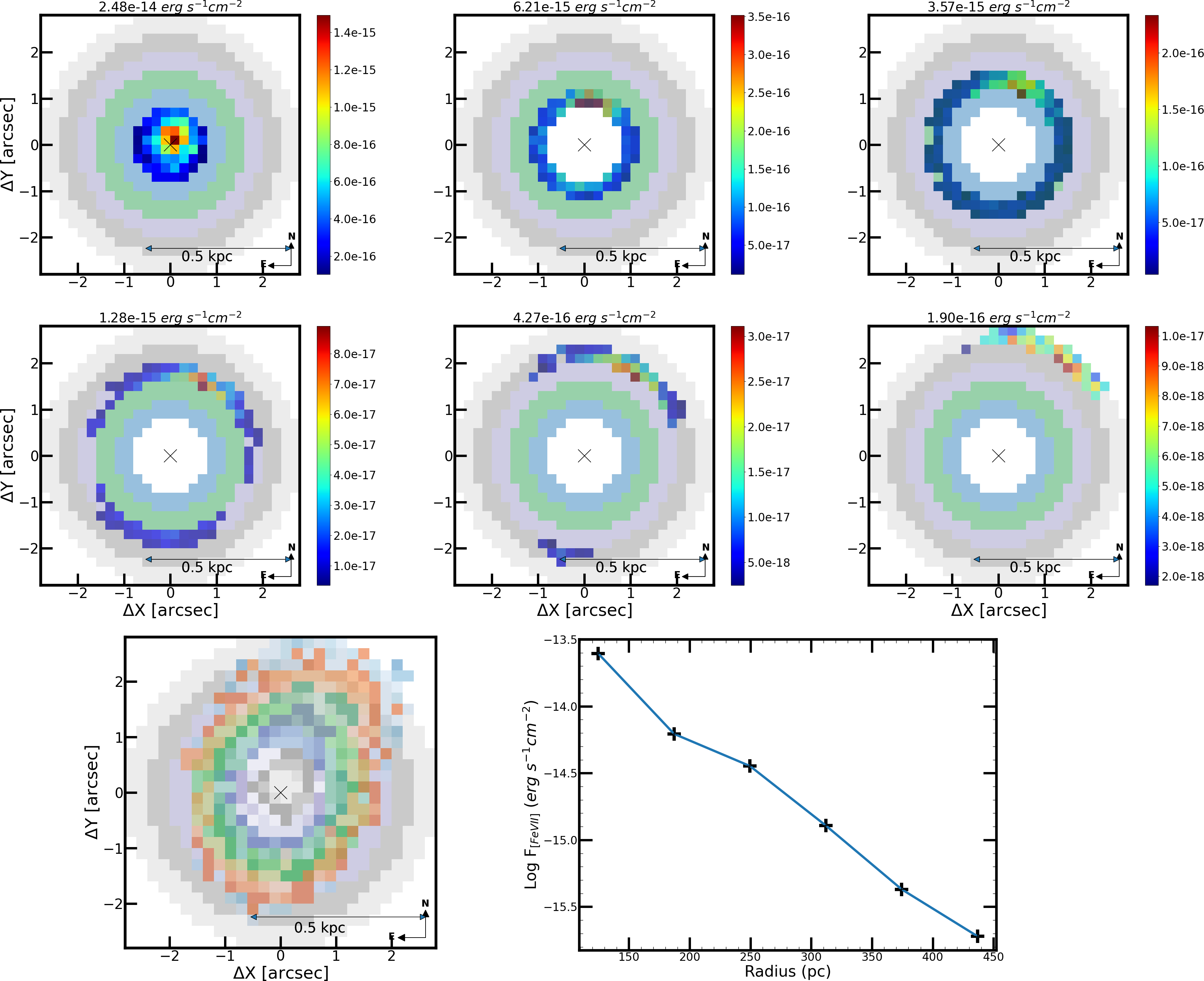}  
    \caption{Same as Figure~\ref{fig:rings_Circinus} but now for the NGC~3081.}
    \label{fig:rings_NGC3081}
\end{figure*}


\begin{figure*}
    \centering
    \includegraphics[width=17cm]{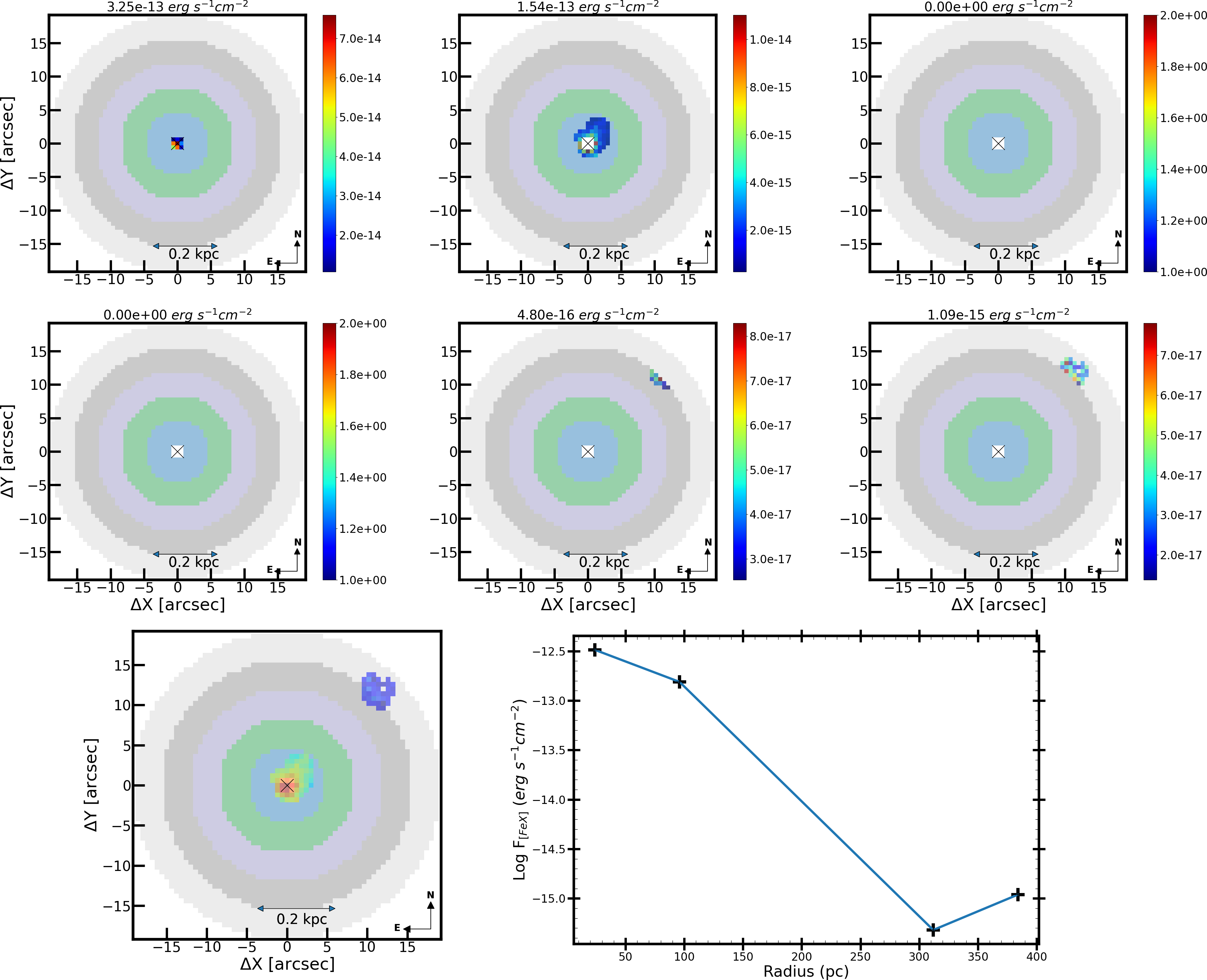}  
    \caption{Same as Figure~\ref{fig:rings_Circinus} but now for the [Fe\,{\sc X}] emission-line in Circinus.}
    \label{fig:rings_FEX_Circinus}
\end{figure*}

\begin{figure*}
    \centering
    \includegraphics[width=17cm]{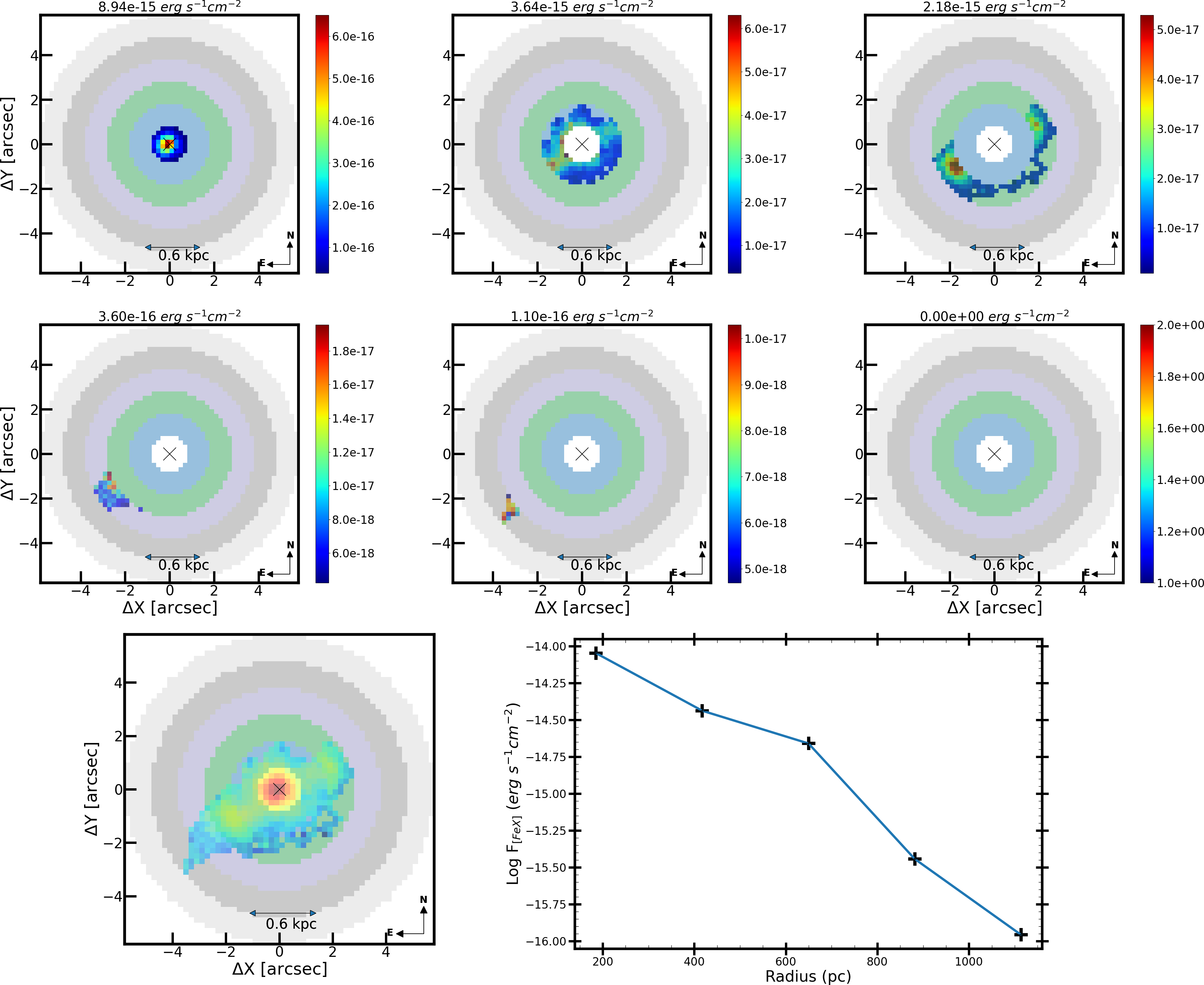}  
    \caption{Same as Figure~\ref{fig:rings_FEX_Circinus} but now for the IC~5063.}
    \label{fig:rings_FEX_IC5063}
\end{figure*}

\begin{figure*}
    \centering
    \includegraphics[width=17cm]{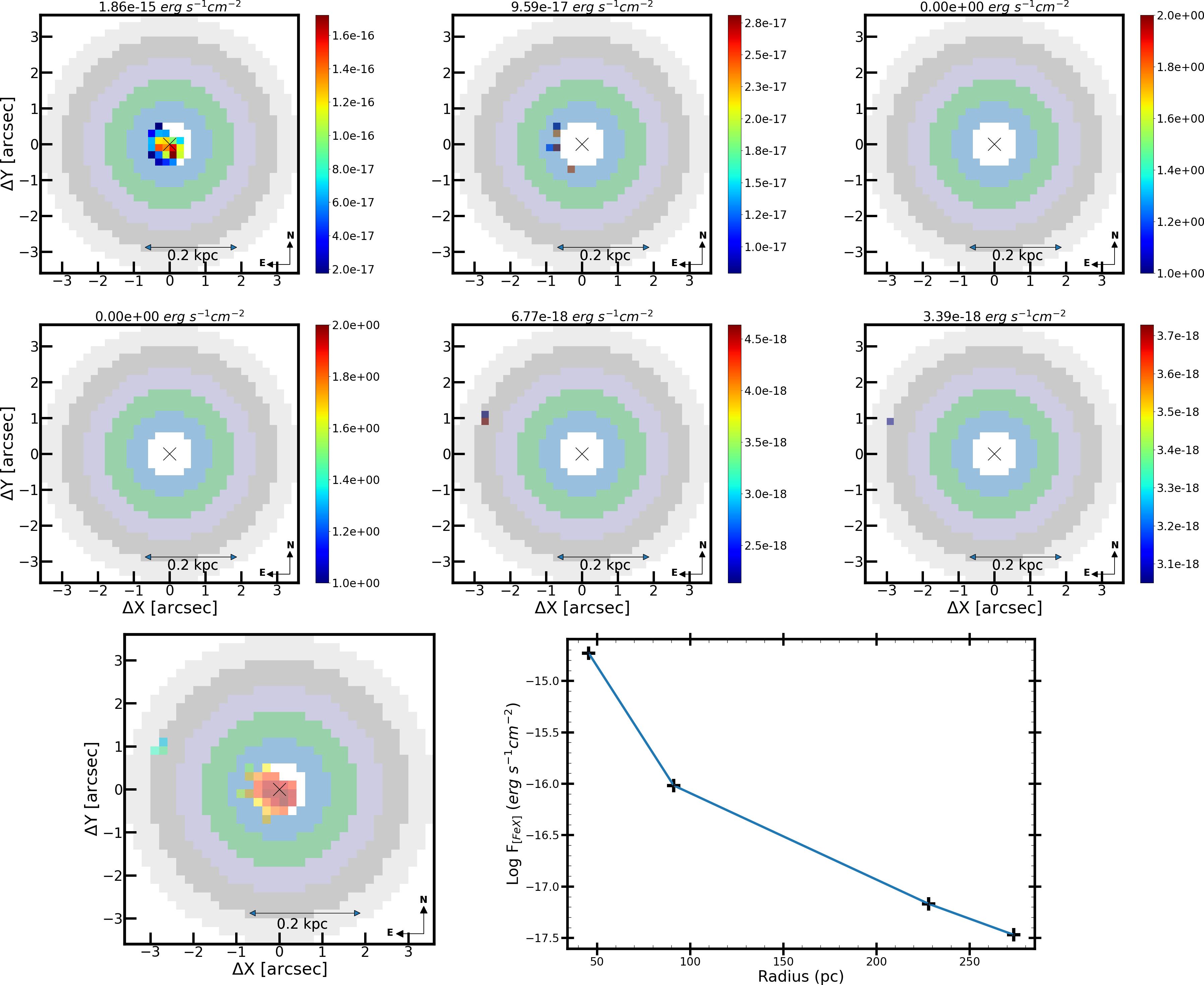}  
    \caption{Same as Figure~\ref{fig:rings_FEX_Circinus} but now for the NGC~5643.}
    \label{fig:rings_FEX_NGC5643}
\end{figure*}

\begin{figure*}
    \centering
    \includegraphics[width=17cm]{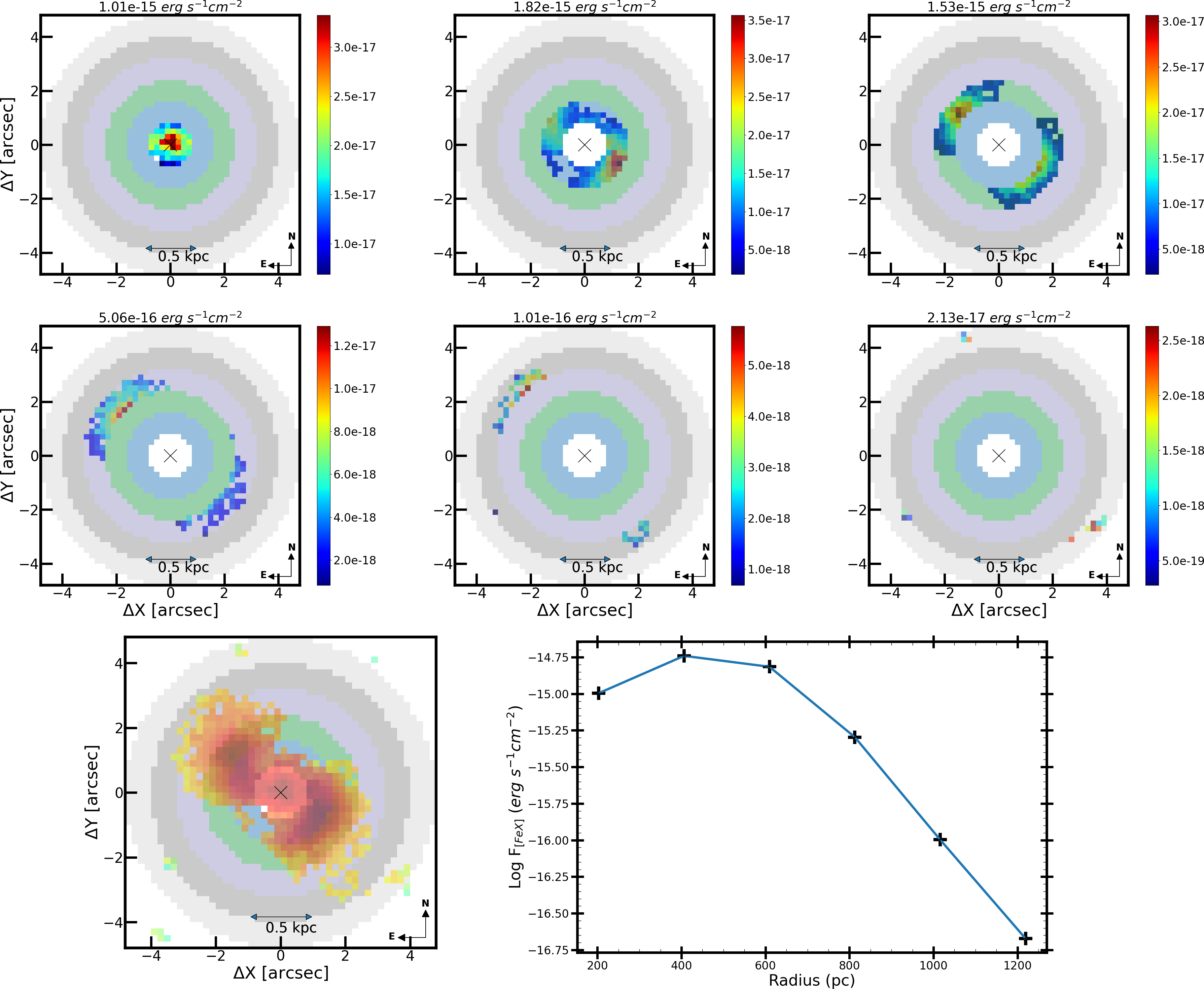}  
    \caption{Same as Figure~\ref{fig:rings_FEX_Circinus} but now for the NGC~3393.}
    \label{fig:rings_FEX_NGC3393}
\end{figure*}

\begin{figure*}
    \centering
    \includegraphics[width=17cm]{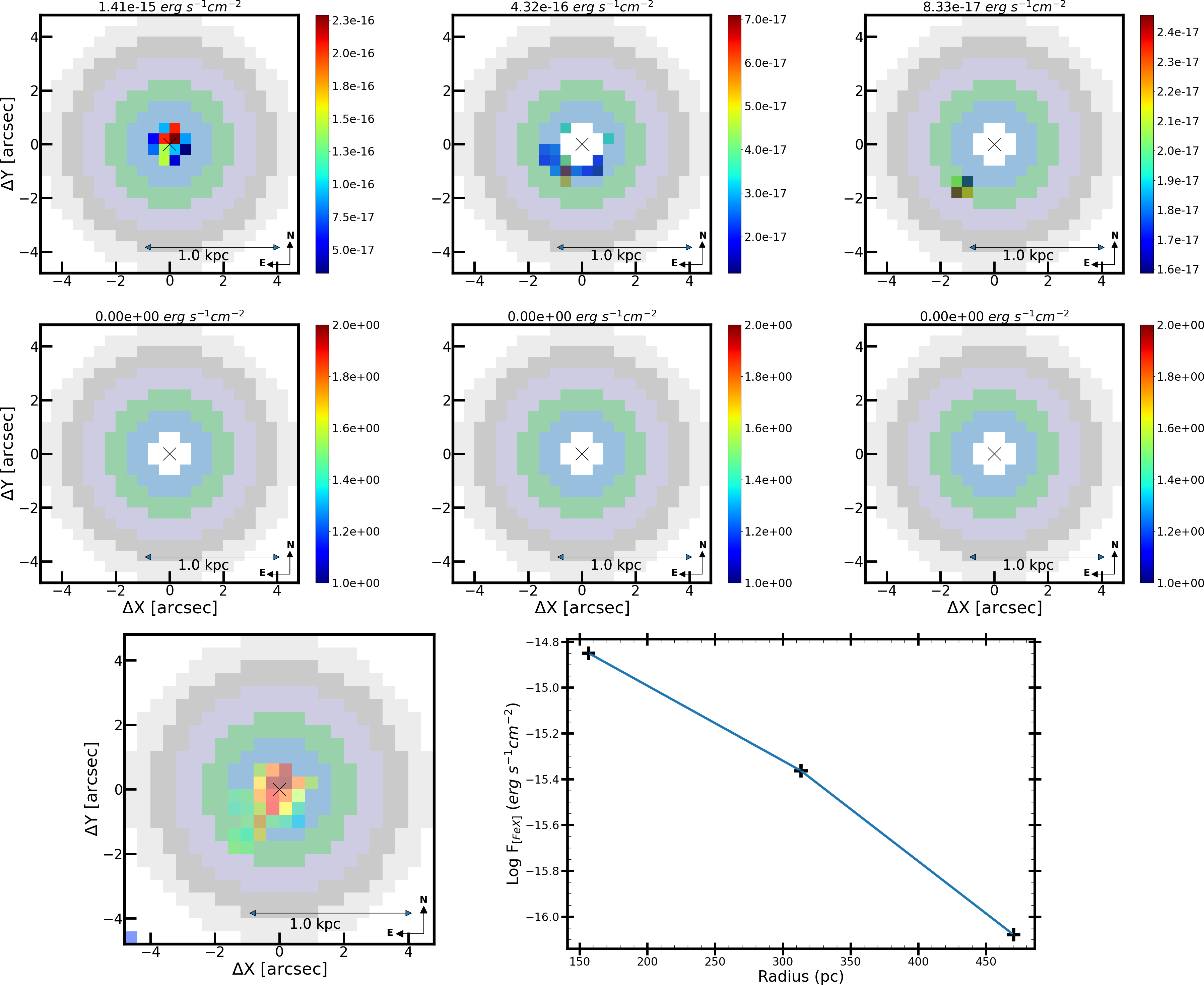}  
    \caption{Same as Figure~\ref{fig:rings_FEX_Circinus} but now for the NGC~5728.}
    \label{fig:rings_FEX_NGC5728}
\end{figure*}

\begin{figure*}
    \centering
    \includegraphics[width=17cm]{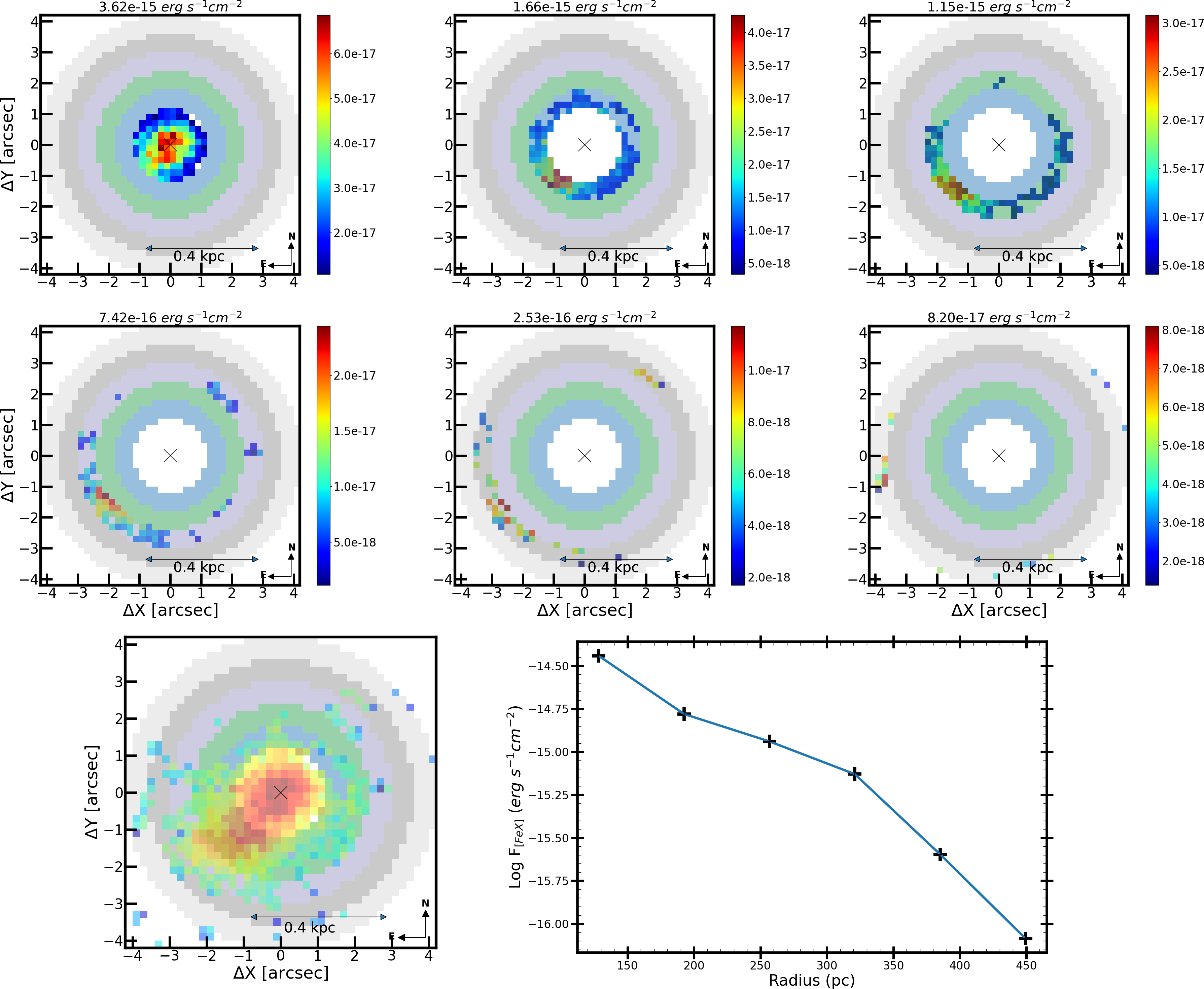}  
    \caption{Same as Figure~\ref{fig:rings_FEX_Circinus} but now for the ESO~428.}
    \label{fig:rings_FEX_ESO428}
\end{figure*}

\begin{figure*}
    \centering
    \includegraphics[width=17cm]{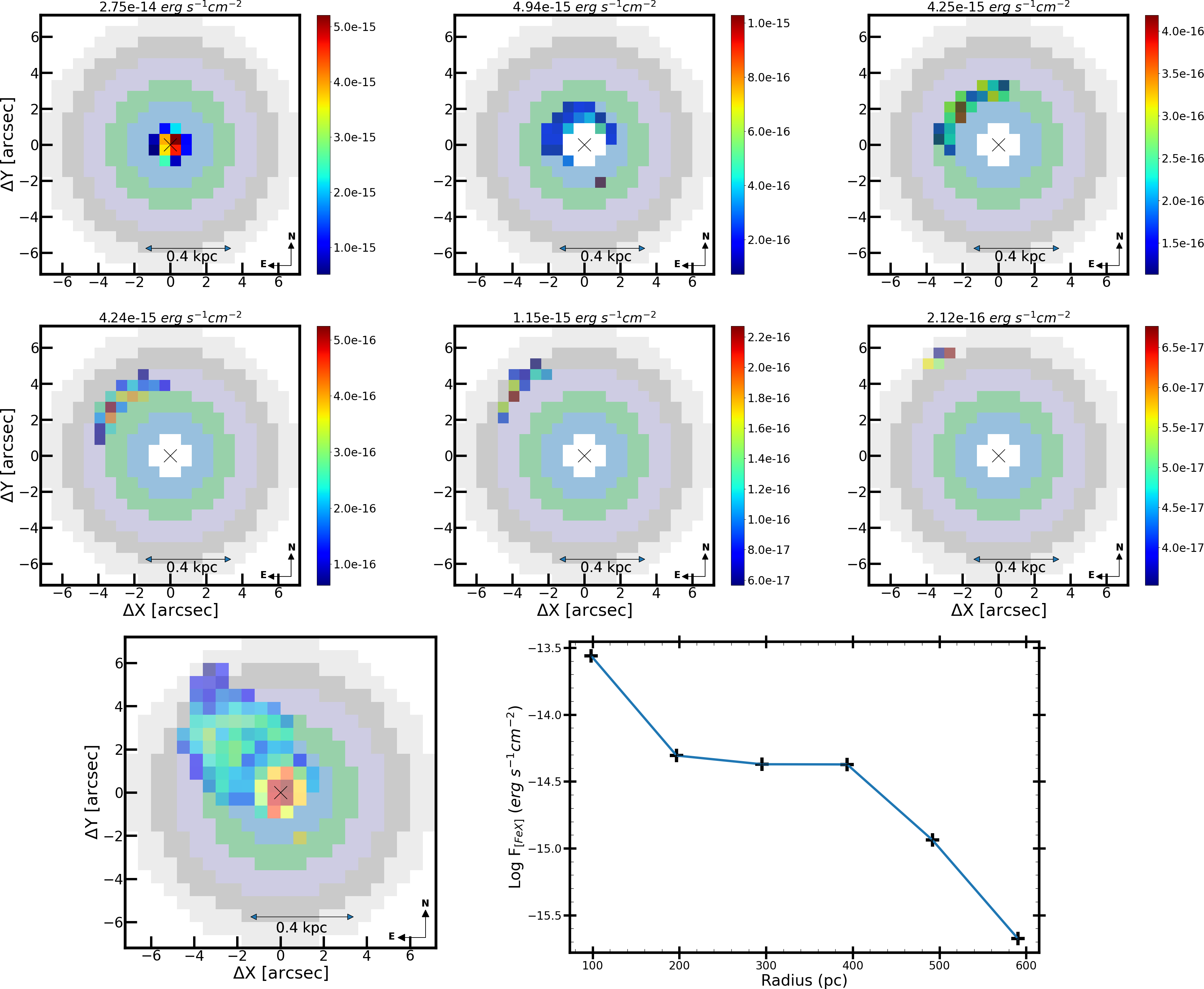}  
    \caption{Same as Figure~\ref{fig:rings_FEX_Circinus} but now for the NGC~1068.}
    \label{fig:rings_FEX_NGC1068}
\end{figure*}

\begin{figure*}
    \centering
    \includegraphics[width=17cm]{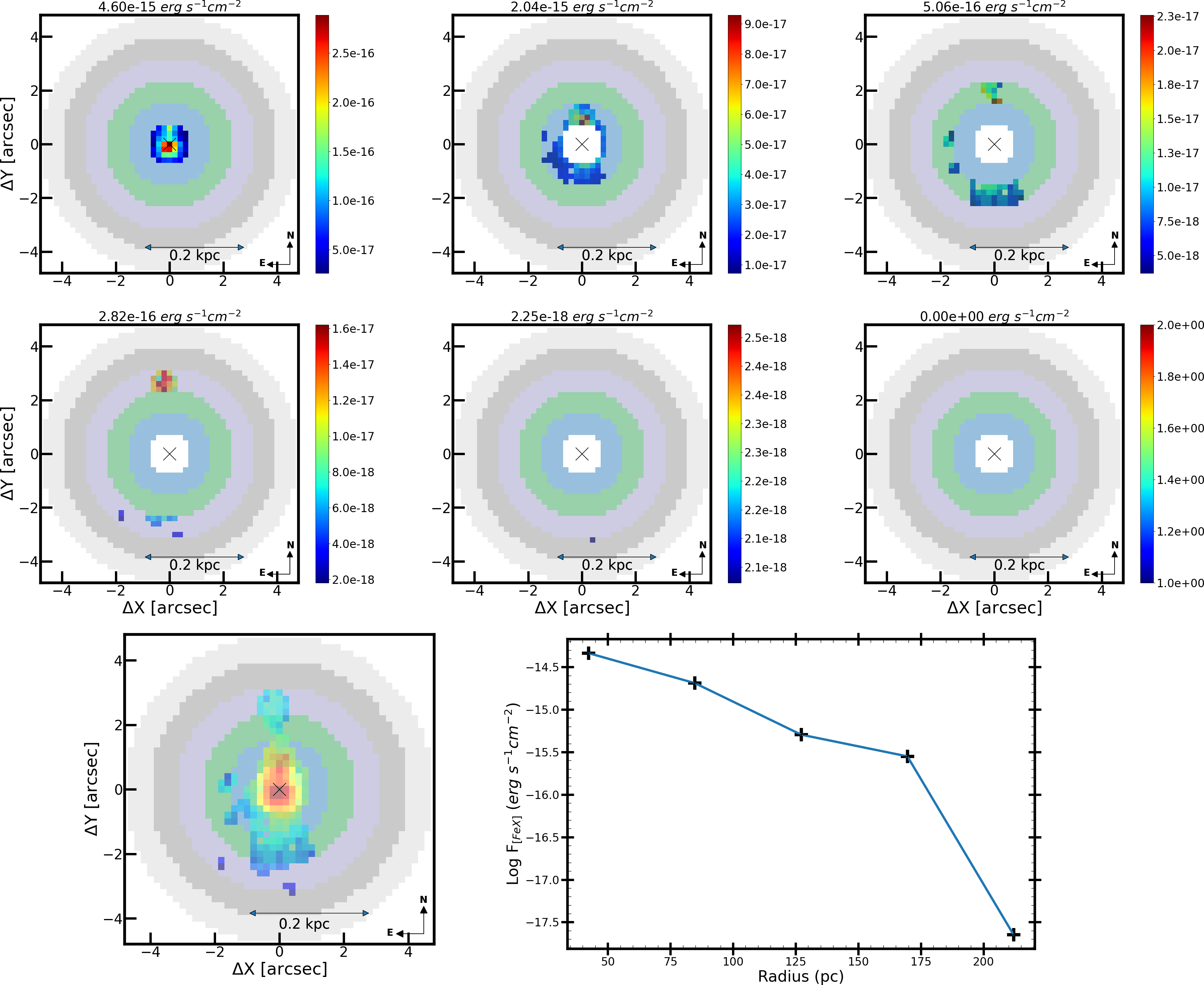}  
    \caption{Same as Figure~\ref{fig:rings_FEX_Circinus} but now for the NGC~1386.}
    \label{fig:rings_FEX_NGC1386}
\end{figure*}

\begin{figure*}
    \centering
    \includegraphics[width=17cm]{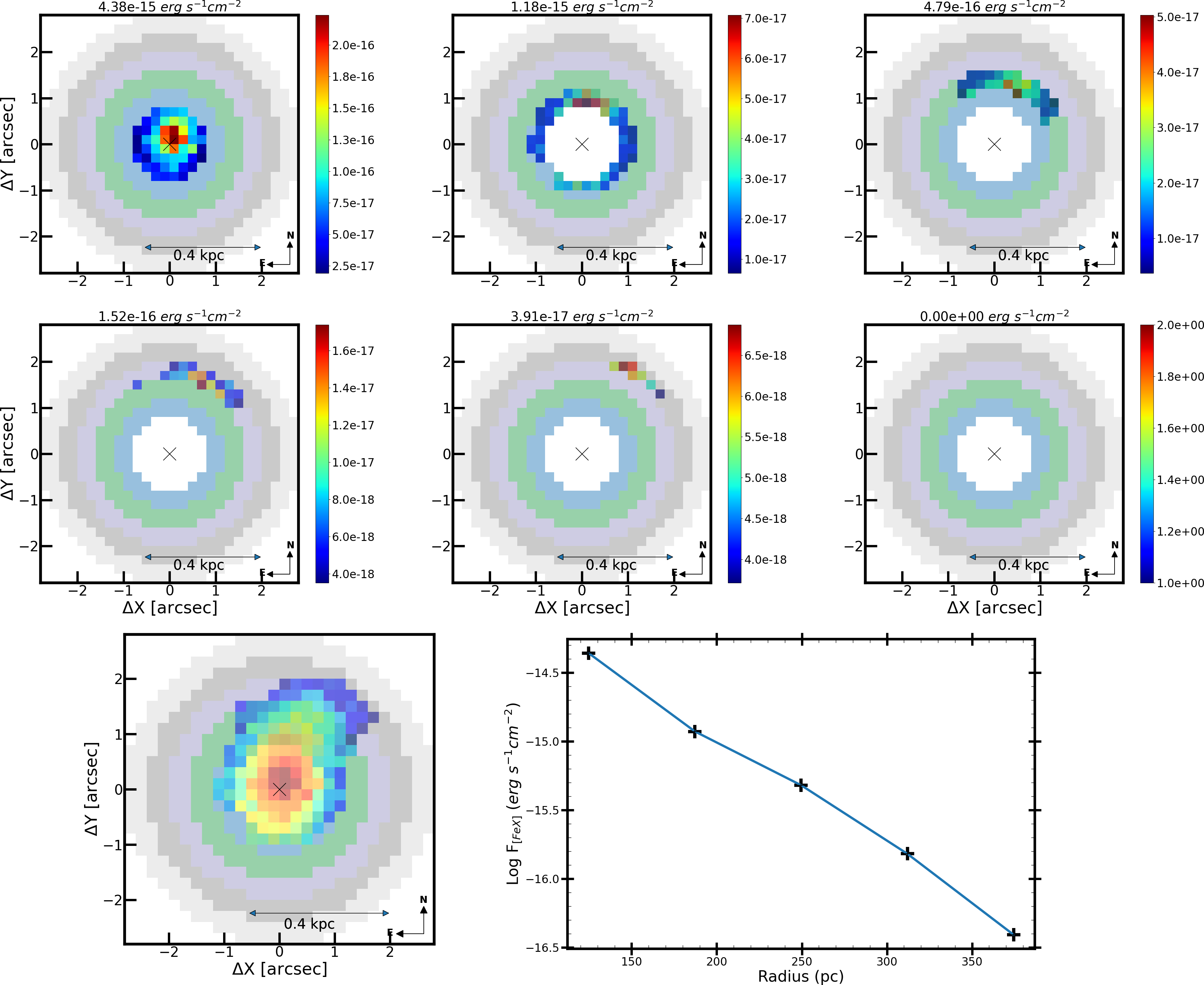}  
    \caption{Same as Figure~\ref{fig:rings_FEX_Circinus} but now for the NGC~3081.}
    \label{fig:rings_FEX_NGC3081}
\end{figure*}


\bsp	
\label{lastpage}
\end{document}